\documentclass[colorlinks=true,linkcolor=red,citecolor=blue]{article}
\usepackage{threeparttable}

\usepackage[utf8]{inputenc}
\usepackage[T1]{fontenc}
\usepackage[table]{xcolor}  
\usepackage{graphicx}
\usepackage{amsmath,amssymb,amsfonts,bm}
\usepackage{booktabs,multirow,multicol,makecell,threeparttable}
\usepackage{longtable,rotating,adjustbox}
\usepackage{caption,subcaption,placeins}
\usepackage{enumitem,setspace,microtype,url,doi}
\usepackage{natbib}

\usepackage{arxiv}
\renewcommand{\shorttitle}{\footnotesize Calibrated Bayesian inference for overall and subgroup-specific causal effects in SW-CRTs}

\usepackage{tikz}
\usetikzlibrary{decorations.pathreplacing, positioning}

\usepackage[utf8]{inputenc} % allow utf-8 input
\usepackage[T1]{fontenc}    % use 8-bit T1 fonts
\usepackage{url}            % simple URL typesetting
\usepackage{booktabs}       % professional-quality tables
\usepackage{amsfonts}       % blackboard math symbols
\usepackage{nicefrac}       % compact symbols for 1/2, etc.
\usepackage{microtype}      % microtypography
\usepackage{lipsum}		% Can be removed after putting your text content
\usepackage{subcaption} % for \subfigure support
\usepackage{makecell} % for \makecell in tables
\usepackage{graphicx}
\usepackage{natbib}
\usepackage{doi}
\usepackage{rotating}   
\usepackage{booktabs}
\usepackage{import}
\usepackage{bm,amsmath,multicol,multirow,amssymb,comment}
\usepackage{graphicx}
\usepackage{longtable,setspace}
\usepackage{enumitem}
\usepackage{caption,placeins,subcaption}
\usepackage{adjustbox}

\newcommand{\E}{\mathbb{E}}
\newcommand{\Var}{\mathbb{V}{\rm ar}}
\newcommand{\I}{\mathbb{I}}

\newtheorem{theorem}{Theorem}%[theorem]

\usepackage{xcolor} 
\usepackage{subcaption}
\usepackage[table]{xcolor}
\definecolor{lightgray}{gray}{0.9}

\title{Calibrated Bayesian inference for overall and subgroup-specific causal effects in stepped-wedge cluster-randomized trials}

\date{\mbox{}}

\author{ {\hspace{1mm} Ruyi Liu} \\
	Department of Biostatistics, Yale School of Public Health, \\
    New Haven, CT, USA\\
	\AND
    {\hspace{1mm} Yuki Ohnishi} \\
	Department of Statistics, University of Florida, \\
    Gainesville, FL, USA\\
	\AND
    {\hspace{1mm} Xi Fang} \\
	Data Science Institute, Medical College of Wisconsin, \\
    Milwaukee, WI, USA\\
	\AND
    {\hspace{1mm} Michael J. Daniels} \\
	Department of Statistics, University of Florida, \\
    Gainesville, FL, USA\\
	\AND
    {\hspace{1mm} Fan Li}\thanks{Corresponding author: Fan Li, Department of Biostatistics, Yale School of Public Health. Email: \href{mailto:fan.f.li@yale.edu}{fan.f.li@yale.edu}.} \\
    Department of Biostatistics, Yale School of Public Health, \\
    New Haven, CT, USA\\
}

\allowdisplaybreaks 

\begin{document}
\maketitle

\newpage
\begin{abstract}
Stepped-wedge cluster-randomized trials (SW-CRTs) randomly assign clusters to sequences that determine when each cluster switches from control to intervention. Conventional Bayesian analyses of SW-CRTs report treatment effect coefficients from a working outcome model, but under model misspecification or informative cluster-period sizes, such coefficients can lack a clear causal interpretation. In this work, we develop Bayesian methods targeting weighted average treatment effect estimands in SW-CRTs, including novel subgroup-specific causal estimands defined by baseline characteristics. We compare model-based g-computation with model-robust standardization, implementing both with linear mixed models and mixed-effects Bayesian additive regression trees as working outcome models, and propose a calibrated variance estimator that augments posterior parameter uncertainty with cluster-level resampling to better capture sampling variability. In simulations with outcome model misspecification and informative cluster-period sizes, model-robust standardization with calibrated inference exhibits negligible bias and near-nominal frequentist coverage for overall and subgroup-specific causal effects, whereas g-computation and uncalibrated intervals become insufficient. We illustrate the proposed methods with a reanalysis of the Trauma Survivors Outcomes and Support trial.
\end{abstract}

\keywords{Bayesian causal inference \and Bayesian additive regression trees \and estimands \and informative cluster-period sizes \and stepped-wedge designs \and subgroup analysis}

\section{Introduction}
\subsection{Background}
Stepped-wedge cluster-randomized trials (SW-CRTs) are a one-way crossover design in which clusters are randomized to different intervention roll-out sequences \citep{turner2017review_p1}. All clusters, such as hospitals, clinics, or community health centers, begin under the control condition and then cross over to intervention according to a randomized, staggered schedule, until all clusters have crossed over \citep{hussey2007design}. This design is often used as an alternative to parallel-arm cluster-randomized trials (CRTs) in health services research and implementation science, because it accommodates phased implementation, such as when training, resource allocation, or infrastructure changes make simultaneous roll-out infeasible, and can be preferable when withholding a potentially beneficial intervention is undesirable \citep{hemming2020reflection}. At the same time, SW-CRTs are analytically more complicated than parallel-arm designs. Because more clusters are exposed in later periods, treatment status is inherently related to calendar time. Valid inference therefore depends critically on how secular trends, within-cluster dependence, and treatment effect heterogeneity across cluster-periods are handled \citep{li2021mixed}. These analytic challenges are closely tied to the broader emphasis on estimands in clinical trials, that is, explicit definitions of the treatment effect parameter a study aims to estimate \citep{iche9r1}. Recently, the CRT-Estimands Framework has extended this perspective to CRTs by emphasizing that the target treatment effect must be specified in relation to the unit of inference \citep{kahan2026crtestimands,kahan2026crtJAMA}. For example, a parallel-arm CRT may be interested in the average effect for the population of individuals, or the average effect for the population of clusters \citep{kahan2023estimands,kahan2024demystifying,bi2025scoping}. This perspective is also relevant for SW-CRTs, where the staggered roll-out, secular trends, and cluster-period structure create multiple plausible estimands that may be targeted by different analysis strategies.

SW-CRTs make the estimands issue more complex because they introduce the cluster-period as a potential unit of observation. In an SW-CRT, outcomes are collected across multiple cluster-periods, and treatment status changes over time within each cluster. Therefore, an average treatment effect can be defined in different ways. One can average over individuals, over clusters, over periods, or over cluster-periods. These different choices can lead to different causal effect definitions, especially when cluster-period sizes are informative \citep{chen2025model,fang2026model}, which means that the number of individuals observed in a cluster-period is associated with the potential outcomes or with the treatment effect within each cluster-period. Recent methodological work has formalized these ideas for SW-CRTs. For example, under different aggregation choices, one can define individual-level, cluster-level, period-level, and cluster-period-level average treatment effects \citep{chen2025model,fang2026model,lee2025estimated}. In particular, \citet{fang2026model} provided a classification of these estimands according to horizontal and vertical aggregation, leading to four estimands: horizontal-individual, horizontal-cluster, vertical-individual, and vertical-cluster effects. These names encode two scientific decisions. The first decision is whether individuals or clusters are the primary units being averaged (the unit of inference). The second decision is how calendar periods enter the aggregation. These more refined definitions of estimands are useful because they force the analysis to match the scientific question, rather than allowing the target effect to be ambiguously determined by a chosen regression model.

In practice, however, the most common analyses of SW-CRTs rely on generalized linear mixed models (GLMMs) and generalized estimating equations (GEEs) \citep{nevins2024adherence}, and then interpret the treatment coefficient as the effect of interest \citep{li2021mixed}. This practice may raise interpretational challenges. First, a treatment coefficient from a working model does not, by itself, specify whether the target effect is individual-level, cluster-level, period-level, or cluster-period-level. In the most concerning case where cluster-period sizes are informative, this ambiguity leads to interpretation challenges because different aggregation choices can correspond to different causal quantities \citep{kahan2024demystifying,lee2025estimated}, some of which may not be easy to decipher. An analogous issue is well documented in parallel-arm CRTs, where the individual-average and cluster-average treatment effects can diverge under informative cluster sizes, and the regression coefficient from a mixed model targets neither one \citep{kahan2023estimands}. Second, even when the estimand is prespecified, standard GLMM- and GEE-based estimators may be biased for that estimand if the outcome mean structure, secular trend, or treatment effect heterogeneity is misspecified; GLMMs additionally rely on assumptions about the random-effects distribution. This concern is especially important when cluster-period sizes are informative. Because regression-based analyses may implicitly assign greater weight to larger cluster-periods, model misspecification can cause the treatment coefficient to converge to a size-weighted effect that differs from the intended estimand \citep{fang2026model}.

\subsection{Objectives and contributions}

The aforementioned issues are salient in trials such as the Trauma Survivors Outcomes and Support (TSOS) study, an SW-CRT conducted at 25 Level I trauma centers across the United States to evaluate stepped collaborative care for post-traumatic stress disorder (PTSD) symptoms and comorbidity \citep{zatzick2021stepped}. Between January 2016 and November 2019, trauma centers were sequentially randomized to transition from enhanced usual care to the intervention at staggered time points. A total of 635 hospitalized injury survivors reporting elevated PTSD symptoms on the PTSD Checklist-Civilian Version (PCL-C) were enrolled and followed after injury. Several features of the trial bring the preceding issues into focus. Cluster sizes varied considerably, from 7 to 40 patients per trauma center, raising the prospect that the individual-weighted and cluster-weighted estimands differ. Patients also differed substantially in baseline risk, including PTSD severity, injury mechanism, and the number of risk factors for enduring PTSD, and the treatment effect may well vary across subpopulations. Indeed, the original trial reported larger symptom reductions among patients with three or more baseline risk factors \citep{zatzick2021stepped}. That original analysis relied on a mixed-effects regression for repeated measures within individuals and clustering of individuals within sites, reporting a treatment coefficient whose causal interpretation depends on the working model and whose implied weighting across individuals and cluster-periods was not pre-specified. The same concern also extends to pre-specified subgroup analysis. These features make TSOS a useful setting for illustrating how estimand choice and outcome model specification may affect the estimation and interpretation of overall and subgroup-specific causal effects in SW-CRTs.

Several recent works have emerged to address the estimands questions in CRTs and SW-CRTs. Within the frequentist paradigm, for parallel-arm CRTs, \citet{su2021model} developed model-assisted estimators under a finite-population framework, and \citet{li2025model} proposed model-robust standardization procedures that achieve consistency under arbitrary working model misspecification. For SW-CRTs, \citet{chen2025model} introduced weighted average treatment effect estimands under a finite-population framework and proposed model-assisted analysis of covariance estimators. More recently, \citet{fang2026model} developed a super-population framework for weighted estimands in SW-CRTs and proposed model-robust standardization methods compatible with a broad class of working outcome models, including linear mixed models, GLMMs, and GEEs, where the fitted outcome model is used only as a working model, rather than as the basis for defining the causal effect. Treatment effect heterogeneity in SW-CRTs has so far been addressed mainly at the design stage, where \citet{li2024planning} derived variance formulas for planning trials powered to detect treatment effect heterogeneity. These prior developments are based on the frequentist paradigm and focus on overall causal effects or on model-based interaction terms, but do not address subgroup-specific causal effect estimands. Within the Bayesian paradigm, \citet{liu2025calibrated} recently developed a calibrated procedure for estimand-aligned analysis of parallel-arm CRTs, targeting the cluster-average and individual-average treatment effects under informative cluster sizes through combining model-robust standardization with a cluster-level resampling correction. This work, however, is confined to parallel-arm CRTs and overall causal effect estimands. Estimand-aligned Bayesian inference for SW-CRTs, including inference targeting subgroup-specific causal effects, remains unexplored to date.

Several considerations motivate the development of estimand-aligned Bayesian inferential strategies for SW-CRTs such as TSOS. First, SW-CRTs can include a limited number of clusters. Frequentist analyses of CRTs with few clusters typically rely on small-sample corrections whose performance can be sensitive to the modeling choices \citep{leyrat2018cluster}, whereas Bayesian approaches can stabilize estimation by incorporating prior information on fixed effects, random effects, and variance components \citep{grantham2022evaluating}. Second, under a correct model formulation, the joint posterior distribution provides a coherent framework for uncertainty quantification and a natural probability-based interpretation of results \citep{jones2021bayesian}, including straightforward interval estimation for complex quantities such as variance components. Third, Bayesian nonparametric methods such as Bayesian additive regression trees (BART) \citep{chipman_bart_2010} offer a flexible approach for modeling nonlinear conditional mean functions, and have been shown to effectively capture nonlinear covariate effects and treatment effect heterogeneity in causal inference applications \citep{hill2011bayesian, dorie2022stan}. Recent extensions incorporating random effects further accommodate multilevel designs \citep{spanbauer2021nonparametric}, making them potentially attractive for SW-CRTs (although this has not been explored in prior work). This flexibility is especially relevant for subgroup analysis, where treatment effect modification may be complex and where correctly specifying all treatment-by-covariate interactions in a linear mixed model can be no easy task. Existing Bayesian analyses of CRTs and SW-CRTs have largely been model-based rather than estimand-based, often reporting posterior summaries of treatment coefficients from a specified outcome regression model. Moreover, when a Bayesian outcome model is used only as a working model, its posterior distribution characterizes uncertainty under that model and need not capture the full repeated sampling variability of an estimand-aligned estimator, particularly under model misspecification. Consequently, nominal posterior credible intervals may fail to attain nominal frequentist coverage, even when the corresponding point estimator is approximately unbiased \citep{antonelli2022causal,li2023bayesian}. This gap motivates us to develop a calibrated Bayesian approach that combines posterior modeling with a resampling correction for sampling variability. Such an approach retains the modeling flexibility of Bayesian inference while improving its frequentist operating characteristics for estimand-aligned analysis of SW-CRTs.

Our contributions are therefore as follows. First, building on the general class of weighted average treatment effect estimands for SW-CRTs recently formalized by \citet{fang2026model}, we expand this framework to subgroup-specific causal estimands defined by baseline characteristics. Second, we develop a Bayesian framework that accommodates both model-based g-computation and model-robust standardization approaches. Our framework allows investigators to leverage hierarchical modeling frameworks, including conventional linear mixed models and mixed-effects BART, while maintaining estimand-aligned inference even under model misspecification or informative cluster-period sizes. Third, we propose a calibrated variance estimation strategy that integrates posterior parameter uncertainty with cluster-level resampling, ensuring that Bayesian credible intervals also achieve valid frequentist coverage. Finally, we conduct extensive simulation studies, comparing model-based g-computation and model-robust standardization approaches under both calibrated and uncalibrated posterior inferential strategies, targeting various weighted estimands under outcome model misspecification and informative cluster-period sizes.

The remainder of this article is organized as follows. Section~\ref{sec:estimands} reviews the causal estimands for SW-CRTs under a super-population perspective, proposes the subgroup-specific weighted average treatment effects, and presents the model-based g-computation and model-robust standardization estimators that target these estimands. Section~\ref{sec:Bayes_SW} develops Bayesian estimation procedures and presents the calibrated variance estimation strategy, along with analytical results. Section~\ref{sec:Simulation studies} reports simulation studies evaluating finite-sample performance across different data-generating mechanisms. Section~\ref{sec:Case Study} illustrates our new methods through a reanalysis of the TSOS trial. Section~\ref{sec:discussion} concludes with a discussion of implications and directions for future research.

\section{Causal effect estimands and model-assisted estimation} \label{sec:estimands}
\subsection{Notation and setup} \label{sec: Notation}

We consider a cross-sectional SW-CRT with $I$ clusters and $J$ periods. Let $i=1, \ldots,I$ represent the cluster index, and $j=1,\ldots,J$ represent the period index. For each cluster-period \((i,j)\), we let \(N_{ij}\) denote the number of individuals included in the study population of that cluster-period, which may vary across clusters and over time. The total number of observations within cluster $i$ is given by $N_i = \sum_{j=1}^J N_{ij}$, the total number of observations within period $j$ is given by $N_j = \sum_{i=1}^I N_{ij}$, and the overall sample size is given by $N = \sum_{i=1}^I \sum_{j=1}^J N_{ij}$. 

\begin{figure}[htbp]
\centering
\begin{tikzpicture}[
    font=\small,
    cell/.style={
        minimum width=2.5cm, minimum height=1.05cm,
        rounded corners=3pt, draw=gray!35, line width=0.5pt,
        align=center
    },
    ctrl/.style={cell, fill=gray!12},
    %ctrl/.style={cell, fill=blue!5!gray!15},
    trt/.style={cell, fill=blue!16},
    colhead/.style={font=\bfseries, align=center},
    rowhead/.style={font=\fontsize{8.5}{11.5}\selectfont, align=center, minimum height=1.05cm}
]
% ---- geometry ----
\def\dx{2.7}    % horizontal spacing between columns
\def\dy{1.25}   % vertical spacing between rows
% ---- column headers (periods) ----
\foreach \j/\lab in {1/{Period 1}, 2/{Period 2}, 3/{Period 3}, 4/{Period 4}, 5/{Period 5}} {
    \node[colhead] at ({\j*\dx}, {\dy}) {\lab};
}
% ---- row headers (cluster groups of two) ----
\node[rowhead, anchor=east] at ({0.02*\dx}, {0*\dy}) {Clusters 1--2\\[-1pt]$(A_i=2)$};
\node[rowhead, anchor=east] at ({0.02*\dx}, {-1*\dy}) {Clusters 3--4\\[-1pt]$(A_i=3)$};
\node[rowhead, anchor=east] at ({0.02*\dx}, {-2*\dy}) {Clusters 5--6\\[-1pt]$(A_i=4)$};
\node[rowhead, anchor=east] at ({0.02*\dx}, {-3*\dy}) {Clusters 7--8\\[-1pt]$(A_i=5)$};
% top-left corner label
\node[colhead, anchor=east] at ({0.02*\dx}, {\dy}) {Clusters};
% ---- Row 1: A_i = 2  (C | T T T T) ----
\node[ctrl] at ({1*\dx}, {0*\dy}) {$N_{1,1},N_{2,1}$};
\node[trt]  at ({2*\dx}, {0*\dy}) {$N_{1,2},N_{2,2}$};
\node[trt]  at ({3*\dx}, {0*\dy}) {$N_{1,3},N_{2,3}$};
\node[trt]  at ({4*\dx}, {0*\dy}) {$N_{1,4},N_{2,4}$};
\node[trt]  at ({5*\dx}, {0*\dy}) {$N_{1,5},N_{2,5}$};
% ---- Row 2: A_i = 3  (C C | T T T) ----
\node[ctrl] at ({1*\dx}, {-1*\dy}) {$N_{3,1},N_{4,1}$};
\node[ctrl] at ({2*\dx}, {-1*\dy}) {$N_{3,2},N_{4,2}$};
\node[trt]  at ({3*\dx}, {-1*\dy}) {$N_{3,3},N_{4,3}$};
\node[trt]  at ({4*\dx}, {-1*\dy}) {$N_{3,4},N_{4,4}$};
\node[trt]  at ({5*\dx}, {-1*\dy}) {$N_{3,5},N_{4,5}$};
% ---- Row 3: A_i = 4  (C C C | T T) ----
\node[ctrl] at ({1*\dx}, {-2*\dy}) {$N_{5,1},N_{6,1}$};
\node[ctrl] at ({2*\dx}, {-2*\dy}) {$N_{5,2},N_{6,2}$};
\node[ctrl] at ({3*\dx}, {-2*\dy}) {$N_{5,3},N_{6,3}$};
\node[trt]  at ({4*\dx}, {-2*\dy}) {$N_{5,4},N_{6,4}$};
\node[trt]  at ({5*\dx}, {-2*\dy}) {$N_{5,5},N_{6,5}$};
% ---- Row 4: A_i = 5  (C C C C | T) ----
\node[ctrl] at ({1*\dx}, {-3*\dy}) {$N_{7,1},N_{8,1}$};
\node[ctrl] at ({2*\dx}, {-3*\dy}) {$N_{7,2},N_{8,2}$};
\node[ctrl] at ({3*\dx}, {-3*\dy}) {$N_{7,3},N_{8,3}$};
\node[ctrl] at ({4*\dx}, {-3*\dy}) {$N_{7,4},N_{8,4}$};
\node[trt]  at ({5*\dx}, {-3*\dy}) {$N_{7,5},N_{8,5}$};
% ---- staggered roll-out guide line ----
\draw[blue!55, line width=1.1pt, rounded corners=2pt]
    ({1.5*\dx}, {0.5*\dy}) --
    ({1.5*\dx}, {-0.5*\dy}) --
    ({2.5*\dx}, {-0.5*\dy}) --
    ({2.5*\dx}, {-1.5*\dy}) --
    ({3.5*\dx}, {-1.5*\dy}) --
    ({3.5*\dx}, {-2.5*\dy}) --
    ({4.5*\dx}, {-2.5*\dy}) --
    ({4.5*\dx}, {-3.5*\dy});
% ---- period-phase braces underneath ----
\def\by{-3.75*\dy}
\draw[decorate, decoration={brace, amplitude=5pt, mirror}, gray!70]
    ({1*\dx-1.25}, {\by}) -- ({1*\dx+1.25}, {\by})
    node[midway, below=6pt, font=\footnotesize\itshape, text=gray!85]{pre-roll-out};
\draw[decorate, decoration={brace, amplitude=5pt, mirror}, gray!70]
    ({2*\dx-1.25}, {\by}) -- ({4*\dx+1.25}, {\by})
    node[midway, below=6pt, font=\footnotesize\itshape, text=gray!85]{roll-out};
\draw[decorate, decoration={brace, amplitude=5pt, mirror}, gray!70]
    ({5*\dx-1.25}, {\by}) -- ({5*\dx+1.25}, {\by})
    node[midway, below=6pt, font=\footnotesize\itshape, text=gray!85]{post-roll-out};
% ---- legend ----
\node[ctrl, minimum width=0.7cm, minimum height=0.5cm] (lc) at ({1.6*\dx}, {-4.75*\dy}) {};
\node[right=2pt of lc, font=\footnotesize] {Control};
\node[trt, minimum width=0.7cm, minimum height=0.5cm, right=2.1cm of lc] (lt) {};
\node[right=2pt of lt, font=\footnotesize] {Treatment};
\end{tikzpicture}
\caption{Schematic of the cross-sectional stepped-wedge design with $I=8$ clusters and $J=5$ periods, where two clusters cross over from control to treatment at each roll-out step. Clusters are randomized to adoption periods $A_i\in\{2,3,4,5\}$, so that period 1 is an all-control baseline, period 5 is an all-treated final period, and the causal estimands are evaluated over the roll-out periods. Each cell shows the cluster-period sizes $N_{ij}$ for the two clusters in that cell, which may vary across clusters and periods under informative cluster-period sizes. The blue line traces the staggered roll-out.}
\label{fig:sw_design}
\end{figure}
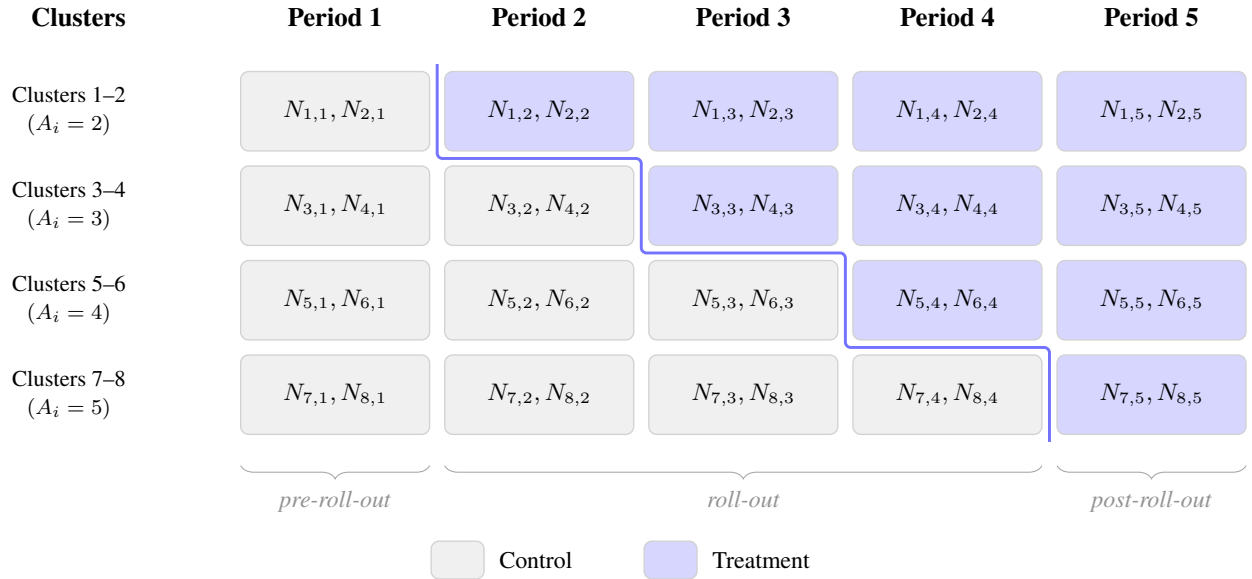

Let \(A_i = a\in \{2,\dots, J\}\) denote the randomized adoption period, or equivalently the treatment sequence, assigned to cluster \(i\). The observed treatment status in period \(j\) is \(Z_{ij}=\mathbb{I}(j\geq A_i)\), where $Z_{ij}=1$ indicates treatment and $Z_{ij}=0$ indicates control. Under the stepped-wedge design, $Z_{ij}$ is non-decreasing in $j$ for each cluster $i$, which means that once a cluster initiates treatment, it remains exposed thereafter. Figure~\ref{fig:sw_design} illustrates this staggered roll-out structure with an example design of $I=8$ clusters and $J=5$ periods, where two clusters cross over from control to treatment at each roll-out step. Following the potential outcomes framework, let $Y_{ijk}(z)$ denote the potential outcome of individual $k$ in cluster $i$ and period $j$ under treatment assignment $z \in \{0,1\}$. Under this notation, we adopt two standard assumptions on the potential outcomes following \citet{chen2025model}. First, we assume no anticipation, meaning that the potential outcome $Y_{ijk}(z)$ depends only on the current treatment condition $Z_{ij} = z$ and not on future treatment assignments $\{Z_{ij'} : j' > j\}$. That is, individuals do not respond to treatment before it is implemented. Second, we assume treatment-duration irrelevance, meaning that the potential outcome $Y_{ijk}(z)$ does not depend on the time at which cluster $i$ first adopted the treatment, so that $Y_{ijk}(1)$ is invariant to the elapsed time since treatment initiation. That is, there is a single version of the treatment, and the treatment effect does not vary with exposure duration. Estimands that instead accommodate duration-dependent (exposure-time-dependent) effects in SW-CRTs have been developed elsewhere \citep{kenny2022analysis, maleyeff2023assessing,wang2024achieve}.

We further assume a cluster-level Stable Unit Treatment Value Assumption (SUTVA), which rules out interference between clusters. Thus, the potential outcomes for any individual in cluster $i$ reflect the cluster-level intervention assigned, but do not depend on treatment in other clusters. Under consistency, the observed outcome satisfies $Y_{ijk} = Z_{ij} Y_{ijk}(1) + (1 - Z_{ij}) Y_{ijk}(0)$. Let \(\bm X_{ijk}\) denote pre-intervention baseline covariates for individual \(k\) in cluster \(i\) and period \(j\), including individual-level and cluster-level characteristics, and let $\bm{X}_{ij} = (\bm{X}_{ij1}, \ldots, \bm{X}_{ijN_{ij}})^{\top}$ collect the baseline covariates in cluster $i$ during period $j$. We treat $N_{ij}$ as a separate covariate from $\bm{X}_{ij}$, to make its role in model formulation more explicit. We assume that cluster-period membership, cluster-period size, and baseline covariates are determined before the current-period outcome and are not affected by the current treatment. Stepped-wedge randomization assigns the treatment sequence at the cluster level. Let $\bm{Y}_{ij}(z) = (Y_{ij1}(z), \ldots, Y_{ijN_{ij}}(z))^{\top}$ denote the vector of potential outcomes in cluster-period \((i,j)\) under treatment condition \(z\). Further, let \(\bm{Y}_i(z) = \{\bm{Y}_{ij}(z): j=1,\ldots,J\}\) denote the collection of potential outcomes across all periods in cluster \(i\) under treatment condition \(z\), and let \(\bm{G}_i = \{(\bm{X}_{ij}, N_{ij}): j=1,\ldots,J\}\) denote the cluster's covariate and cluster-period size history. Under randomization, the assigned adoption sequence \(A_i\) is independent of the cluster's potential outcome trajectories conditional on the covariate and cluster-period size history, namely \(A_i \perp \{\bm{Y}_i(0), \bm{Y}_i(1)\} \mid \bm{G}_i\). Combined with the no-anticipation and treatment-duration irrelevance assumptions, this implies the cluster-period-level conditional ignorability \(Z_{ij} \perp \{\bm{Y}_{ij}(0), \bm{Y}_{ij}(1)\} \mid \{\bm{X}_{ij}, N_{ij}\}\) used for identification. For the roll-out periods \(j=2,\ldots, J-1\), both treatment conditions occur with positive probability $P(Z_{ij} = z \mid \bm{X}_{ij}, N_{ij}) > 0$ for $z \in \{0,1\}$ by design, to ensure the identifiability of the causal estimands. The all-control baseline period and all-treated final period are excluded from the causal estimands as positivity of treatment assignment fails to hold in those periods \citep{chen2025model,fang2026model}. 

\subsection{Causal estimands for overall and subgroup-specific causal effects} \label{sec: Causal estimands}

Following \citet{fang2026model}, we consider a general class of weighted average treatment effects, $\Delta_{\omega} \equiv f(\mu_{\omega}(1),\mu_{\omega}(0))$, where $f$ is a predetermined smooth function that defines the scale of the effect measure. Common choices include $f(x,y) = x-y$ for the risk difference, $f(x,y) = x/y$ for the risk ratio, and $f(x,y) = x(1-y)/\{y(1-x)\}$ for the odds ratio. $\mu_{\omega}(z)$ is the weighted average potential outcome under treatment condition $z$, and is defined as
\begin{equation}
    \mu_{\omega}(z) = \frac{\mathbb{E} \left[ \sum_{j=2}^{J-1} \omega_{ij} \, \overline{Y}_{ij}(z) \right]}{\mathbb{E} \left[ \sum_{j=2}^{J-1} \omega_{ij} \right]},
\label{weighted estimand}
\end{equation}
where $\overline{Y}_{ij}(z)$ denotes the weighted cluster-period mean potential outcome $\overline{Y}_{ij}(z) = \left(\sum_{k=1}^{N_{ij}} \omega_{ijk} Y_{ijk}(z)\right)\big/ \left(\sum_{k=1}^{N_{ij}} \omega_{ijk}\right)$. Here $\omega_{ijk}$ is a pre-specified non-negative individual-level weight that determines the contribution of each individual to the overall estimand, and $\omega_{ij} = \sum_{k=1}^{N_{ij}} \omega_{ijk}$ represents the total weight at the cluster-period level. These weights are part of the estimand definition, and typically depend on cluster-period sizes. The expectation in $\mu_{\omega}(z)$ is defined with respect to the super-population of clusters. The summation is restricted to roll-out periods $j = 2,\ldots,J-1$ because the cluster-level propensity score equals zero in the first period (where no clusters are treated) and one in the last period (where all clusters are treated), violating positivity assumption that is required to conceptualize the causal effect. 

These weights can further be aggregated to the cluster level as $\omega_i = \sum_{j=1}^{J} \omega_{ij}$ and to the period level as $\omega_j = \sum_{i=1}^{I} \omega_{ij}$. \citet{fang2026model} introduced four distinct estimands based on different weighting schemes across individuals, clusters, and periods: (1) the horizontal-individual average treatment effect (h-iATE), which assigns equal weight to each individual across all cluster-periods, with $\omega_{ijk} = 1$ and $\omega_{ij} = N_{ij}$, (2) the horizontal-cluster average treatment effect (h-cATE), which gives equal weight to each cluster, using $\omega_{ijk} = 1 / \sum_{s=2}^{J-1} N_{is}$ and $\omega_{ij} = N_{ij} / \sum_{s=2}^{J-1} N_{is}$, (3) the vertical-individual average treatment effect (v-iATE), which allocates equal weight across periods, adjusting for expected cluster-period size, with $\omega_{ijk} = 1 / \{I \cdot \mathbb{E}[N_{ij}]\}$ and $\omega_{ij} = N_{ij} / \{I \cdot \mathbb{E}[N_{ij}]\}$, and (4) the vertical-cluster average treatment effect (v-cATE), which places equal weight on each cluster-period, setting $\omega_{ijk} = 1 / N_{ij}$ and $\omega_{ij} = 1$. These four estimands correspond to conceptually distinct population-level causal effects, each addressing a different inferential goal. 

\begin{table}[htbp]
\centering
\caption{Summary of the four subgroup-specific weighted average treatment effect estimands for subgroup $r$. The subgroup membership indicator is $S_{ijk,r}\in\{0,1\}$, and $N_{ij}^{(r)}=\sum_{k=1}^{N_{ij}}S_{ijk,r}$ denotes the number of subgroup-$r$ members in cluster-period $(i,j)$. Cluster-period weights satisfy $\omega_{ij}^{(r)}=\sum_{k=1}^{N_{ij}}\omega_{ijk}^{(r)}$. The subgroup-specific causal estimand is $\Delta_\omega^{(r)}=f(\mu_\omega^{(r)}(1),\mu_\omega^{(r)}(0))$. Setting $S_{ijk,r}=1$ for all $(i,j,k)$ recovers the overall estimands. The weights are set to zero in cluster-periods when a subgroup-specific cluster mean is not defined. The h-cATE definition simplifies as shown when every cluster contains at least one subgroup-$r$ member across the roll-out periods, and the v-cATE definition simplifies as shown when every roll-out cluster-period contains at least one subgroup-$r$ member.}
\label{tab:subgroup_estimands}
\newcommand{\rowstrut}{\rule[-1cm]{0pt}{2cm}}
\setlength{\tabcolsep}{10pt}
\begin{adjustbox}{width=\textwidth}
\begin{threeparttable}
\begin{tabular}{llll}
\toprule
\textbf{Aggregation} & \textbf{Estimand} & \textbf{Subgroup-restricted weights} & \textbf{Definition} \\
\midrule
\multirow{6}{*}{Horizontal}
& h-iATE
& $\displaystyle \rowstrut \omega_{ijk}^{(r)}=S_{ijk,r},\quad \omega_{ij}^{(r)}=N_{ij}^{(r)}$
& $\displaystyle \mu_{\omega}^{(r)}(z)=\frac{\mathbb{E}\!\left[\sum_{j=2}^{J-1} N_{ij}^{(r)}\,\overline{Y}_{ij}^{(r)}(z)\right]}{\mathbb{E}\!\left[\sum_{j=2}^{J-1} N_{ij}^{(r)}\right]}$ \\
\cmidrule(l){2-4}
& h-cATE
& $\displaystyle \rowstrut \omega_{ijk}^{(r)}=\frac{S_{ijk,r}}{\sum_{s=2}^{J-1} N_{is}^{(r)}},\quad \omega_{ij}^{(r)}=\frac{N_{ij}^{(r)}}{\sum_{s=2}^{J-1} N_{is}^{(r)}}$
& $\displaystyle \mu_{\omega}^{(r)}(z)=\mathbb{E}\!\left[\sum_{j=2}^{J-1} \dfrac{N_{ij}^{(r)}}{\sum_{s=2}^{J-1} N_{is}^{(r)}}\,\overline{Y}_{ij}^{(r)}(z)\right]$ \\
%$\displaystyle \mu_{\omega}^{(r)}(z)=\frac{\mathbb{E}\!\left[\sum_{j=2}^{J-1} \dfrac{N_{ij}^{(r)}}{\sum_{s=2}^{J-1} N_{is}^{(r)}}\,\overline{Y}_{ij}^{(r)}(z)\right]}{\mathbb{E}\!\left[\sum_{j=2}^{J-1} \dfrac{N_{ij}^{(r)}}{\sum_{s=2}^{J-1} N_{is}^{(r)}}\right]}$ \\
\midrule
\multirow{6}{*}{Vertical}
& v-iATE
& $\displaystyle \rowstrut \omega_{ijk}^{(r)}=\frac{S_{ijk,r}}{I\,\mathbb{E}[N_{ij}^{(r)}]},\quad \omega_{ij}^{(r)}=\frac{N_{ij}^{(r)}}{I\,\mathbb{E}[N_{ij}^{(r)}]}$
& $\displaystyle \mu_{\omega}^{(r)}(z)=\mathbb{E}\!\left[\sum_{j=2}^{J-1} \dfrac{N_{ij}^{(r)}}{I\,\mathbb{E}[N_{ij}^{(r)}]}\,\overline{Y}_{ij}^{(r)}(z)\right]\bigg/ \; \mathbb{E}\!\left[\sum_{j=2}^{J-1} \dfrac{N_{ij}^{(r)}}{I\,\mathbb{E}[N_{ij}^{(r)}]}\right]$ \\
\cmidrule(l){2-4}
& v-cATE
& $\displaystyle \rowstrut \omega_{ijk}^{(r)}=\frac{S_{ijk,r}}{N_{ij}^{(r)}},\quad \omega_{ij}^{(r)}=1$
& $\displaystyle \mu_{\omega}^{(r)}(z)=\frac{1}{J-2}\mathbb{E}\!\left[\sum_{j=2}^{J-1} \overline{Y}_{ij}^{(r)}(z)\right]$ \\
\bottomrule
\end{tabular}
\end{threeparttable}
\end{adjustbox}
\end{table}

In many applications, investigators are also interested in treatment effects within pre-specified subpopulations, such as patient subgroups defined by baseline severity or risk score. To accommodate this, we provide an extension of the weighted treatment effect estimands in SW-CRTs to represent subgroup-specific causal effects. Let $S_{ijk,r}\in\{0,1\}$ indicate whether individual $k$ in cluster $i$ during period $j$ belongs to subgroup $r\in\{1,\ldots,R\}$, where $S_{ijk,r}=1$ if the individual satisfies the pre-specified baseline criterion defining subgroup $r$, and $S_{ijk,r}=0$ otherwise. The subgroup indicators are deterministic functions of pre-treatment baseline covariates, such as age above a clinical threshold, baseline severity category, or comorbidity count above a cutoff. Because these indicators are defined before treatment assignment, they can be incorporated into the estimand through subgroup-restricted weights. Subgroups may be mutually exclusive or overlapping, where each subgroup $r$ defines a separate target population and therefore a separate subgroup-specific causal estimand. Specifically, for subgroup $r \in \{1, \ldots, R\}$, the subgroup-restricted weights corresponding to the four estimands are defined as: (1) for the h-iATE, $\omega_{ijk}^{(r)} = S_{ijk,r}$; (2) for the h-cATE, $\omega_{ijk}^{(r)} = S_{ijk,r} / \left(\sum_{s=2}^{J-1} \sum_{l=1}^{N_{is}} S_{isl,r}\right)$ if $\sum_{s=2}^{J-1} \sum_{l=1}^{N_{is}} S_{isl,r} > 0$, and $\omega_{ijk}^{(r)}=0$ otherwise; (3) for the v-iATE, $\omega_{ijk}^{(r)} = S_{ijk,r} / \left\{I \cdot \mathbb{E}\left[\sum_{k=1}^{N_{ij}} S_{ijk,r}\right]\right\}$; and (4) for the v-cATE, $\omega_{ijk}^{(r)} = S_{ijk,r} / \sum_{k=1}^{N_{ij}} S_{ijk,r}$ if $\sum_{k=1}^{N_{ij}} S_{ijk,r} > 0$, and $\omega_{ijk}^{(r)} = 0$ otherwise. The zero-weight conventions in (2) and (4) address clusters and cluster-periods that contain no members of subgroup $r$, for which the subgroup-specific weights would otherwise have a zero denominator. For the subgroup h-cATE, a cluster with no subgroup-$r$ members in any roll-out period receives zero weight. For the subgroup v-cATE, a cluster-period with no subgroup-$r$ members receives zero weight. The subgroup h-cATE and v-cATE therefore average over clusters and cluster-periods, respectively, that contain at least one subgroup-$r$ member. The corresponding cluster-period weight is given by $\omega_{ij}^{(r)} = \sum_{k=1}^{N_{ij}} \omega_{ijk}^{(r)}$, and the weighted cluster-period mean potential outcome for subgroup $r$ under treatment $z$ is defined as 
$$\overline{Y}_{ij}^{(r)}(z) = \frac{\sum_{k=1}^{N_{ij}} \omega_{ijk}^{(r)} Y_{ijk}(z)}{\sum_{k=1}^{N_{ij}} \omega_{ijk}^{(r)}}.$$ 
Throughout, we adopt the convention that $\omega_{ij}^{(r)}\overline{Y}_{ij}^{(r)}(z)=0$ whenever $\omega_{ij}^{(r)}=0$, so that cluster-periods with no subgroup-$r$ members are omitted from all weighted sums; the same convention applies to the observed means $\overline{Y}_{ij}^{(r)}$ and $\overline{\bm X}_{ij,\omega}^{(r)}$. The subgroup-specific weighted average potential outcome is then given by
\begin{equation}
\mu_\omega^{(r)}(z) = \frac{\mathbb{E} \left[ \sum_{j=2}^{J-1} \omega_{ij}^{(r)} \overline{Y}_{ij}^{(r)}(z) \right]}{\mathbb{E} \left[ \sum_{j=2}^{J-1} \omega_{ij}^{(r)} \right]}.
\label{subgroup-specific weighted estimand}
\end{equation}
The subgroup-specific causal effect estimand is subsequently defined as $\Delta_\omega^{(r)} \equiv f\left( \mu_\omega^{(r)}(1), \mu_\omega^{(r)}(0) \right)$, for a chosen contrast function $f$. Table~\ref{tab:subgroup_estimands} summarizes the subgroup-restricted weights and the resulting weighted average potential outcomes for the four subgroup-specific causal estimands. To ensure that $\Delta_\omega^{(r)}$ is well-defined and identifiable from observed data, we further require the subgroup positivity assumption: $0 < P\left(Z_{ij} = z \mid \bm{X}_{ij}, N_{ij}\right) < 1$ for all cluster-periods with $\omega_{ij}^{(r)} > 0$, each $z \in \{0, 1\}$, subgroup $r \in \{1, \ldots, R\}$, and period $j \in \{2, \ldots, J-1\}$. Note that taking $S_{ijk,r} = 1$ for all $(i,j,k)$ recovers the original weights for the overall causal effect, and hence the overall causal effect estimands are obtained as a special case.

\subsection{Model-based g-computation} \label{sec:Model-based g-computation}
The estimands in Section \ref{sec: Causal estimands} are defined in terms of potential outcomes and therefore are not directly observable. We now focus on their identification from the observed data. Under stepped-wedge randomization, the law of iterated expectations gives an identification formula through which the estimands of interest can be expressed as functions of the observed data:
\begin{equation}
\mu_{\omega} (z) = \frac{\mathbb{E} \left[ \sum_{j=2}^{J-1} \omega_{ij}  \mathbb{E} \left[ \overline{Y}_{ij}  \mid  Z_{ij}=z, \bm{X}_{ij}, N_{ij}\right]\right]}{\mathbb{E} \left[ \sum_{j=2}^{J-1} \omega_{ij}  \right]},
\label{eq:expression for mu_omega}
\end{equation}
and becomes $\mu_{\omega}^{(r)}(z) = \mathbb{E} \left[ \sum_{j=2}^{J-1} \omega_{ij}^{(r)} \mathbb{E} \left[ \overline{Y}_{ij}^{(r)} \mid  Z_{ij}=z, \bm{X}_{ij}, N_{ij}\right]\right]\Big/\mathbb{E} \left[ \sum_{j=2}^{J-1} \omega_{ij}^{(r)} \right]$ when considering the subgroup-specific causal effect for subgroup $r$. This identification result requires that the conditional mean of $\overline{Y}_{ij}$ depends only on the contemporaneous treatment, covariates, and cluster-period size, and not on the covariates from other periods within the same cluster. Equation~\eqref{eq:expression for mu_omega} shows that estimation of $\mu_\omega(z)$ can proceed by estimating the conditional mean of the relevant weighted cluster-period outcome under treatment level $z$, and then averaging those fitted values over the target population encoded by the weights. In practice, we replace $\mathbb{E}\left[\overline{Y}_{ij} \mid Z_{ij}=z, \bm{X}_{ij}, N_{ij}\right]$ by an estimated conditional mean from a working outcome model denoted as $\widehat{\mathbb{E}}\left[\overline{Y}_{ij} \mid Z_{ij}=z, \bm{X}_{ij}, N_{ij}\right]$, and we marginalize over the cluster-level super-population by replacing outer expectations in Equation~\eqref{eq:expression for mu_omega} with their empirical averages over the $I$ observed clusters, giving the model-based g-computation estimator:
\begin{equation}
\widehat\mu_{\omega}^{\mathrm{mb}}(z)=
\frac{\sum_{i=1}^{I}\sum_{j=2}^{J-1}
\omega_{ij} \widehat{\mathbb{E}}\left[\overline{Y}_{ij} \mid Z_{ij}=z, \bm{X}_{ij}, N_{ij}\right]}{
\sum_{i=1}^{I}\sum_{j=2}^{J-1}
\omega_{ij}},
\qquad
\widehat\Delta_{\omega}^{\mathrm{mb}}=f\{\widehat\mu_{\omega}^{\mathrm{mb}}(1),\widehat\mu_{\omega}^{\mathrm{mb}}(0)\}.
\label{eq:gcomp_general}
\end{equation}

Importantly, the choice of weight $\omega_{ijk}$ enters identification through the marginalization but not through the conditional regression itself. This separation enables a single working outcome model to target multiple estimands within the class. The corresponding subgroup-specific g-computation estimator $\widehat{\mu}_\omega^{\,\mathrm{mb},(r)}(z)$ is obtained by replacing $\omega_{ij}$ and $\overline{Y}_{ij}$ with their subgroup-restricted counterparts $\omega_{ij}^{(r)}$ and $\overline{Y}_{ij}^{(r)}$. When the working model is fitted at the individual level, the same fitted model can be used for multiple estimands by changing only the weighted averaging step. 

Next, we exemplify the model-based g-computation with a familiar working outcome model. We focus on continuous outcomes and anchor the discussion in a linear mixed model that allows period-specific fixed effects and treatment effects. Although the target estimands are defined only over the roll-out periods $j=2,\ldots,J-1$, the working model may be fitted using observations from all $J$ periods. Accordingly, we write the working model as 
\begin{align}
Y_{ijk}
= \beta_{0j} + \theta_{j} Z_{ij} + \bm{\beta}^{\top}_{X}\bm{X}_{ijk} + \beta_N N_{ij}
+ \alpha_i + \delta_{ij} + \epsilon_{ijk},
\label{eq:LMM}
\end{align}
where $\alpha_i \sim \mathcal N(0,\sigma_\alpha^2)$ are cluster-level random intercepts, $\delta_{ij} \sim \mathcal N(0,\sigma_\delta^2)$ are cluster-period random effects \citep{hooper2016sample}, and $\epsilon_{ijk} \sim \mathcal N(0,\sigma_\epsilon^2)$ are individual-level errors. The period-specific intercepts $\beta_{0j}$ accommodate secular trends, and the period-specific treatment effects $\theta_j$, for the roll-out periods $j=2,\ldots,J-1$, allow the treatment contrast to vary across calendar periods. We set $\theta_1=\theta_J=0$, because the treatment indicator is constant within the all-control first period and the all-treated final period, so that these contrasts are not identified and are absorbed into the intercepts $\beta_{01}$ and $\beta_{0J}$. $\bm{\beta}_{X}$ denotes the vector of regression coefficients for $\bm{X}_{ijk}$, while $\beta_{N}$ is the regression coefficient for the cluster-period size $N_{ij}$. Entering covariate terms linearly can be restrictive, and a more flexible specification, such as a spline or a set of cluster-period size categories, could be used instead. We retain the linear form here, and note that because this is a working model, any such misspecification affects only the efficiency, and not the consistency, of the model-robust estimator. The mixed-effects BART specification of Section~\ref{sec:Bayes_BART} additionally accommodates nonlinear dependence on cluster-period size without explicit pre-specification.

In practice, SW-CRTs are also commonly analyzed using a model with a single cluster-level random intercept, corresponding to a simple exchangeable within-cluster correlation structure, as reflected in a recent systematic review \citep{nevins2024adherence}. We therefore consider this reduced model and explore the implications for estimation and inference for our estimands. Specifically, such a reduced model is given by
\begin{align}
Y_{ijk}
= \beta_{0j} + \theta_{j} Z_{ij} +\bm{\beta}^{\top}_{X}\bm{X}_{ijk} + \beta_N N_{ij}
+ \alpha_i + \epsilon_{ijk}.
\label{eq:LMM_reduced}
\end{align}
The choice between the two random-effects structures could in principle be informed by predictive model selection criteria such as Watanabe-Akaike Information Criterion (WAIC) or leave-one-out cross-validation. We report results under both specifications throughout, and focus on whether the proposed inference remains valid regardless of which structure is adopted. In both models, the random effects are integrated out when forming the super-population marginal estimand. Thus, empirical Bayes predictions of $\alpha_i$ or $\delta_{ij}$ are not included in $\mathbb{E}\left[\overline{Y}_{ij} \mid Z_{ij}=z, \bm{X}_{ij}, N_{ij}\right]$, where the random effects contribute to modeling the correlation structure and variance components, but the marginal mean uses their mean-zero distribution. Define the weighted covariate mean within cluster-period $(i,j)$ by \(\overline{\bm X}_{ij,\omega}=\sum_{k=1}^{N_{ij}}\omega_{ijk}\bm X_{ijk}/\sum_{k=1}^{N_{ij}}\omega_{ijk}\). Either of the two models can imply the expected cluster-period mean outcome, conditional on the treatment and covariates, which is \(\mathbb E \left[\overline Y_{ij}\mid Z_{ij}=z,{\bm X}_{ij},N_{ij}\right]= \beta_{0j} + \theta_j z +\bm\beta_X^{\top}\overline{\bm X}_{ij,\omega}+\beta_N N_{ij}\).
Replacing the unknown parameters by their estimates gives
\begin{equation}
\widehat m_{ij}(z)=\widehat\beta_{0j}+\widehat\theta_j z+\widehat{\bm\beta}_X^{\top}\overline{\bm X}_{ij,\omega}+ \widehat\beta_N N_{ij},
\label{eq:lmm_prediction}
\end{equation}
and substituting Equation~\eqref{eq:lmm_prediction} into Equation~\eqref{eq:gcomp_general} yields the covariate-adjusted model-based g-computation estimator as $\widehat{\Delta}_{\omega}^{\text{mb}} = f \left(\widehat{\mu}_{\omega}^{\text{mb}}(1), \widehat{\mu}_{\omega}^{\text{mb}}(0)\right)$, where 
\[
\widehat\mu_{\omega}^{\text{mb}}(z)
= \frac{\sum_{i=1}^{I}\sum_{j=2}^{J-1} \omega_{ij} \cdot
\left(\widehat\beta_{0j}+\widehat\theta_{j} z+\widehat{\bm\beta}_X^{\top}\overline{\bm X}_{ij,\omega}
+\widehat\beta_N N_{ij}\right)}
{\sum_{i=1}^{I}\sum_{j=2}^{J-1} \omega_{ij}},
\]
where $\omega_{ij}$ and $\omega_{ijk}$ are the pre-specified cluster-period and individual-level weights, which define the target estimand. 

Noticeably, the same construction extends to subgroup-specific causal estimands by incorporating subgroup-restricted weights and, when appropriate, treatment-by-subgroup interactions in the working outcome model to express the covariate-induced heterogeneity. Define
\(\overline{\bm X}_{ij,\omega}^{(r)}=\sum_{k=1}^{N_{ij}}\omega_{ijk}^{(r)}\bm X_{ijk}/\sum_{k=1}^{N_{ij}}\omega_{ijk}^{(r)}\)
for cluster-periods with $\omega_{ij}^{(r)}>0$. If the working model does not include treatment-by-subgroup interactions, then the subgroup-specific prediction is obtained from Equation \eqref{eq:lmm_prediction} by replacing $\overline{\bm X}_{ij,\omega}$ with $\overline{\bm X}_{ij,\omega}^{(r)}$. This approach targets the subgroup estimand through subgroup-restricted standardization, but it can typically require the working model to describe treatment effect heterogeneity correctly. For a parametric subgroup analysis based on a specific covariate $X_{ijk}^{(s)}$, which may be the subgroup indicator itself or the covariate used to define the subgroup, we consider augmenting the working model with a treatment-by-subgroup interaction \citep{li2024planning}. The full model for subgroup analysis is specified as
\begin{align}
Y_{ijk}= \beta_{0j} + \theta_{j} Z_{ij} +\bm{\beta}^{\top}_{X}\bm{X}_{ijk} + \beta_N N_{ij}+ \theta_j^{(s)} Z_{ij}X_{ijk}^{(s)} + \alpha_i + \delta_{ij} +\epsilon_{ijk},
\label{eq:subgroup_LMM}
\end{align}
where $\theta_j^{(s)}$ quantifies the period-specific differential treatment effect across levels of the subgroup-defining covariate $X_{ijk}^{(s)}$. The corresponding reduced model for subgroup analysis is
\begin{align}
Y_{ijk}= \beta_{0j} + \theta_{j} Z_{ij} +\bm{\beta}^{\top}_{X}\bm{X}_{ijk} + \beta_N N_{ij}+ \theta_j^{(s)} Z_{ij}X_{ijk}^{(s)} + \alpha_i +\epsilon_{ijk}.
\label{eq:subgroup_LMM_reduced}
\end{align}
Then, based on the subgroup working outcome model in \eqref{eq:subgroup_LMM} or \eqref{eq:subgroup_LMM_reduced}, the subgroup-specific conditional mean implied by the augmented working model is $\mathbb{E} \left[ \overline{Y}_{ij}^{(r)} \mid  Z_{ij}=z, \bm{X}_{ij}, N_{ij}\right] = \beta_{0j} + \theta_j z + \bm\beta_X^{\top}\overline{\bm X}_{ij,\omega}^{(r)}
 + \beta_N N_{ij} +  \theta_j^{(s)} z \overline {\bm X}_{ij,\omega}^{(s,r)}$, where $\overline {\bm X}_{ij,\omega}^{(s,r)} = \sum_{k=1}^{N_{ij}} \omega_{ijk}^{(r)} {\bm X}_{ijk}^{(s)}/\sum_{k=1}^{N_{ij}}\omega_{ijk}^{(r)}$. The resulting estimator takes the form $\widehat{\Delta}_{\omega}^{\text{mb},(r)} = f(\widehat{\mu}_{\omega}^{\text{mb},(r)}(1), \widehat{\mu}_{\omega}^{\text{mb},(r)}(0))$, where 
\[
\widehat\mu_{\omega}^{\text{mb},(r)}(z)
=
\frac{
\sum_{i=1}^{I}\sum_{j=2}^{J-1} \omega_{ij}^{(r)} \cdot
\left(
\widehat\beta_{0j}
+\widehat\theta_{j} z
+\widehat{\bm\beta}_X^{\top}\overline{\bm X}_{ij,\omega}^{(r)}
+\widehat\beta_N N_{ij} 
+ \widehat\theta_j^{(s)}z\overline {\bm X}_{ij,\omega}^{(s,r)}
\right)
}{
\sum_{i=1}^{I}\sum_{j=2}^{J-1} \omega_{ij}^{(r)}
}.
\]
%where $\omega_{ijk}^{(r)}$ and $\omega_{ij}^{(r)}=\sum_{k=1}^{N_{ij}}\omega_{ijk}^{(r)}$ are the subgroup-restricted weights. 
The model-based g-computation estimator is estimand-aligned in the sense that the final averaging step is determined by the pre-specified estimand weights. However, it remains model-based, because consistency requires that the working outcome model be correctly specified.
%The validity of this g-computation estimator, as well as its subgroup counterpart, hinges on correct specification of the working outcome model. 
Misspecification of the conditional regression, including omission of relevant nonlinear terms or treatment effect modifiers, may introduce bias, particularly under informative cluster-period sizes or subgroup-specific causal effect heterogeneity. Under informative cluster-period sizes, bias may also arise from misspecifying the random-effects structure; our simulations will examine this issue by fitting both exchangeable and nested exchangeable structures.

\subsection{Model-robust standardization}\label{sec:Model-robust standardization}
To address the potential bias of g-computation under outcome model misspecification, we adopt the model-robust standardization estimator developed in \citet{fang2026model}, which augments g-computation with a residual correction term and remains consistent for the target estimand under outcome model misspecification, while attaining efficiency gains from covariate adjustment when the working model is close to the conditional mean \citep{wang2024model,li2025model,fang2026model}. In Section~\ref{sec:Bayes_point} and Section 1 of the Web Appendix, we further establish root-$I$ consistency of the proposed Bayesian point estimator (to be introduced in Section \ref{sec:Bayes_point}) under arbitrary working model misspecification.

For brevity, we present the estimator using the subgroup notation, because the overall estimand is obtained as the special case \(S_{ijk,r}=1\) for all individuals. For subgroup $r$, the proposed model-robust estimator is given by $\widehat{\Delta}_{\omega}^{\text{mr},(r)} = f\left( \widehat{\mu}_{\omega}^{\text{mr},(r)}(1), \widehat{\mu}_{\omega}^{\text{mr},(r)}(0)\right)$, where
\begin{equation}
\begin{aligned}
\widehat{\mu}_{\omega}^{\text{mr},(r)}(z)
= \frac{1}{\sum_{j=2}^{J-1}\omega_j^{(r)}} \sum_{j=2}^{J-1}\omega_j^{(r)}
\left\{
\frac{\sum_{i=1}^{I}\omega_{ij}^{(r)}\,\widehat{m}_{ij}^{(r)}(z)}{\omega_j^{(r)}} +
\frac{\sum_{i=1}^{I}\omega_{ij}^{(r)}\,\mathbb{I}(Z_{ij}=z)\left[\overline{Y}_{ij}^{(r)}-\widehat{m}_{ij}^{(r)}(z)\right]}{\sum_{i=1}^{I}\mathbb{I}(Z_{ij}=z)\,\omega_{ij}^{(r)}}
\right\},
\end{aligned}
\label{eq:mr_estimator}
\end{equation}
with $\omega_j^{(r)} = \sum_{i=1}^I \omega_{ij}^{(r)}$ denoting the period-level subgroup-restricted weights, %$\overline{Y}_{ij}^{(r)}$ the subgroup-restricted weighted cluster-period mean outcome, 
and $\widehat{m}_{ij}^{(r)}(z) = \widehat{\mathbb{E}} \left[\overline{Y}_{ij}^{(r)}\mid Z_{ij}=z,\bm{X}_{ij},N_{ij}\right]$ the predicted subgroup-restricted cluster-period mean from a working outcome model fitted on the observed data. The expression in braces decomposes the period-$j$ contribution into a g-computation term, $\sum_i \omega_{ij}^{(r)} \widehat{m}_{ij}^{(r)}(z) / \omega_j^{(r)}$, and a residual correction term that averages observed outcome residuals over treated (or control) clusters in period $j$. Setting $S_{ijk,r}=1$ for all $(i,j,k)$ recovers the model-robust estimator $\widehat{\mu}_\omega^{\,\mathrm{mr}}(z)$ for the overall causal effect estimands, so that all subsequent theoretical and inferential results developed under the subgroup-specific notation apply directly to the overall causal effect as a special case.

Equivalently, the period-specific contribution in Equation \eqref{eq:mr_estimator} can be rewritten as a weighted sum of the unadjusted period-$j$ estimator augmented by a regression-based correction \citep{fang2026model}:
\begin{equation}
\begin{aligned}
\widehat{\mu}_{\omega,j}^{\text{mr},(r)}(z) = \frac{\sum_i \omega_{ij}^{(r)}\mathbb{I}(Z_{ij}=z)\overline{Y}_{ij}^{(r)}}{\sum_i \omega_{ij}^{(r)}\mathbb{I}(Z_{ij}=z)} + \left\{\frac{\sum_i \omega_{ij}^{(r)} \widehat{m}_{ij}^{(r)}(z)}{\omega_j^{(r)}} - \frac{\sum_i \omega_{ij}^{(r)}\mathbb{I}(Z_{ij}=z) \widehat{m}_{ij}^{(r)}(z)}{\sum_i \omega_{ij}^{(r)}\mathbb{I}(Z_{ij}=z)}\right\}.
\end{aligned}
\label{eq:mr_estimator_2}
\end{equation}
The first term is the unadjusted period-$j$ estimator, and the second term in braces is the regression-based augmentation. Under stepped-wedge randomization, this augmentation converges in probability to zero for any fixed working function \(\widehat m_{ij}^{(r)}(z)\), so it does not change the probability limit of the unadjusted estimand-aligned estimator. When the working outcome model is prognostic for the cluster-period outcome, the augmentation can improve efficiency by adjusting for baseline covariate imbalance within periods. This representation clarifies the role of the working model. The working model is used to improve precision, not to define the causal estimand. Therefore, consistency of \(\widehat{\mu}_{\omega}^{\mathrm{mr},(r)}(z)\) does not require correct specification of \(\widehat m_{ij}^{(r)}(z)\). If the working model is correctly specified, the estimator can achieve improved large-sample efficiency. \citet{fang2026model} showed that the asymptotic variance is minimized when $\widehat{m}_{ij}^{(r)}(z)$ approaches the true conditional mean $\mathbb{E}[\overline{Y}_{ij}^{(r)} \mid Z_{ij}=z, \bm{X}_{ij}, N_{ij}]$. On the other hand, if the working model is misspecified, the residual correction protects against asymptotic bias. This is the key model-robustness property established for SW-CRTs by \citet{fang2026model}, which we extend in Section~\ref{sec:Bayes_point} to accommodate the Bayesian point estimator, where posterior draws of the working model are standardized to target the estimand and the resulting point estimator is shown to retain root-$I$ consistency under working model misspecification. 

As in g-computation, the weight $\omega_{ijk}^{(r)}$ enters the standardization step but not the working model itself, so a single fitted outcome model can assist in targeting multiple estimands within the weighted ATE class by simply substituting different weights, and their subgroup-specific counterparts by substituting the corresponding pre-specified weights. %The model-robust framework accommodates a broad range of working outcome models ${m}_{ij}^{(r)}(z)$, including parametric models such as linear mixed models and flexible nonparametric models such as Bayesian additive regression trees. Importantly, the working model used in Equation~\eqref{eq:mr_estimator} can be the same conditional regression introduced in Section~\ref{sec:Model-based g-computation}, so any outcome model adequate for g-computation extends directly to the model-robust estimator without modification. 
To make the estimator concrete, consider the linear mixed model in Equation \eqref{eq:subgroup_LMM} as the working outcome model, so that $\widehat m_{ij}^{(r)}(z) = \widehat\beta_{0j} + \widehat\theta_j z + \widehat{\bm\beta}_X^{\top}\overline{\bm X}_{ij,\omega}^{(r)} + \widehat\beta_N N_{ij} + \widehat\theta_j^{(s)} z\,\overline X_{ij,\omega}^{(s,r)}$, with the random effects integrated out under marginalization. Substituting this working model into Equation~\eqref{eq:mr_estimator_2} and cancelling the terms that are constant across clusters within each period gives the explicit form of the model-robust estimator under the linear mixed working model,
\begin{equation}
\begin{aligned}
\widehat{\mu}_{\omega}^{\text{mr},(r)}(z)
= \frac{1}{\sum_{j=2}^{J-1}\omega_j^{(r)}} \sum_{j=2}^{J-1}\omega_j^{(r)}
\Bigg\{&
\frac{\sum_{i=1}^{I}\omega_{ij}^{(r)}\,\mathbb{I}(Z_{ij}=z)\overline{Y}_{ij}^{(r)}}{\sum_{i=1}^{I}\omega_{ij}^{(r)}\mathbb{I}(Z_{ij}=z)} + 
\frac{\sum_{i=1}^{I}\omega_{ij}^{(r)}\left(\widehat{\bm\beta}_X^{\top}\overline{\bm X}_{ij,\omega}^{(r)}+\widehat\beta_N N_{ij}+\widehat\theta_j^{(s)} z\,\overline X_{ij,\omega}^{(s,r)}\right)}{\omega_j^{(r)}} \\
&-
\frac{\sum_{i=1}^{I}\omega_{ij}^{(r)}\,\mathbb{I}(Z_{ij}=z)\left[\widehat{\bm\beta}_X^{\top}\overline{\bm X}_{ij,\omega}^{(r)}+\widehat\beta_N N_{ij}+\widehat\theta_j^{(s)} z\,\overline X_{ij,\omega}^{(s,r)}\right]}{\sum_{i=1}^{I}\mathbb{I}(Z_{ij}=z)\,\omega_{ij}^{(r)}}
\Bigg\}.
\end{aligned}
\label{eq:mr_lmm}
\end{equation}
Here $z$ is a fixed treatment condition imposed in the standardization, so the intercept term $\widehat\beta_{0j}$ and the main treatment effect term $\widehat\theta_j z$ do not vary across clusters within period $j$. They take the same value under the full-period and condition-$z$ weightings and hence cancel in the regression-based augmentation, as reflected in Equation~\eqref{eq:mr_lmm}. The augmentation therefore depends on the working model only through the terms that vary across clusters, namely $\widehat{\bm\beta}_X^{\top}\overline{\bm X}_{ij,\omega}^{(r)}+\widehat\beta_N N_{ij}+\widehat\theta_j^{(s)} z\,\overline X_{ij,\omega}^{(s,r)}$, which adjust for within-period imbalance in baseline covariates, cluster-period size, and subgroup-specific treatment effect modification.

\section{Calibrated Bayesian inference for weighted estimands in SW-CRTs}\label{sec:Bayes_SW}

We now embed the estimand-aligned estimators in a Bayesian framework. The key idea is to use posterior draws from a working outcome model to generate draw-specific predictions, and then apply the same standardization formulas described in Sections~\ref{sec:Model-based g-computation} and~\ref{sec:Model-robust standardization}. The Bayesian model supplies a posterior distribution for the nuisance outcome regression, while the final standardization step continues to determine the causal estimand.

\subsection{Point estimation with a Bayesian working outcome model}\label{sec:Bayes_point}
To make the idea more concrete, we anchor Bayesian inference in the linear mixed models introduced in Section~\ref{sec:Model-based g-computation}. Let
\[
\Theta  =  \Big\{\{\beta_{0j}\}_{j=1}^J, \{\theta_j\}_{j=2}^{J-1}, \bm{\beta}^{\top}_{X},\beta_N, \{\alpha_i\}_{i=1}^I, \{\delta_{ij}\}_{i=1}^I{}_{j=1}^J, \sigma_\epsilon^2, \sigma_\alpha^2, \sigma_\delta^2\Big\}
\]
collect the unknown parameters in the full model \eqref{eq:LMM}; for the reduced model \eqref{eq:LMM_reduced} the cluster-period random effects \(\delta_{ij}\) and their variance component \(\sigma_\delta^2\) are omitted. Conditional on \(\Theta\), the likelihood implied by \eqref{eq:LMM} (or \eqref{eq:LMM_reduced}) is Gaussian. We assume the following weakly informative, conjugate priors for the fixed effects and variance components. For the full model \eqref{eq:LMM}, the period-specific intercepts and treatment effects are assigned independent normal priors: $\beta_{0j} \sim \mathcal{N}(0,1000)$ for $j = 1, \ldots, J$ and $\theta_j \sim \mathcal{N}(0,1000)$ for $j = 2, \ldots, J-1$. The covariate coefficients and cluster-period size coefficient share a common prior: $(\bm{\beta}_X^\top, \beta_N)^\top \sim \mathcal{N}(\bm{0}, 1000\bm{I})$. %, where $\bm{\beta}_X$ denotes the vector of regression coefficients for baseline covariates $\bm{X}_{ijk}$. 
The variance components are assigned inverse-gamma priors, $\sigma_{\alpha}^2 \sim \text{IG}(0.001, 0.001)$, $\sigma_{\delta}^2 \sim \text{IG}(0.001, 0.001)$, and $\sigma_{\epsilon}^2 \sim \text{IG}(0.001, 0.001)$, where $\text{IG}(\cdot,\cdot)$ denotes the inverse-gamma distribution parameterized by a shape parameter and a scale parameter. Inverse-gamma priors are chosen here for conjugacy and computational efficiency. Half-Cauchy priors on the standard deviations \citep{gelman2006prior} are a common alternative when conjugacy is not required. For the reduced model \eqref{eq:LMM_reduced}, the same priors are used for $(\beta_{0j}, \theta_j)^\top$, $(\bm{\beta}_X^\top, \beta_N)^\top$, $\sigma_{\alpha}^2$, and $\sigma_{\epsilon}^2$, with $\sigma_{\delta}^2$ omitted. For the subgroup working models \eqref{eq:subgroup_LMM} and \eqref{eq:subgroup_LMM_reduced}, $\Theta$ additionally includes the treatment-by-subgroup coefficients $\{\theta_j^{(s)}\}_{j=2}^{J-1}$, which are assigned independent $\mathcal{N}(0,1000)$ priors. Let \(\{\Theta_b\}_{b=1}^B\) denote posterior draws. For each draw \(b\), and each period \(j=2,\dots, J-1\), define the draw-specific marginal prediction for the weighted cluster-period mean (averaged to the cluster-period using the pre-specified individual weights \(\omega_{ijk}\)) as
\[
m_{ij;b}(z)  =  \beta_{0j;b} + \theta_{j;b} z + \bm\beta_{X;b}^{\top}\overline{\bm X}_{ij,\omega} +  \beta_{N;b} N_{ij},
\]
since the random effects have mean zero. The posterior draw-specific model-based mean for the general weighted estimand is then
\[
\widehat\mu_{\omega;b}^{\text{mb}}(z)
= \frac{\sum_{i=1}^{I}\sum_{j=2}^{J-1} \omega_{ij}\, \widehat m_{ij;b}(z)}
{\sum_{i=1}^{I}\sum_{j=2}^{J-1} \omega_{ij}},
\qquad
\widehat\Delta_{\omega;b}^{\text{mb}}  =  f\left(\widehat\mu_{\omega;b}^{\text{mb}}(1), \widehat\mu_{\omega;b}^{\text{mb}}(0)\right).
\]
The Bayesian point estimator is given by the posterior mean $\widehat\Delta_{\omega}^{\text{mb}}
 =  \frac{1}{B}\sum_{b=1}^B \widehat\Delta_{\omega;b}^{\text{mb}}$. Subgroup-specific posterior estimators are obtained by replacing \(\omega_{ijk}\), \(\omega_{ij}\), \(\omega_j\), \(\overline Y_{ij}\), and \(\overline{\bm X}_{ij,\omega}\) with their subgroup-restricted counterparts, which gives \(\widehat\Delta_{\omega}^{\text{mb},(r)}\). 

For the model-robust standardization of Section~\ref{sec:Model-robust standardization}, define for each draw \(b\) the plug-in regression \(m_{ij;b}(z)\). %and write \(\widehat{\mathbb E}_{;b}[\overline Y_{ij}\mid Z_{ij} = z,\bm X_{ij},N_{ij}]=\widehat m_{ij;b}(z)\). 
Then, the model-robust point estimator is
\[
\widehat\mu_{\omega;b}^{\text{mr}}(z)
=\frac{1}{\sum_{j=2}^{J-1}\omega_j}\sum_{j=2}^{J-1}\omega_j
\left\{
\frac{\sum_{i=1}^I \omega_{ij}\,\mathbb I(Z_{ij}=z)\,\big(\overline Y_{ij}-\widehat m_{ij;b}(z)\big)}{\sum_{i=1}^I \omega_{ij}\,\mathbb I(Z_{ij}=z)}
+ \frac{\sum_{i=1}^I \omega_{ij}\, \widehat m_{ij;b}(z)}{\omega_j}
\right\},
\]
with \(\omega_j=\sum_{i=1}^I \omega_{ij}\). The model-robust posterior draw of the effect is
\(
\widehat\Delta_{\omega;b}^{\text{mr}} = f \left(\widehat\mu_{\omega;b}^{\text{mr}}(1),\widehat\mu_{\omega;b}^{\text{mr}}(0)\right)
\),
and the Bayesian point estimator is 
$$\widehat\Delta_{\omega}^{\text{mr}}=\frac{1}{B}\sum_{b=1}^B \widehat\Delta_{\omega;b}^{\text{mr}}.$$
%Subgroup versions follow by replacing \(\left(\omega_{ijk},\omega_{ij}\right)\) with \(\left(\omega_{ijk}^{(r)},\omega_{ij}^{(r)}\right)\). 
For the point estimator derived from the model-robust standardization, we prove in the following Theorem~\ref{prop:mr_consistency} (Section 1 of the Web Appendix), via empirical-process arguments, that this estimator is model-robust and root-$I$ consistent, even under outcome model misspecification. While \citet{fang2026model} established consistency of the corresponding frequentist estimator via convergence in probability, we provide an explicit root-$I$ convergence rate with a formal proof for Bayesian working models. We use $\widehat{\mu}_{\omega}^{\text{mr}}(z)$ to denote the model-robust estimator of the arm-specific mean potential outcomes. This estimator, $\widehat{\mu}_{\omega}^{\text{mr}}(z)$, uses the working models to estimate the conditional means $\mathbb{E} \left[ \overline{Y}_{ij}  \mid  Z_{ij}=z, \bm{X}_{ij}, N_{ij}\right]$ before standardization. Furthermore, $\widehat{\mu}_\omega^{\text{mr},(r)}(z)$ denotes the model-robust estimator of the weighted mean potential outcome under treatment level $z$ for subgroup $r \in \{1, \ldots, R\}$, and $\mu_{\omega}^{(r)}(z)$ represents the true subgroup-specific weighted mean potential outcome. The following result applies to the subgroup estimator and, by taking subgroup membership indicator \(S_{ijk,r}=1\), to the overall causal effect estimator as a special case.
\begin{theorem}\label{prop:mr_consistency}
Assuming the stepped-wedge randomization scheme, bounded design propensity $\pi_{z,j} \in (0,1)$ for $j \in \{2,\ldots,J-1\}$, and bounded cluster-period weights. Then, under additional mild regularity conditions stated in Section 1 of the Web Appendix, the model-robust point estimator $\widehat{\mu}_\omega^{\text{mr},(r)}(z)$ is root-$I$ consistent:
$$\widehat{\mu}_\omega^{\text{mr},(r)}(z) - \mu_{\omega}^{(r)}(z) = O_p(I^{-1/2}),$$
even under working outcome model misspecification.
\end{theorem}
Although we provide an illustration of the Bayesian point estimator with linear mixed models, Theorem~\ref{prop:mr_consistency} is not tied to any given parametric working model. The regularity conditions require that the fitted conditional mean functions lie, with probability tending to one, in a Donsker class with square-integrable envelope, which covers more complicated mixed-effects BART in Section \ref{sec:Bayes_BART} as well as linear mixed models.

\subsection{Calibrated variance estimation}\label{sec:Bayes_corr}
Let $\widehat\Delta_{\omega}$ serve as a general notation for the point estimator of the weighted average treatment effect $\Delta_{\omega}$, derived using either the model-based g-computation approach in Section~\ref{sec:Model-based g-computation} or the model-robust standardization approach in Section~\ref{sec:Model-robust standardization}. 
$\widehat\Delta_{\omega}$ is used to estimate one of the following estimands with pre-specified weighting schemes: h-iATE, h-cATE, v-iATE, or v-cATE. 

A classical \(95\%\) credible interval for $\Delta_{\omega}$ is obtained by taking the empirical \(2.5\)th and \(97.5\)th percentiles of the $B$ posterior draws of $\widehat\Delta_{\omega}$. However, for model-robust standardization, the estimator depends not only on the posterior draws of the conditional mean model but also directly on the observed cluster-period residuals. Therefore, posterior parameter uncertainty alone does not generally represent the full repeated-sampling variability of the estimator. This issue is especially important under working model misspecification, where standard posterior credible intervals need not achieve nominal frequentist coverage. The model-robust formulations in Section~\ref{sec:Model-robust standardization} depend jointly on the parameter vector $\Theta$ from the conditional mean model and on aggregated observed outcomes at the cluster-period level, thereby motivating a correction to the posterior variance to account for additional sampling variability and to recover nominal frequentist coverage. To improve coverage while retaining Bayesian modeling benefits, below we extend the variance calibration strategy of \citet{antonelli2022causal} (who focused on independent data) and of \citet{liu2025calibrated} (who focused on parallel-arm CRTs) to the SW-CRT setting. 

Let $\bm D=\{(Y_{ijk},Z_{ij},\bm X_{ijk},N_{ij}) : i=1,\dots,I;  j=1,\dots,J;  k=1,\dots,N_{ij}\}$ denote the observed data. We quantify uncertainty in $\widehat\Delta_{\omega}$ by combining (i) posterior variability of parameters given \(\bm D\) and (ii) sampling variability of $\bm D$ via a cluster-level resampling scheme that preserves the stepped-wedge roll-out. The sampling variability is estimated by a cluster-level bootstrap. For \(q=1,\ldots, Q\), generate bootstrap datasets \(\bm D^{(1)},\dots,\bm D^{(Q)}\) by sampling clusters \(i\) with replacement from \(\{1,\dots,I\}\). For each selected cluster, include its full trajectory across all periods, including treatment assignments, covariates, cluster-period sizes, and outcome. This preserves each cluster’s randomized roll-out path (non-decreasing \(Z_{ij}\) in \(j\)) and the within-cluster dependence structure. For each $q$-th bootstrap dataset, compute 
$$\widehat\Delta_{\omega}(\bm D^{(q)}) \approx \frac{1}{B}\sum_{b=1}^B \widehat\Delta_{{\omega};b}(\bm D^{(q)}),$$ where \(\widehat\Delta_{{\omega};b}(\bm D^{(q)})\) denotes the plug-in estimate of $\Delta_{\omega}$ on dataset \(\bm D^{(q)}\) using the \(b\)th posterior draw \(\Theta_b\). In each case, we reuse the posterior draws \(\{\Theta_b\}_{b=1}^B\) obtained from the original \(\bm D\) without refitting within the bootstrap samples, and compute the corresponding standardized means exactly as in Section~\ref{sec:Bayes_point} but replacing empirical sums over \((i,j,k)\) by those in \(\bm D^{(q)}\).

The calibrated variance estimator is then given by
\begin{equation}\label{eq:SW_calibrated_var}
\operatorname{Var}\left(\widehat{\Delta}_{\omega}\right)=\underbrace{\operatorname{Var}_{\bm D^{(q)}} \left\{\,\mathbb E_{\Theta\mid \bm D}\big[\Delta_{\omega}(\bm D^{(q)},\Theta)\big]\,\right\}}_{\text{cluster-level sampling variability of } \bm D}
 + \underbrace{\operatorname{Var}_{\Theta\mid \bm D} \left[\Delta_{\omega}(\bm D,\Theta)\right]}_{\text{posterior parameter variability}}.
\end{equation}
The first term is the empirical variance across the \(Q\) posterior mean bootstrap estimates \(\left\{\widehat\Delta_{\omega}(\bm D^{(q)})\right\}_{q=1}^Q\), and the second term is the empirical variance across posterior draw-level estimates on the original data. Throughout the remainder of the paper, we refer to credible intervals based on the posterior percentiles as \emph{uncalibrated} inference, and to intervals constructed from the calibrated variance estimator in Equation~\eqref{eq:SW_calibrated_var} via a normal approximation as \emph{calibrated} inference. Subgroup-specific intervals are constructed analogously by using the subgroup-specific weights throughout. Note that the cluster bootstrap must resample whole clusters with their period trajectories to preserve the stepped-wedge roll-out and within-cluster dependence. The following Theorem~\ref{calibrated_var_consistency} establishes the consistency of the calibrated variance estimator under correct specification of the outcome regression model, whose posterior contracts at a sufficiently fast rate. The proof is provided in Section 2 of the Web Appendix. 

\begin{theorem}\label{calibrated_var_consistency}
Assuming stepped-wedge randomization, and that the outcome model contracts at a rate faster than $I^{-1/4}$. Then, under mild regularity conditions on the data-generating process and the posterior distribution, the variance estimator $\widehat{V}=\operatorname{Var}_{\boldsymbol{D}^{(q)}}\left\{E_{\Theta\mid\boldsymbol{D}}\left[\Delta_{\omega}(\boldsymbol{D}^{(q)},\Theta)\right]\right\}+\operatorname{Var}_{\Theta\mid\boldsymbol{D}}\left[\Delta_{\omega}(\boldsymbol{D},\Theta)\right]$ is consistent, i.e., $\widehat{V} - V = o_p(I^{-1})$, where $V = \operatorname{Var}_{\boldsymbol{D}}\left\{E_{\Theta \mid \boldsymbol{D}}\left[\Delta_{\omega}\left(\boldsymbol{D}, \Theta\right)\right]\right\}$.
\end{theorem}

Theorem~\ref{calibrated_var_consistency} requires correct specification of the outcome model. Under working model misspecification, the calibrated variance estimator is no longer guaranteed to be consistent, but it tends to be conservative for the model-robust estimator, in the sense that it overestimates the true variance. In line with this, the calibrated standard errors of the model-robust estimators are comparable to or larger than the Monte Carlo standard deviations, as we empirically demonstrate in Section~\ref{sec:Simulation studies}. Moreover, valid coverage depends on the point estimator as well as the variance estimator. A consistent variance estimator alone cannot restore coverage if the point estimator is substantially biased, which is why we pair the calibrated variance with the model-robust point estimator. We emphasize that, although Theorems~\ref{prop:mr_consistency} and \ref{calibrated_var_consistency} parallel results established for parallel-arm CRTs by \citet{liu2025calibrated}, their proofs require a fundamentally different structure rather than a routine adaptation (Sections 1 and 2 of the Web Appendix). First, whereas the parallel-arm augmentation term involves the known randomization probability as a fixed denominator, the estimator in Equation~\eqref{eq:mr_estimator} is self-normalized within each period by the random quantity $\sum_{i}\mathbb{I}(Z_{ij}=z)\,\omega_{ij}^{(r)}$. We therefore decompose the error term into two parts: one that uses the design propensity $\pi_{z,j}$, and a remainder driven by the gap between the realized weighted fraction of clusters under condition $z$ in period $j$, $\widehat{\pi}_{z,j}$, and $\pi_{z,j}$. Second, each cluster contributes to the estimator in every roll-out period, and these period-specific contributions are correlated within the same cluster, so they cannot be treated as independent. We sum each cluster's contributions across periods into a single cluster-level term, so that clusters remain the independent units, and define one function class over $(\bm{X}_{ij},N_{ij},j)$ that covers all periods. The empirical process argument is then applied at the cluster level. Third, in the proof of Theorem~\ref{calibrated_var_consistency}, the key step is to show that posterior uncertainty contributes negligibly to the variance. In the parallel-arm CRT case of \citet{liu2025calibrated}, this follows because each cluster's contribution is multiplied by the deviation of its treatment indicator from the known randomization probability, which has mean zero under randomization. In SW-CRTs, our estimator does not carry this mean-zero factor directly. Instead, using Equation~\eqref{eq:mr_estimator_2}, we show that in each period this contribution is governed by the difference between two weighted averages of the derivative of the conditional mean from the working outcome model with respect to its parameters, one over all clusters and one over clusters under condition $z$. Because randomization makes treatment status independent of covariates, we show that the two averages converge to the same limit, and their difference is of order $I^{-1/2}$. These technical developments are all specific to the stepped-wedge design and have no obvious counterpart in the theory developed for parallel-arm CRTs in \citet{liu2025calibrated}.

\subsection{Leveraging mixed-effects BART as a more flexible working outcome model}\label{sec:Bayes_BART}
 
The Bayesian standardization procedures developed above require posterior draws of the conditional mean function, but do not require this function to be represented by a linear mixed model. In principle, any working outcome model that yields predictions under an imposed treatment condition \(Z_{ij}=z\) can be used in the model-based g-computation and model-robust standardization estimators. Linear mixed models provide a familiar parametric choice, but they require investigators to specify the functional form of covariate effects and, for subgroup analyses, the relevant treatment-by-subgroup interactions. This may be no easy task given the context of the SW-CRT.

In comparison, BART \citep{chipman_bart_2010} provides a more flexible nonparametric ensemble approach that models the outcome mean as a sum of regression trees, with a regularization prior to prevent overfitting. Mixed-effects BART extends this framework to clustered data by incorporating cluster- and cluster-period-specific random effects \citep{spanbauer2021nonparametric}. In SW-CRTs, we further consider the following mixed-effects BART working model:
\begin{align}
Y_{ijk} = \sum_{l=1}^L h\left(\bm{P}_{j}, Z_{ij}, \bm{X}_{ijk}, N_{ij}; T_l, M_l\right) + \alpha_i + \delta_{ij} + \epsilon_{ijk},
\label{eq:mixedBART}
\end{align}
where $\bm{P}_{j}$ denotes the $J$-dimensional vector of period indicator variables, \(T_l\) is the \(l\)-th regression tree, and \(M_l\) is the corresponding set of terminal node parameters. We include $\bm{P}_{j}$ in the BART mean function to construct the design matrix based on period dummy variables. The function $h(\cdot)$ assigns a terminal node value from $M_l$ to $\left(\bm{P}_{j},Z_{ij}, \bm{X}_{ijk},N_{ij}\right)$ according to the structure of the binary tree $T_l$. Although the causal estimands are defined over intermediate periods $j = 2, \ldots, J-1$, the working outcome model is fitted using observations from all $J$ periods to improve estimation of secular trends and variance components. The sum-of-trees term flexibly captures secular trends, nonlinear covariate effects, treatment-by-period interactions, and treatment effect heterogeneity induced by baseline covariates or cluster-period size. A default of $L=50$ trees is commonly used for adequate flexibility \citep{bleich2014variable}. The random effect \(\alpha_i\) accounts for time-invariant cluster heterogeneity, and \(\delta_{ij}\) accounts for additional cluster-period-level variation. One could alternatively allow $\delta_{ij}$ to be correlated across periods, for example through an autoregressive structure, to reflect the temporal ordering of the cluster-periods (this is known as the discrete-time decay correlation structure in the SW-CRT literature \citep{kasza2019impact}). We retain the independent specification because the random effects enter only through the working correlation structure, and because this specification is easily implemented in standard software. Analogous to the reduced model \eqref{eq:LMM_reduced}, we also fit a reduced mixed-effects BART that omits cluster-period random effects:
\begin{align}
Y_{ijk}
= \sum_{l=1}^L h \left(\bm{P}_{j},Z_{ij}, \bm{X}_{ijk}, N_{ij}; T_l, M_l\right)
+ \alpha_i + \epsilon_{ijk}.
\label{eq:mixedBART_nodelta}
\end{align}
Comparing Equations~\eqref{eq:mixedBART} and~\eqref{eq:mixedBART_nodelta} also allows us to assess the need to additionally differentiate the between-period correlation from the within-period correlation, given that the conditional mean structure is already represented by a more flexible nonparametric component. Given posterior draws from either model specification, the model-based g-computation and model-robust standardization estimators in Section~\ref{sec:Bayes_point} are obtained by replacing the estimated conditional mean with the corresponding sum-of-trees prediction averaged to the cluster-period level, and calibrated intervals follow from the resampling procedure of Section~\ref{sec:Bayes_corr}. Mixed-effects BART models are implemented using the \verb|mxbart| function from the \verb|mxBART| \citep{mxBART} package. The relative performance of linear mixed models versus mixed-effects BART under varying degrees of outcome model misspecification is examined in the simulation studies of Section~\ref{sec:Simulation studies}.

\section{Simulation studies}\label{sec:Simulation studies}
\subsection{Simulation design}
We conduct simulation studies under different data-generating models to compare the performance of Bayesian model-robust standardization and g-computation for point estimation, as well as to assess how combining each method with uncalibrated and calibrated posterior variance estimation affects frequentist coverage and efficiency. We focus on a cross-sectional design, and each simulated dataset comprises $I = 30$ clusters observed over $J = 4$ time periods. Clusters are randomized in equal numbers to adoption periods \(A_i\in\{2,3,4\}\), so that period $1$ is an all-control baseline period, period $4$ is an all-treated final period, and the causal estimands are aggregated over the roll-out periods \(\mathcal J=\{2,3\}\). Two individual-level covariates are generated: a binary covariate $X_{ijk,1} \sim B(0.5)$ and a continuous covariate $X_{ijk,2} \sim N(0, 0.01)$, where $B(p)$ denotes a Bernoulli distribution with success probability $p$ and $N(\mu, \sigma^2)$ denotes a normal distribution with mean $\mu$ and variance $\sigma^2$. The potential outcomes follow the linear mixed model (LMM):
\[
Y_{ijk}(z) = \beta_{0j} + \theta_{ijk} z + \beta_{1j} X_{ijk,1} + \beta_{2j} X_{ijk,2}^2 + \alpha_i + \delta_{ij} + \epsilon_{ijk}, \quad z \in \{0,1\},
\]
where $\beta_{0j} = 0.25 + \frac{0.02}{J - 1} (j - 1)$ denotes a linear secular trend, increasing from $0.25$ at $j=1$ to $0.27$ at $j=J$, and the remaining coefficients are specified as $\beta_{1j} = 3j/2, \ \beta_{2j} = j/J, \ j = 1, \dots, J$. We specify simulation scenarios of different levels of complexity to evaluate the impact of informative cluster size and treatment effect heterogeneity. Random effects are incorporated at multiple levels: cluster effects $\alpha_i \sim N(0, 0.05)$, cluster-period effects $\delta_{ij} \sim N(0, 0.05)$, and individual-level errors $\epsilon_{ijk} \sim N(0, 0.9)$. These components are assumed mutually independent, yielding a within-period ICC of $0.10$ and a between-period ICC of $0.05$. Cluster-period sizes $N_{ij}$ are drawn from a discrete uniform distribution $\text{Unif}(N_{\min}, N_{\max})$ over the integers $N_{\min}, \ldots, N_{\max}$. The detailed specifications of the two scenarios are provided below.

\begin{itemize}
    \item[(i)] $\theta_{ijk} = 1 + \sin(X_{ijk,1}) + \exp(-X_{ijk,2}), \quad
 N_{ij} \sim \text{Unif}(10,50)$
 \item[(ii)] $\theta_{ijk} = 1 + j \ \sin(X_{ijk,1}) - j^2 \ \exp(-X_{ijk,2}) 
+ \displaystyle\frac{\sqrt{N_{ij}}}{\frac{1}{IJ} \sum_{i=1}^I \sum_{j=1}^J N_{ij}} 
+ \displaystyle\frac{3\log(N_{ij}) N_{ij}^2}{\mathbb{E}[N_{i1}]^2}, \quad
 N_{ij} \sim \text{Unif}(5+5j, \ 45+5j)$
\end{itemize}
Scenario (i) is a baseline case in which the treatment effect depends only on covariates and is independent of cluster-period size and calendar time. With the distribution of $\theta_{ijk}$ constant over time and independent of $N_{ij}$, the values of the different weighted estimands are identical, providing a benchmark for evaluating estimator performance under non-informative cluster-period sizes. Scenario (ii) introduces informative cluster-period sizes, with the treatment effect additionally depending on calendar time and on cluster-period size. Scenario (ii) is intended as a stress test in which the dependence of the treatment effect on cluster-period size is strong, so that the weighted estimands take clearly distinct values and the consequences of targeting the wrong one are visible. The cluster-period sizes used here are also larger than those in TSOS.

To focus the presentation, we report results for the h-iATE and h-cATE only. The vertical estimands, v-iATE and v-cATE, differ solely through the pre-specified weights and enter the estimation and inference procedures in exactly the same way, so the comparisons of interest are completely analogous and we omit them for brevity. For each of the estimands in the two scenarios, we consider two outcome modeling approaches each with two specifications: (1) LMM in Equations~\eqref{eq:LMM} and \eqref{eq:LMM_reduced}, and (2) mixed-effects BART in Equations~\eqref{eq:mixedBART} and \eqref{eq:mixedBART_nodelta}, each fitted with $L=50$ and $L=10$ trees. We compare BART-based estimators with their LMM counterparts under both g-computation and model-robust frameworks, each combined with either calibrated or uncalibrated posterior inference. For LMM, posterior inference is based on a total of $3{,}000$ Markov chain Monte Carlo (MCMC) iterations with the first $1{,}000$ discarded as burn-in. For mixed-effects BART, we perform $10{,}000$ MCMC iterations with the first $5{,}000$ discarded as burn-in. In the simulations, convergence is monitored using the Geweke diagnostic across replicates, with traceplots inspected for a subset of replicates.

Aside from overall causal effect estimation, we also estimate subgroup-specific causal effects defined by the individual-level binary covariate $X_{ijk,1}$, which may represent, for example, a medical history indicator or a comorbidity indicator in practice. For instance, $X_{ijk,1} = 1$ may denote individuals with a history of hypertension, while $X_{ijk,1} = 0$ denotes those without. In this setting, the binary covariate serves as the subgroup membership indicator $S_{ijk,r} = \mathbb{I} (X_{ijk,1} = r)$, for $r \in \{0,1\}$. For subgroup analysis, in addition to the four working outcome models described above, we fit LMMs that include a treatment-by-subgroup interaction for the subgroup-defining covariate. The full and reduced specifications are given in Equations~\eqref{eq:subgroup_LMM} and \eqref{eq:subgroup_LMM_reduced}, respectively. We estimate and report the causal estimands for the subgroup with $X_{ijk,1}=1$, using the subgroup-specific weighted estimand given in Equation~\eqref{subgroup-specific weighted estimand}. The estimation procedure parallels that of the overall causal effect, except that the weights are restricted to the subgroup of interest. 

We evaluate finite-sample performance of each method using relative bias, Monte Carlo standard deviation (MCSD), average estimated standard error (AESE), and empirical coverage. Let $u=1,\ldots,U$ index simulation replicates. For a given estimand $\Delta_{\omega}$, let $\widehat{\Delta}_{\omega;b;u}$ denote the draw-level estimate computed from the $u$-th simulated dataset using the $b$-th posterior draw. The final point estimate is the average of the posterior mean across all replicates, defined as $\widehat{\Delta}_{\omega}=U^{-1}\sum_{u=1}^U \widehat{\Delta}_{\omega;;u}$, where $\widehat{\Delta}_{\omega;;u} = B^{-1}\sum_{b=1}^B \widehat{\Delta}_{\omega;b;u}$. Relative bias is given by $(\widehat{\Delta}_{\omega}-\Delta_{\omega})/\Delta_{\omega}$, where the true estimand value $\Delta_{\omega}$ is approximated from a simulated super-population of $300{,}000$ clusters. Smaller absolute relative bias indicates more accurate estimation. Variability across Monte Carlo replications is summarized by the MCSD, given by $\{(U-1)^{-1}\sum_{u=1}^U(\widehat{\Delta}_{\omega;;u}-\widehat{\Delta}_{\omega})^2\}^{1/2}$. Posterior uncertainty is summarized by the AESE, which is the average across simulation replicates of the posterior standard deviation within each replicate, defined as $U^{-1}\sum_{u=1}^U\{(B-1)^{-1}\sum_{b=1}^B(\widehat{\Delta}_{\omega;b;u}-\widehat{\Delta}_{\omega;;u})^2\}^{1/2}$. For uncalibrated inference, AESE is computed as the average posterior standard deviation across simulation replicates. For calibrated inference, AESE is computed as the average calibrated standard error from Equation \eqref{eq:SW_calibrated_var}. Comparable values of MCSD and AESE indicate that the estimated posterior uncertainty appropriately reflects sampling variability. Empirical coverage is computed as the proportion of Monte Carlo replicates for which the corresponding $95\%$ credible interval contains the true estimand value, considering both uncalibrated and calibrated posterior credible intervals constructed as described in Section~\ref{sec:Bayes_corr}. All simulations were implemented in R version 4.3.1. Because the simulations are computationally intensive, all simulations are based on \(U=100\) replicates.

\subsection{Results for overall causal effect estimands}
\begin{table}[htbp]
  \centering
  \caption{Estimation of the overall h-cATE under Scenario (i) with non-informative cluster-period sizes and Scenario (ii) with informative cluster-period sizes, based on $100$ simulation replications. Working outcome models include linear mixed models (LMM) and mixed-effects Bayesian additive regression trees (BART), each fit under exchangeable and nested exchangeable random-effects structures; for BART, $L$ denotes the number of trees. For the calibrated variance estimation method, $100$ resampled datasets are used within each simulation via the proposed cluster bootstrap procedure. ``RB'' indicates the relative bias of the estimated value. ``Coverage'' refers to the $95$\% credible interval coverage across all simulated datasets. ``MCSD'' denotes Monte Carlo standard deviation. ``AESE'' denotes the average estimated standard error.}
\begin{adjustbox}{width=\textwidth}
\begin{tabular}{cclcrrrrrrrr}
\toprule
& & & & \multicolumn{4}{c}{\textbf{Exchangeable}} & \multicolumn{4}{c}{\textbf{Nested exchangeable}} \\
\cmidrule(lr){5-8} \cmidrule(lr){9-12}
\textbf{Scenario} & \textbf{Working model} & \textbf{Estimator} & \textbf{Posterior Inference}
& \textbf{RB (\%)} & \textbf{Coverage (\%)} & \textbf{MCSD} & \textbf{AESE}
& \textbf{RB (\%)} & \textbf{Coverage (\%)} & \textbf{MCSD} & \textbf{AESE} \\
\midrule

\multirow{12}{*}{(i)}
  & \multirow{4}{*}{LMM}
    & \multirow{2}{*}{g-computation}
      & Uncalibrated & -0.1 & 91 & 0.101 & 0.086 & -0.2 & 97 & 0.099 & 0.109 \\
  &&& Calibrated     & -0.1 & 91 & 0.101 & 0.086 & -0.2 & 97 & 0.099 & 0.109 \\
  \cmidrule{3-12}
  && \multirow{2}{*}{Model-robust}
      & Uncalibrated & -0.2 & 9  & 0.112 & 0.007 & -0.2 & 10 & 0.112 & 0.009 \\
  &&& Calibrated     & -0.2 & 93 & 0.112 & 0.109 & -0.2 & 92 & 0.112 & 0.109 \\
  \cmidrule{2-12}
  & \multirow{4}{*}{BART ($L=50$)}
    & \multirow{2}{*}{g-computation}
      & Uncalibrated & 0.3 & 78 & 0.102 & 0.068 & 0.1 & 89 & 0.093 & 0.093 \\
  &&& Calibrated     & 0.3 & 80 & 0.102 & 0.070 & 0.1 & 91 & 0.093 & 0.093 \\
  \cmidrule{3-12}
  && \multirow{2}{*}{Model-robust}
      & Uncalibrated & -0.1 & 34 & 0.115 & 0.027 & -0.1 & 30 & 0.108 & 0.017 \\
  &&& Calibrated     & -0.1 & 92 & 0.115 & 0.105 & -0.1 & 93 & 0.108 & 0.106 \\
  \cmidrule{2-12}
  & \multirow{4}{*}{BART ($L=10$)}
    & \multirow{2}{*}{g-computation}
      & Uncalibrated & 0.3 & 75 & 0.101 & 0.059 & 0.1 & 91 & 0.085 & 0.083 \\
  &&& Calibrated     & 0.3 & 78 & 0.101 & 0.061 & 0.1 & 91 & 0.085 & 0.084 \\
  \cmidrule{3-12}
  && \multirow{2}{*}{Model-robust}
      & Uncalibrated & -0.1 & 23 & 0.110 & 0.013 & -0.1 & 11 & 0.108 & 0.005 \\
  &&& Calibrated     & -0.1 & 90 & 0.110 & 0.107 & -0.1 & 93 & 0.108 & 0.106 \\
\midrule

\multirow{12}{*}{(ii)}
  & \multirow{4}{*}{LMM}
    & \multirow{2}{*}{g-computation}
      & Uncalibrated & 4.8 & 28 & 2.118 & 0.356 & -11.4 & 89 & 1.528 & 2.016 \\
  &&& Calibrated     & 4.8 & 27 & 2.118 & 0.365 & -11.4 & 90 & 1.528 & 2.018 \\
  \cmidrule{3-12}
  && \multirow{2}{*}{Model-robust}
      & Uncalibrated & 0.2 & 2  & 1.700 & 0.028 & 0.0 & 14 & 1.621 & 0.154 \\
  &&& Calibrated     & 0.2 & 96 & 1.700 & 1.818 & 0.0 & 98 & 1.621 & 1.811 \\
  \cmidrule{2-12}
  & \multirow{4}{*}{BART ($L=50$)}
    & \multirow{2}{*}{g-computation}
      & Uncalibrated & -1.1 & 11 & 1.709 & 0.155 & -1.6 & 31 & 1.540 & 0.281 \\
  &&& Calibrated     & -1.1 & 91 & 1.709 & 1.654 & -1.6 & 96 & 1.540 & 1.668 \\
  \cmidrule{3-12}
  && \multirow{2}{*}{Model-robust}
      & Uncalibrated & -1.2 & 8  & 1.690 & 0.130 & -1.5 & 13 & 1.540 & 0.156 \\
  &&& Calibrated     & -1.2 & 92 & 1.690 & 1.659 & -1.5 & 96 & 1.540 & 1.676 \\
  \cmidrule{2-12}
  & \multirow{4}{*}{BART ($L=10$)}
    & \multirow{2}{*}{g-computation}
      & Uncalibrated & -1.8 & 16 & 1.705 & 0.120 & -3.4 & 64 & 1.664 & 0.997 \\
  &&& Calibrated     & -1.8 & 96 & 1.705 & 1.673 & -3.4 & 93 & 1.664 & 1.836 \\
  \cmidrule{3-12}
  && \multirow{2}{*}{Model-robust}
      & Uncalibrated & -1.8 & 11 & 1.669 & 0.084 & -1.4 & 19 & 1.661 & 0.278 \\
  &&& Calibrated     & -1.8 & 97 & 1.669 & 1.678 & -1.4 & 97 & 1.661 & 1.754 \\
\bottomrule
\end{tabular}
\end{adjustbox}
  \label{tab:overall_effect_hcATE_table}
\end{table}

The main text focuses on the h-cATE, which aligns with the cluster-level decision-making context of many SW-CRTs, including the TSOS trial analyzed in Section~\ref{sec:Case Study}. Full numerical results for both h-cATE and h-iATE under the overall causal effect setting are reported in Section 3 of the Web Appendix, with subgroup results in Section 4 of the Web Appendix. The patterns under h-iATE are similar to those reported here for h-cATE. Table~\ref{tab:overall_effect_hcATE_table} presents the relative bias, coverage, MCSD, and AESE of the h-cATE estimators under both scenarios. The key finding is that model-robust standardization with calibrated inference is the only approach that consistently achieves negligible bias and close to nominal coverage across both scenarios, regardless of the working outcome model specification.

Under Scenario (i), with non-informative cluster-period sizes, all method combinations achieve negligible bias for the h-cATE. Coverage performance, however, varies substantially across inferential strategies. Only model-robust standardization with calibrated inference achieves close to nominal coverage uniformly across all working outcome model specifications. By contrast, coverage of all other method combinations is sensitive to the random-effects specification. For example, g-computation with LMM achieves nominal coverage under the correctly specified nested exchangeable random-effects structure, but coverage drops substantially when the random-effects structure is misspecified as simple exchangeable. Furthermore, model-robust standardization with uncalibrated inference yields severe undercoverage across all outcome model specifications, reflecting that posterior parameter variability alone substantially underestimates the true sampling variability. This pattern highlights the importance of calibration when using Bayesian model-robust standardization. Because the model-robust formulation incorporates both the parameter vector from the conditional expectation and aggregated cluster-period observed outcomes, posterior variance must be calibrated to account for the additional data variability. By contrast, for LMM g-computation, the calibrated and uncalibrated results are essentially identical, because the estimator depends on the data only through the fitted model parameters, so the cluster-level resampling term contributes negligibly.

Under Scenario (ii), with informative cluster-period sizes, g-computation with LMM shows bias due to model misspecification and informative cluster sizes, and the bias varies in both magnitude and direction across random-effects specifications. BART g-computation reduces bias compared to LMM and yields more stable performance across correctly specified and misspecified random-effects structures. Most notably, model-robust standardization maintains negligible bias under both LMM and BART working outcome models, demonstrating that it provides protection against both informative cluster-period sizes and random-effects structure misspecification. Regarding coverage, calibrated inference remains essential for model-robust standardization estimators, as uncalibrated posterior uncertainty fails to capture sampling variability. Across both correctly and incorrectly specified random-effects structures, model-robust standardization with calibrated inference is the only approach that achieves uniformly near-nominal coverage, regardless of the working outcome model. Regarding the BART specification, results with $L=10$ and $L=50$ trees are broadly similar. Under Scenario (ii) with the exchangeable structure, BART with $L=10$ trees yields coverage slightly closer to the nominal level than $L=50$. A possible explanation is that, with the limited number of covariates in our design, a smaller ensemble already provides adequate flexibility, suggesting that BART complexity may be tuned based on the number of available covariates in practice.

We further note a general consideration regarding BART under SW-CRT designs. Because the h-iATE and h-cATE estimands have distinct true values under informative cluster-period sizes, reliably distinguishing between them via g-computation requires the working outcome model to closely approximate the true conditional expectation. For BART, this may require a larger number of clusters than the $I=30$ setting considered here, which may explain the bias observed for BART g-computation under Scenario (ii). The same limitation also affects frequentist coverage. When the working outcome model does not yet closely approximate the truth, posterior parameter uncertainty no longer fully captures the sampling variability of the resulting estimator, leading to undercoverage of uncalibrated credible intervals. This gap is particularly pronounced for BART g-computation under Scenario (ii). As a flexible nonparametric model, BART produces stable predictions across posterior draws, yielding small posterior parameter uncertainty (small AESE). Yet under informative cluster-period sizes, BART predictions are sensitive to the composition of clusters in the sample, generating substantial sampling variability (large MCSD) that the posterior alone cannot capture. Calibration helps narrow this gap, which explains the substantial coverage improvement observed for BART g-computation under Scenario (ii). Still, this improvement is not guaranteed to reach the nominal level in general, and we recommend model-robust standardization with calibrated inference as the safest approach across settings.

Section 7 of the Web Appendix compares the proposed calibrated Bayesian model-robust standardization against its frequentist counterpart of \citet{fang2026model} with leave-one-cluster-out jackknife inference. Both frameworks show negligible bias, while the jackknife standard error tends to exceed the Monte Carlo standard deviation under informative cluster-period sizes, yielding more conservative interval coverage compared to our proposed Bayesian counterparts.

\begin{figure}[htbp]
  \centering
  \adjustbox{max width=0.95\textwidth, max height=\textheight}{%
    \includegraphics{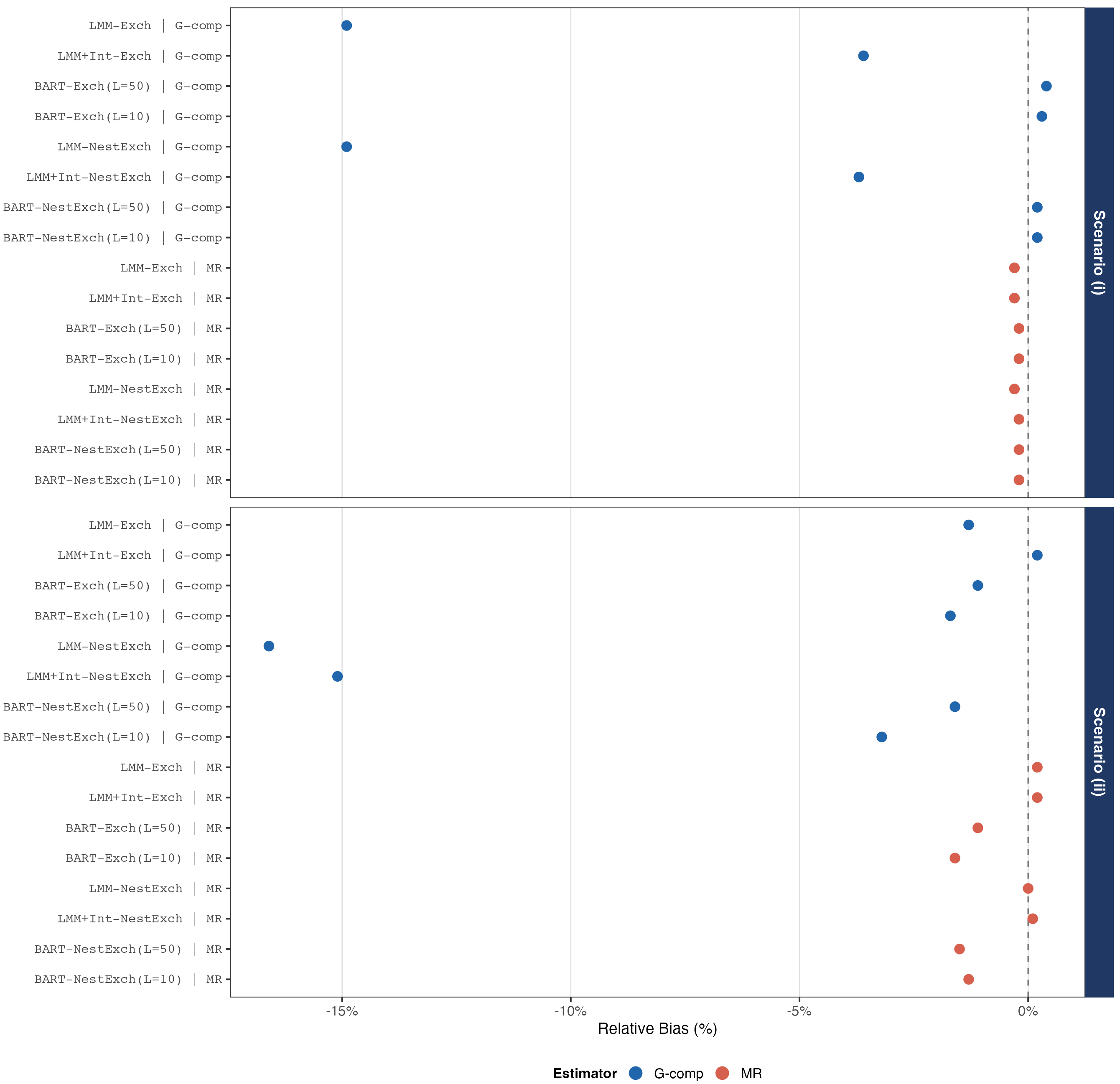}
  }
  \caption{Relative bias of the h-cATE estimator for the subgroup causal effect ($X_{ijk,1}=1$), under Scenario (i) with non-informative cluster-period sizes and Scenario (ii) with informative cluster-period sizes. Results are based on 100 simulation replications with 30 clusters and 4 periods. Working outcome models include linear mixed models without a treatment-by-subgroup interaction (LMM), linear mixed models with a treatment-by-subgroup interaction (LMM+Int), and mixed-effects Bayesian additive regression trees (BART). Each model is fit under exchangeable (Exch) and nested exchangeable (NestExch) random-effects structures; for BART, $L$ denotes the number of trees. ``G-comp'' denotes the model-based g-computation estimator and ``MR'' denotes the model-robust standardization estimator. The dashed vertical line indicates zero relative bias.}
  \label{fig:subgroup_RB_cATE}
\end{figure}

\begin{figure}[htbp]
  \centering
  \adjustbox{max width=0.9\textwidth, max height=\textheight}{%
\includegraphics{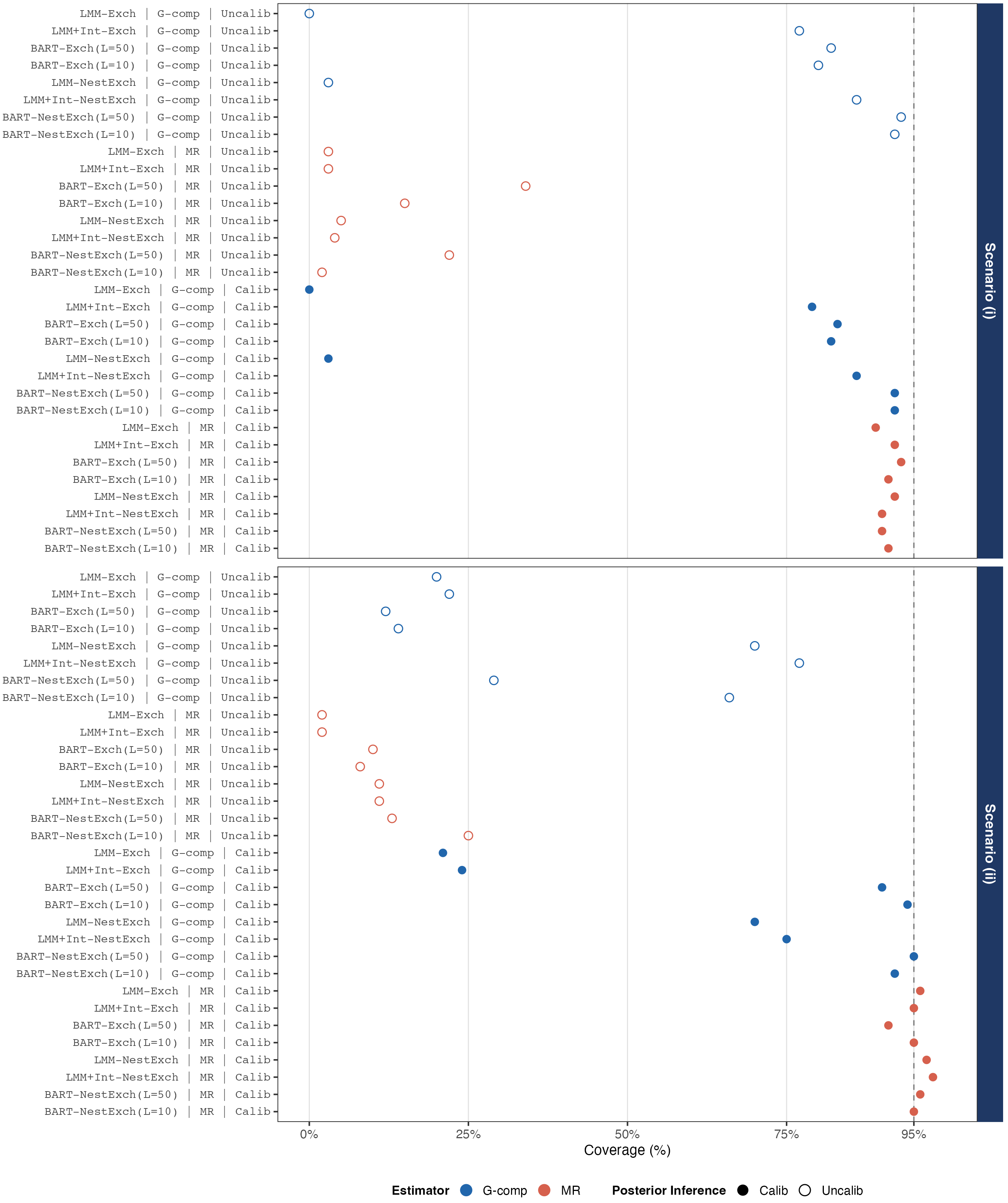}
  }
  \caption{Empirical coverage of $95\%$ credible intervals for the h-cATE estimator of the subgroup causal effect ($X_{ijk,1}=1$), under Scenario (i) with non-informative cluster-period sizes and Scenario (ii) with informative cluster-period sizes. Results are based on 100 simulation replications with 30 clusters and 4 periods. Working outcome models include linear mixed models without a treatment-by-subgroup interaction (LMM), linear mixed models with a treatment-by-subgroup interaction (LMM+Int), and mixed-effects Bayesian additive regression trees (BART). Each model is fit under exchangeable (Exch) and nested exchangeable (NestExch) random-effects structures; for BART, $L$ denotes the number of trees. ``G-comp'' denotes the model-based g-computation estimator and ``MR'' denotes the model-robust standardization estimator. ``Uncalib'' denotes uncalibrated posterior credible intervals and ``Calib'' denotes calibrated credible intervals from the proposed cluster bootstrap procedure with 100 resampled datasets. The dashed vertical line indicates the nominal $95\%$ coverage level.}
  \label{fig:subgroup_Coverage_cATE}
\end{figure}

\subsection{Results for subgroup-specific causal effects}
Figures~\ref{fig:subgroup_RB_cATE} and~\ref{fig:subgroup_Coverage_cATE} present the relative bias and empirical coverage for the subgroup-specific h-cATE under both scenarios. A central message of this analysis is that subgroup estimation introduces an additional layer of model specification challenge beyond what is required for overall causal effects. When the working outcome model fails to capture differential treatment effects across subgroups, g-computation suffers from substantial bias. Model-robust standardization with calibrated inference, by contrast, continues to provide robust subgroup-specific estimates and near-nominal coverage, regardless of how the working model encodes or omits treatment effect heterogeneity.

This dependence on working model specification is evident even in the most favorable setting, Scenario (i), where cluster-period sizes are non-informative. The LMM g-computation without a treatment-by-subgroup interaction term shows a relative bias of approximately $-15\%$ and near-zero coverage. Including the treatment-by-subgroup interaction term (Equations~\eqref{eq:subgroup_LMM} and \eqref{eq:subgroup_LMM_reduced}) substantially reduces bias and improves coverage, but does not fully resolve the issue. In sharp contrast, model-robust standardization yields negligible bias across all working models, LMM with or without treatment-by-subgroup interaction, and BART under either random-effects structure, demonstrating its robustness to a wide range of model misspecification. As before, calibrated inference is essential for model-robust estimators to achieve near-nominal coverage.

Under Scenario (ii), with informative cluster-period sizes, the limitations of LMM g-computation are further highlighted. That is, even with the treatment-by-subgroup interaction term included, bias remains substantial under both random-effects structures, with relative bias reaching nearly $-15\%$ even under the correctly specified nested exchangeable random-effects structure. This indicates that in LMM, including treatment-by-subgroup interactions alone is insufficient when the true treatment effect also depends nonlinearly on time, covariates, and cluster-period size. BART g-computation reduces bias relative to LMM and exhibits a notable practical advantage for subgroup analysis. By automatically capturing nonlinear relationships and interaction effects through its sum-of-trees structure, BART obviates the need to pre-specify a treatment-by-subgroup interaction term, a task that becomes infeasible in trials with multiple candidate subgroups.

In terms of bias alone, BART g-computation performs comparably to model-robust standardization. However, this comparability does not extend to frequentist coverage in SW-CRTs. For example, BART g-computation falls below the nominal level under some specifications, whereas calibrated model-robust standardization with any working outcome model remains uniformly close to nominal. Reliable interval coverage thus requires model-robust standardization paired with calibrated inference. Across all settings, working models, and inclusion or exclusion of treatment-by-subgroup interaction terms, this combination achieves negligible bias and near-nominal coverage, mirroring the overall causal effect results.

\section{An illustrative application to the TSOS stepped-wedge cluster-randomized trial} \label{sec:Case Study}
\subsection{Study background} 
We apply the proposed methods to data from the TSOS trial, a pragmatic SW-CRT conducted at 25 Level I trauma centers across the United States \citep{zatzick2021stepped}. The study enrolled 635 hospitalized injury survivors who reported elevated PTSD symptoms on the PCL-C, with a score of 35 or higher indicating high levels of posttraumatic distress. Trauma centers were randomized per the stepped-wedge protocol to transition from enhanced usual care to the collaborative care intervention, and participants were enrolled under the condition in effect at their center at the time of enrollment. The trial spanned $J=5$ periods, and each participant was assigned to the cluster-period $(i,j)$ corresponding to the center and period of enrollment. The intervention combined proactive injury case management with evidence-based pharmacotherapy and psychotherapeutic elements targeting PTSD symptoms. Cluster sizes as enrolled ranged from 7 to 40 participants, with a mean of 25. The primary outcome of interest is the PCL-C score assessed at 6 months postinjury. Baseline covariates included demographic characteristics, injury severity, prior trauma history, and comorbid conditions. The original analysis identified significant PTSD symptom reductions at 6 months for intervention patients but not at 12 months, with larger effects observed among patients with three or more baseline risk factors for enduring PTSD. The purpose of this section is to demonstrate the proposed estimators on real data under a common analytic setup, rather than to provide a definitive reanalysis of the trial. 

\subsection{Model specification and implementation}
All analyses adjust for age, gender, and baseline PTSD risk, categorized as fewer than three versus three or more surgical ward risk factors for enduring PTSD. The 6-month PCL-C score was missing for 144 of the 635 enrolled participants (23\%). For illustrative purposes, we impute the missing outcomes once using multilevel joint modeling imputation. The imputation model includes a center random intercept and, as fixed effects, the period indicators, the treatment-by-period interactions for the roll-out periods, baseline PTSD risk group, gender, age, the cluster-period size, and a treatment-by-risk-group interaction, so that it accommodates the subgroup-specific causal effects targeted by the subgroup analyses. Imputed values are truncated to the admissible PCL-C range of 17 to 85. The same imputed dataset is used for every working model and estimator, which keeps the analytic sample identical across all analyses. We use a single imputed dataset so that all working models and estimators are compared on identical data, at the cost of not propagating imputation uncertainty into the reported intervals, which are therefore likely to be narrower than those from a multiple imputation analysis. Since our aim is to compare estimators under a common analytic setup rather than to provide a definitive reanalysis, the estimates should not be read as a final assessment of the intervention's effect.

Subgroup-specific causal effects are estimated separately for patients with fewer than three risk factors and those with three or more risk factors. We apply the proposed methods using both LMM and mixed-effects BART as working outcome models. For LMM, we fit models corresponding to Equations~\eqref{eq:LMM} and \eqref{eq:LMM_reduced} for overall causal effect estimation, and Equations~\eqref{eq:LMM}, \eqref{eq:LMM_reduced}, \eqref{eq:subgroup_LMM}, and \eqref{eq:subgroup_LMM_reduced} for subgroup-specific causal effect estimation. For mixed-effects BART, we fit models corresponding to Equations~\eqref{eq:mixedBART} and \eqref{eq:mixedBART_nodelta} with $L=50$ trees for both overall and subgroup-specific causal effects. Each estimation task is performed using both model-based g-computation and model-robust standardization, combined with either uncalibrated or calibrated posterior inference. As in the simulation studies, the target estimands are the h-iATE and h-cATE.

For each model, we perform 30,000 MCMC iterations with the first 20,000 discarded as burn-in. Calibrated variance estimation is based on 100 cluster-level bootstrap resamples. Convergence is assessed using traceplots and the Geweke diagnostic. All analyses were implemented in R version 4.3.1. For conciseness, the main paper presents only graphical summaries of the TSOS analyses under the nested exchangeable random-effects structure. The corresponding numerical results are provided in Section 5 of the Web Appendix. Results under a simple exchangeable random-effects structure are reported in Section 6 of the Web Appendix. The main findings and patterns are consistent with those obtained under the nested exchangeable specification.

\subsection{Overall causal effects}
\begin{figure}[htbp]
  \centering
  \adjustbox{max width=\textwidth, max height=\textheight}{%
    \includegraphics{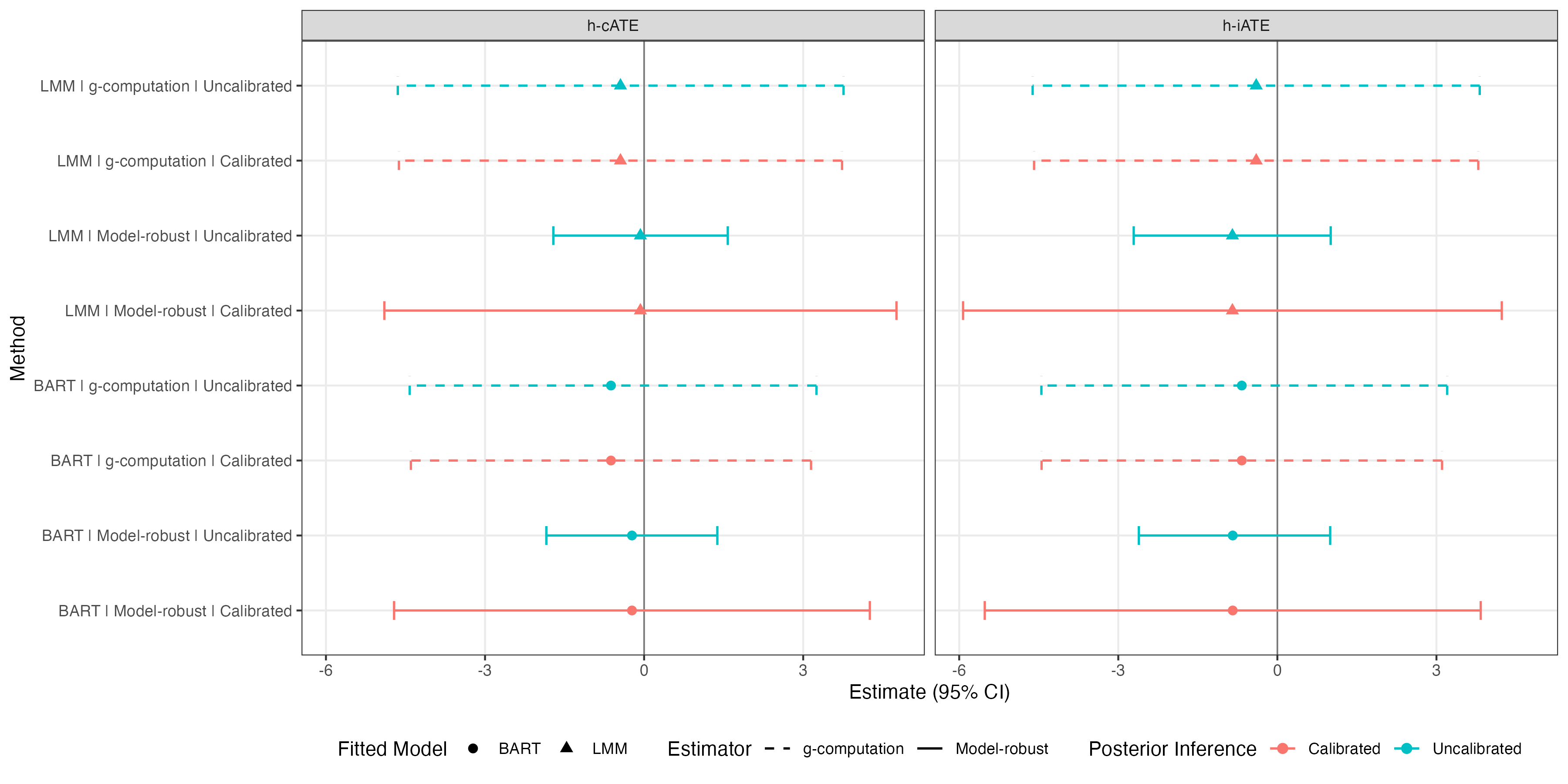}
  }
  \caption{Estimated overall causal effects on PCL-C score at 6 months postinjury in the TSOS trial under different combinations of working outcome models, estimation approaches, and posterior inference methods. Point estimates are shown with 95\% credible intervals. Dashed lines indicate g-computation estimators; solid lines indicate model-robust standardization estimators. Triangles denote linear mixed models (LMM); circles denote mixed-effects BART. Teal intervals represent uncalibrated posterior inference; coral intervals represent calibrated posterior inference.}
  \label{fig:overall_effect_TSOS}
\end{figure}

Figure~\ref{fig:overall_effect_TSOS} presents estimated overall causal effects on PCL-C score at 6 months postinjury. All point estimates are negative, with negative values indicating symptom reduction favoring the intervention, but they are small relative to the interval width. All 95\% credible intervals include zero. Consistent with simulation findings, calibrated inference yields wider credible intervals than uncalibrated inference for model-robust estimators, reflecting the additional sampling variability captured by the cluster bootstrap procedure. By contrast, the calibrated and uncalibrated intervals are nearly indistinguishable for the g-computation estimators, because these estimators depend on the data only through the fitted model parameters, so the cluster-level resampling term contributes negligibly to the variance estimation.

Our finding of no statistically significant overall causal effect differs from the original TSOS analysis, which reported a statistically significant 6-month intervention effect \citep{zatzick2021stepped}. This discrepancy likely reflects a difference in the target estimand. The original analysis reported the treatment effect coefficient from a mixed-effects regression model for repeated measures but without an explicit weighting scheme under the potential outcomes framework. In contrast, our analysis targets the 6-month h-iATE and h-cATE through an estimand-aligned approach, providing causal interpretations under explicit weighting schemes.

\subsection{Subgroup-specific causal effects}
% \begin{figure}[htbp]
%   \centering
%   \adjustbox{max width=\textwidth, max height=\textheight}{%
%     \includegraphics{Documents/data_analysis_plots/TSOS_plot_subgroups_nested_Sep16.png}
%   }
%   \caption{Estimated subgroup-specific causal effects on PCL-C score at 6 months postinjury, stratified by baseline PTSD risk (upper panel: $\geq 3$ risk factors; lower panel: $< 3$ risk factors). Point estimates are shown with 95\% credible intervals. Dashed lines indicate g-computation estimators; solid lines indicate model-robust standardization estimators. Triangles denote linear mixed models (LMM) without treatment-by-subgroup interaction; squares denote LMM with treatment-by-subgroup interaction; circles denote mixed-effects BART. Teal intervals represent uncalibrated posterior inference; coral intervals represent calibrated posterior inference.}
%   \label{fig:subgroup_effect_TSOS}
% \end{figure}

\begin{sidewaysfigure}
  \centering
  \includegraphics[
    width=\linewidth,
    height=0.8\textwidth,
    keepaspectratio
  ]{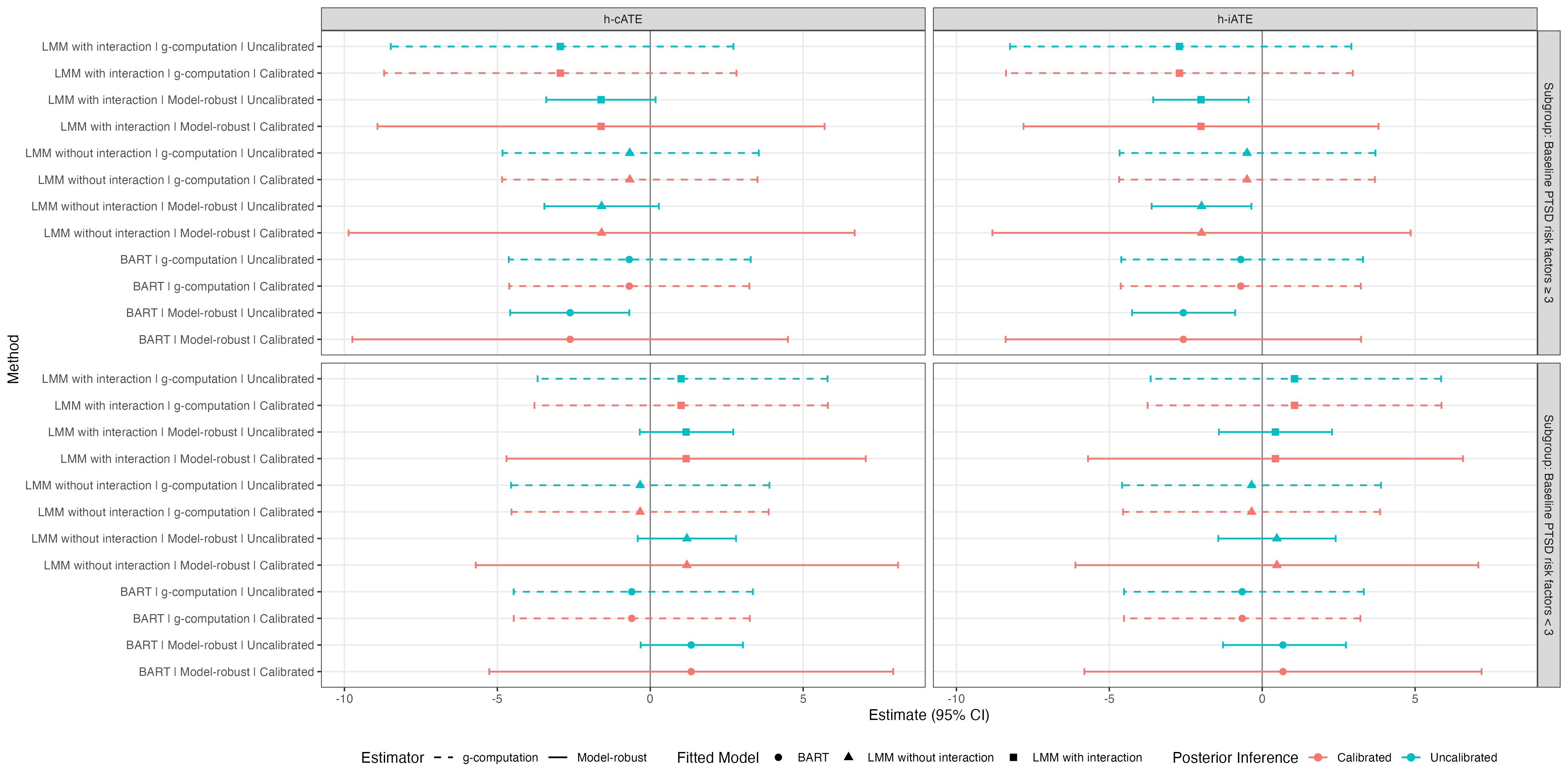}
  \caption{Estimated subgroup-specific causal effects on PCL-C score at 6 months postinjury, stratified by baseline PTSD risk (upper panel: $\geq 3$ risk factors; lower panel: $< 3$ risk factors). Point estimates are shown with 95\% credible intervals. Dashed lines indicate g-computation estimators; solid lines indicate model-robust standardization estimators. Triangles denote linear mixed models (LMM) without treatment-by-subgroup interaction; squares denote LMM with treatment-by-subgroup interaction; circles denote mixed-effects BART. Teal intervals represent uncalibrated posterior inference; coral intervals represent calibrated posterior inference.}
  \label{fig:subgroup_effect_TSOS}
\end{sidewaysfigure}

Figure~\ref{fig:subgroup_effect_TSOS} presents subgroup-specific causal effects estimation results on PCL-C score at 6 months postinjury, stratified by baseline PTSD risk. We observe a clear pattern of treatment effect heterogeneity. For patients with three or more baseline risk factors, point estimates are negative across all methods, indicating symptom reduction favoring the intervention. For patients with fewer than three risk factors, the model-robust estimates are positive, indicating no benefit of intervention. Moreover, a notable finding is that model-robust standardization produces clearly different estimates for h-cATE versus h-iATE, while g-computation yields nearly identical estimates across the two estimands. This pattern is evident in the results for the subgroup with fewer than three risk factors, and it highlights the importance of model-robust estimation when the target estimand matters, as g-computation can fail to distinguish between individual-weighted and cluster-weighted average treatment effects simply due to working model construction.

The choice of working model also has implications for subgroup analysis. For g-computation, point estimates vary considerably depending on whether the LMM includes a treatment-by-subgroup interaction and whether the working outcome model is an LMM or a mixed-effects BART model. For example, including the treatment-by-subgroup interaction in the LMM specification shifts g-computation estimates, particularly in the high-risk subgroup, underscoring the importance of appropriate model specification when using g-computation for subgroup inference. In contrast, model-robust standardization yields point estimates that vary far less across working model specifications within each estimand than those from g-computation, demonstrating its robustness to outcome model choice.

Furthermore, BART-based estimators under model-robust standardization yield the most pronounced negative estimates in the high-risk subgroup. Since the truth is unknown in the application, this pattern alone cannot establish which working model better captures treatment effect heterogeneity. The formal comparison comes from the simulation studies, where BART g-computation reduces bias relative to LMM without requiring a treatment-by-subgroup interaction to be pre-specified. Consistent with the patterns observed for overall causal effects and in simulation studies, calibrated inference continues to produce wider credible intervals for model-robust estimators compared to uncalibrated inference because it captures additional sampling variability beyond posterior parameter uncertainty. This difference is important for interpretation. That is, in the simulations, calibrated inference achieved coverage close to the nominal level, whereas the corresponding uncalibrated intervals exhibited substantial undercoverage. Thus, the narrower uncalibrated intervals can be anti-conservative; when interval exclusion of 0 is used as the decision rule, they are more likely to reject a true null effect and may therefore overstate the evidence for a treatment effect. By contrast, the calibrated approach provides a more reliable and appropriately cautious assessment of uncertainty.

\section{Discussion} \label{sec:discussion}
In this work, we develop Bayesian procedures for estimand-aligned, model-robust inference in SW-CRTs. Our work builds on recent frequentist developments that clarify weighted average treatment effect estimands in CRTs and SW-CRTs \citep{kahan2023estimands, chen2025model, fang2026model}, and extends the model-robust standardization framework \citep{li2025model, fang2026model} to a Bayesian setting. In doing so, we draw on the calibrated Bayesian variance estimation strategy of \citet{antonelli2022causal} (restricted to independent data setting) and \citet{liu2025calibrated} (restricted to simpler parallel-arm CRTs), extending it to the stepped-wedge design through a cluster-level resampling scheme that preserves the randomized roll-out. Together, these contributions enable estimand-aligned posterior inference with valid frequentist coverage, and, to our knowledge, provide the first estimand-aligned Bayesian framework for subgroup-specific causal effect estimation in SW-CRTs under informative cluster-period sizes. Our central message is that model-robust standardization is essential when targeting well-defined causal estimands under potential outcome model misspecification or informative cluster-period sizes. At the same time, calibrated posterior inference is equally essential for achieving valid frequentist coverage.

In the analysis of SW-CRTs, the choice of the working outcome model warrants careful consideration. When using linear mixed models, investigators must attend to model specification, including the inclusion of treatment-by-subgroup interactions for subgroup analysis and the appropriate random-effects structure. Our simulations show that omitting the treatment-by-subgroup interaction term leads to substantial bias for g-computation in subgroup estimation, while model-robust standardization remains robust to this specification choice. These complications are amplified when multiple subgroups defined by different covariates are of interest, as the proliferation of interaction terms can lead to identifiability issues and model instability, particularly in SW-CRTs with a limited number of clusters. Mixed-effects BART \citep{spanbauer2021nonparametric} offers an attractive alternative without the need for explicit interaction specification. As a flexible nonparametric approach, it automatically captures nonlinear relationships and treatment-by-subgroup interactions through its sum-of-trees structure, while its regularization prior helps prevent overfitting. For investigators uncertain about the functional form of treatment effect heterogeneity or interested in multiple subgroup comparisons, mixed-effects BART combined with model-robust standardization and calibrated inference provides a robust and flexible approach for estimand-aligned inference in SW-CRTs. More broadly, mixed-effects BART is one instance of a wider class of Bayesian nonparametric working models that could be used within the proposed framework \citep{daniels2023bayesian}. For example, Dirichlet process mixture priors on the random-effects distribution would relax the normality assumption on the cluster and cluster-period effects \citep{kleinman1998semi}. Since the standardization step is separate from the working model, any such model that yields posterior draws of the conditional mean can in theory be substituted directly. A parallel extension concerns the posterior calibration. Our cluster-level resampling scheme could be replaced by a Bayesian bootstrap, which places Dirichlet weights on the observed clusters rather than resampling them, and would keep the entire procedure within the Bayesian paradigm \citep{yiu2025semiparametric}. Our proof of Theorem~\ref{calibrated_var_consistency} relies on the resampled clusters being independent units and would require modification under the normalized Dirichlet weights. Whether such an approach can also improve the finite-sample coverage in SW-CRTs remains an open question and will be explored in future research.

The following two directions emerging from this work merit future investigation. First, we have focused on cross-sectional SW-CRTs, in which each period enrolls a distinct set of individuals, primarily for simplicity of exposition. In principle, the proposed methodology carries over to closed-cohort designs, and the general model-robust standardization framework accommodating closed-cohort SW-CRTs has been outlined in Web Appendix A of \citet{fang2026model}. Operationalizing this extension within our calibrated Bayesian framework, however, requires more sophisticated working outcome models with suitable working correlation structures that jointly capture the within-cluster-period dependence and the within-individual serial dependence induced by repeated measurements, and developing the details with empirical simulation validation would be a worthwhile avenue for future research. Second, our estimands accommodate informative cluster-period sizes and at most allow treatment effects to vary flexibly with calendar time, but rest on the treatment-duration irrelevance assumption. In practice, treatment effects in SW-CRTs may also vary with exposure time, that is, the time elapsed since a cluster crossed over to the intervention, and a growing literature has developed estimands and estimators that explicitly target such duration-specific effects \citep{kenny2022analysis,maleyeff2023assessing,wang2024achieve,lee2025analysis}. Bridging the literature on informative cluster-period sizes with that on exposure-time-dependent treatment effects is not a trivial task and, to our knowledge, has not been pursued under the model-robust standardization framework. Developing estimand-aligned, calibrated Bayesian inference for duration-specific effects under informative cluster-period sizes therefore represents a direction that we plan to explore in future work. More broadly, the recent shift toward estimand-aligned, model-robust inference in CRTs has unfolded largely outside the Bayesian paradigm \citep{li2025model,fang2026model}. By extending model-robust standardization and calibrated variance estimation to the stepped-wedge setting, the proposed framework allows investigators to retain the benefits of Bayesian modeling while targeting well-defined causal estimands with valid frequentist coverage. We hope such developments broaden the applicability of the model-robust standardization framework and encourage wider adoption of estimand-aligned inference in SW-CRTs, echoing the principles articulated in the recent consensus-based CRT-Estimands Framework \citep{kahan2026crtestimands, kahan2026crtJAMA}.

%%%%%%%%%%% main text

\section*{Acknowledgements}
Research in this article was supported by a Patient-Centered Outcomes Research Institute Award\textsuperscript{\textregistered} (PCORI\textsuperscript{\textregistered} Award ME-2022C2-27676). The statements presented in this article are solely the responsibility of the authors and do not necessarily represent the official views of PCORI\textsuperscript{\textregistered}, its Board of Governors, or the Methodology Committee.

\section*{Data Availability Statement}
The sample R code to implement the proposed methods can be found at the GitHub website at \url{https://github.com/Ruyi-Liu/Calibrated-Bayes-analysis-of-stepped-wedge-cluster-randomized-trials}. The Trauma Survivors Outcomes and Support (TSOS) trial data analyzed in this article are not publicly available. They were obtained from the TSOS study investigators under a data use agreement and may be made available upon reasonable request to the TSOS study team, subject to their approval.

\bibliographystyle{apalike}
\bibliography{references}

\clearpage
\newgeometry{left=0.7in, right=0.7in, top=1in, bottom=1in} % or any other margin size

\appendix

\setcounter{figure}{0}
\renewcommand{\thefigure}{A\arabic{figure}}

\setcounter{table}{0}
\renewcommand{\thetable}{\Alph{table}}

\renewcommand{\theequation}{A\arabic{equation}}
\setcounter{equation}{0}

\renewcommand{\shorttitle}{\footnotesize Supplementary Materials}
\title{Supplementary Materials}
%\section{Proof} \label{supp:sec:thm1}
% \allowdisplaybreaks
% \raggedbottom

\section*{Preliminaries: Estimators and Assumptions}

Following the decomposition and approach of \citet{antonelli2022causal}, we adapt the proof to the SW-CRT setting. Under a more general representation, the model-robust estimator for the weighted estimand $\Delta_\omega$ takes the form of
\[
{\widehat \Delta}_{\omega}^{\mathrm{mr}} = f({\widehat \mu}_{\omega}^{\mathrm{mr}}(1), {\widehat \mu}_{\omega}^{\mathrm{mr}}(0)),
\]
with
\[\widehat{\mu}_{\omega}^{\mathrm{mr}}(z)=\frac{1}{\sum_{j=2}^{J-1}\omega_j}\sum_{j=2}^{J-1}\omega_j\left\{\frac{\sum_{i=1}^{I}\omega_{ij}\mathbb{I}(Z_{ij}=z)\left(\overline{Y}_{ij}-\widehat{\mathbb{E}}[\overline{Y}_{ij}\mid Z_{ij}=z,\bm{X}_{ij},N_{ij}]\right)}{\sum_{i=1}^{I}\mathbb{I}(Z_{ij}=z)\omega_{ij}}+\frac{\sum_{i=1}^{I}\omega_{ij}\,\widehat{\mathbb{E}}[\overline{Y}_{ij}\mid Z_{ij}=z,\bm{X}_{ij},N_{ij}]}{\omega_j}\right\},
\]
where $\omega_{ijk}$ is a pre-specified non-negative individual-level weight that determines the contribution of each individual to the overall estimand, $\omega_{ij} = \sum_{k=1}^{N_{ij}} \omega_{ijk}$ represents the total weight at the cluster-period level, and these weights can be aggregated to the period level as $\omega_j = \sum_{i=1}^{I} \omega_{ij}$. The horizontal-individual ATE uses $\omega_{ijk}=1$ and hence $\omega_{ij}=N_{ij}$. The horizontal-cluster ATE sets $\omega_{ijk}=(\sum_{s=2}^{J-1}N_{is})^{-1}$, yielding $\omega_{ij}=N_{ij}/\sum_{s=2}^{J-1}N_{is}$. The vertical-individual ATE uses $\omega_{ijk}=\{I \cdot \mathbb{E}[N_{ij}]\}^{-1}$ and $\omega_{ij}=N_{ij}/\{I \cdot \mathbb{E}[N_{ij}]\}$. The vertical-cluster ATE sets $\omega_{ijk}=N_{ij}^{-1}$, so that $\omega_{ij}=1$. For simplicity, we will use the general weighted form of the estimand in the following.

The above formulation extends to subgroup-specific analysis by allowing the individual-level weights to depend on pre-specified subgroup membership. Let $S_{ijk,r}\in\{0,1\}$ denote the indicator that individual $k$ in cluster $i$ during period $j$ belongs to subgroup $r\in\{1,\ldots,R\}$. For subgroup $r$, we define subgroup-restricted weights $\omega_{ijk}^{(r)}$ by incorporating $S_{ijk,r}$ into the individual-level weighting scheme, with induced cluster-period weights $\omega_{ij}^{(r)}=\sum_{k=1}^{N_{ij}}\omega_{ijk}^{(r)}$ and period-level weights $\omega_j^{(r)}=\sum_{i=1}^I\omega_{ij}^{(r)}$. Replacing $\omega_{ijk}$, $\omega_{ij}$, and $\omega_j$ by their subgroup-specific counterparts yields the subgroup-specific model-robust estimator
$$
\begin{aligned}
\widehat{\mu}_{\omega}^{\mathrm{mr},(r)}(z)
=
\frac{1}{\sum_{j=2}^{J-1}\omega_j^{(r)}}
\sum_{j=2}^{J-1}\omega_j^{(r)} &
\left\{
\frac{\sum_{i=1}^{I}\omega_{ij}^{(r)}\mathbb{I}(Z_{ij}=z)\left(\overline{Y}_{ij}^{(r)}-\widehat{\mathbb{E}} \left[\overline{Y}_{ij}^{(r)}\mid Z_{ij}=z,\bm{X}_{ij},N_{ij}\right]\right)}
{\sum_{i=1}^{I}\mathbb{I}(Z_{ij}=z)\omega_{ij}^{(r)}}\right. \\
&+
\left.\frac{\sum_{i=1}^{I}\omega_{ij}^{(r)}\,\widehat{\mathbb{E}}\left[\overline{Y}_{ij}^{(r)}\mid Z_{ij}=z,\bm{X}_{ij},N_{ij}\right]}
{\omega_j^{(r)}}
\right\},
\end{aligned}
$$
where $\overline{Y}_{ij}^{(r)}$ denotes the weighted cluster-period mean outcome for subgroup $r$. The corresponding model-robust subgroup-specific treatment effect estimator is defined as $\widehat{\Delta}_{\omega}^{\mathrm{mr},(r)}=f(\widehat{\mu}_{\omega}^{\mathrm{mr},(r)}(1),\widehat{\mu}_{\omega}^{\mathrm{mr},(r)}(0))$. When the subgroup indicator satisfies $S_{ijk,r}=1$ for all individuals, the subgroup-specific weighting scheme reduces to the original weighting scheme, and the estimator $\widehat{\mu}_{\omega}^{\mathrm{mr},(r)}(z)$ coincides with the corresponding non-subgroup estimator $\widehat{\mu}_{\omega}^{\mathrm{mr}}(z)$. Accordingly, we adopt the subgroup-specific notation as a general framework in the remainder of the proof, with all theoretical results applying directly to the non-subgroup analysis as a special case. We focus on the proof for the arm-specific means $\mu_{\omega}(z)$. For the difference $f(x,y)=x-y$, used throughout our paper, the results extend directly to $\Delta_\omega$ because averaging over posterior draws commutes with a linear contrast. For nonlinear $f$, an additional delta-method argument with $f$ smooth near $(\mu_\omega(1),\mu_\omega(0))$ would be required. Then, 
$$
\begin{aligned}
\widehat{\mu}_{\omega}^{\mathrm{mr}, (r)}(z) - {\mu}^{(r)}_{\omega}(z) = \left( \sum_{j=2}^{J-1}\omega_j^{(r)} \right)^{-1} \sum_{j=2}^{J-1}\omega_j^{(r)}  \left\{
\frac{\sum_{i=1}^{I}\omega_{ij}^{(r)}\mathbb{I}(Z_{ij}=z)\left(\overline{Y}_{ij}^{(r)}-\widehat{m}^{(r)}_{z,ij}\right)}
{\sum_{i=1}^{I}\mathbb{I}(Z_{ij}=z)\omega_{ij}^{(r)}} + \frac{\sum_{i=1}^{I}\omega_{ij}^{(r)}\widehat{m}^{(r)}_{z,ij}}
{\omega_j^{(r)}} - {\mu}^{(r)}_{\omega}(z)
\right\},
\end{aligned}
$$
where $\widehat{m}^{(r)}_{z,ij} = \widehat{\mathbb{E}}\left[\overline{Y}_{ij}^{(r)}\mid Z_{ij}=z,\bm{X}_{ij},N_{ij}\right]$ is the estimated conditional mean (from the outcome model), and ${m}^{(r)}_{z,ij}$ is the true conditional mean. The proof of Theorem 1 does not require a parametric working model, and instead relies on the Donsker conditions stated there. For Theorem 2, we assume a parametric working model for the outcome regression and apply Taylor series expansions for $\widehat{m}^{(r)}_{z,ij}$.

We use the following decomposition:
\[
\widehat{\mu}_{\omega}^{\mathrm{mr}, (r)}(z) - {\mu}^{(r)}_{\omega}(z) = A + B,
\]
where
\begin{align*}
A &= \left( \sum_{j=2}^{J-1}\omega_j^{(r)} \right)^{-1} \sum_{j=2}^{J-1}\omega_j^{(r)}  \left\{\frac{\sum_{i=1}^{I}\omega_{ij}^{(r)}\left(\widehat{m}^{(r)}_{z,ij} - {m}^{(r)}_{z,ij}\right)}
{\omega_j^{(r)}} -\frac{\sum_{i=1}^{I}\omega_{ij}^{(r)}\mathbb{I}(Z_{ij}=z)\left(\widehat{m}^{(r)}_{z,ij} - {m}^{(r)}_{z,ij}\right)}
{\sum_{i=1}^{I}\mathbb{I}(Z_{ij}=z)\omega_{ij}^{(r)}}\right\} \\
B &= \left( \sum_{j=2}^{J-1}\omega_j^{(r)} \right)^{-1} \sum_{j=2}^{J-1}\omega_j^{(r)}  \left\{
\frac{\sum_{i=1}^{I}\omega_{ij}^{(r)}\mathbb{I}(Z_{ij}=z)\left(\overline{Y}_{ij}^{(r)}-{m}^{(r)}_{z,ij}\right)}
{\sum_{i=1}^{I}\mathbb{I}(Z_{ij}=z)\omega_{ij}^{(r)}} + \frac{\sum_{i=1}^{I}\omega_{ij}^{(r)}{m}^{(r)}_{z,ij}}
{\omega_j^{(r)}} - {\mu}^{(r)}_{\omega}(z)
\right\}.
\end{align*}

\paragraph{Assumptions}
Let $P_0 \in \mathcal{P}$ denote the true (unknown) probability distribution that generates the observed data. In our setting, this corresponds to the distribution of the observed cluster trajectories: $(Z_{ij}, \bm{X}_{ij}, N_{ij}, \bm{Y}_{ij})$, where $Z_{ij}$ is the treatment assignment, $\bm{X}_{ij}$ denotes the matrix of covariates for cluster $i$ period $j$, $N_{ij}$ is the cluster-period size, and $\bm{Y}_{ij}$ is the vector of observed outcomes for individuals in cluster $i$ period $j$. Consider the collection of all random variables in cluster $i$,
\[
\bm{\widetilde{O}}_i^{\text{full}}
=
\left\{
\bm{Y}_i(\bm{z}),
\bm{N}_i,
\bm{X}_i
:
\bm{z}\in\mathcal{Z}
\right\},
\]
where $\bm{Y}_i(\bm{z})=\{\bm Y_{ij}(z_j): j=1,\ldots,J\}$ denotes the vector of potential outcomes for cluster $i$ across periods under a treatment sequence $\bm{z}=(z_1,\ldots,z_J)$, $\mathcal{Z}$ denotes the set of treatment sequences admissible under the stepped-wedge design (i.e., monotone sequences with a single crossover from control to treatment), $\bm{N}_i=\{N_{ij}: j=1,\ldots,J\}$ denotes the cluster-period sizes, and $\bm{X}_i=\{\bm{X}_{ij}: j=1,\ldots,J\}$ denotes the corresponding covariates. The complete data consist of $\{(\bm{\widetilde{O}}_i^{\text{full}}, \bm{Z}_i): i=1,\ldots,I\}$, where $\bm{Z}_i\in\mathcal{Z}$ is the treatment sequence for cluster $i$, and only the components of $\bm{Y}_i(\bm{Z}_i)$ corresponding to the realized sequence are observed.

\noindent \textbf{Assumption 1 (Super-population)}:
\begin{itemize}
    \item[(a)] Random variables $ \bm{\widetilde{O}}_1^{full}, \ldots, \bm{\widetilde{O}}_I^{full} $ are mutually independent.
    \item[(b)] The total number of individuals in cluster $i$ during period $j$, $N_{ij}$, follows an unknown distribution $\mathcal{P}^N$ over a finite support on $\mathbb{N}^{+}$.
    \item[(c)] $\sup_{P_0} \mathbb{E}_{P_0} \left[ \left( \overline{Y}_{ij}^{(r)}-{m}_{z,ij}^{(r)} \right)^2 \right] \leq K_y < \infty$.
    \item[(d)] $\sup_{P_0} \mathbb{E}_{P_0} \left[ \left( m^{(r)}_{z,ij} \right)^2 \right] \leq \widetilde{K}_y < \infty$.
\end{itemize}

\noindent \textbf{Assumption 2 (Bound on posterior dispersion of the outcome regression)}: $\sup_{P_0} \mathbb{E}_{P_0} \operatorname{Var} \left( \widehat{m}_{z,ij}^{(r)} - {m}_{z,ij}^{(r)} \mid \boldsymbol{D} \right) \leq K_m < \infty$.

It states that the posterior variance of the cluster-period conditional mean is uniformly bounded. Because $m_{z,ij}^{(r)}$ is fixed, this is a bound on posterior dispersion. It is a mild regularity condition in applications.

\noindent \textbf{Assumption 3 (Posterior contraction of the outcome model).}
There exists a sequence $\epsilon_{I} \to 0$ and a constant $Q > 0$, independent of $\epsilon_{I}$, such that
\[
\sup_{P_0} \mathbb{E}_{P_0} \mathbb{P}_I \left(
\frac{1}{\sqrt{I}} \left\| \widehat{\bm m}_z^{(r)} - \bm m_z^{(r)} \right\|
> Q \epsilon_{I}
\;\middle|\; \boldsymbol{D}
\right) \to 0,
\]
where $I$ denotes the number of clusters, $\widehat{\bm m}_z^{(r)} = \{\widehat m_{z,ij}^{(r)}\}_{i=1,\ldots,I;\, j=2,\ldots,J-1}$ and $\bm m_z^{(r)} = \{ m_{z,ij}^{(r)} \}_{i=1,\ldots,I;\, j=2,\ldots,J-1}$ collect the estimated and true cluster-period conditional mean outcomes under treatment level $z$, and $\|\cdot\|$ denotes the Euclidean norm taken over the cluster-period entries. This assumption states that the posterior distribution of the outcome model contracts to the true conditional mean function at rate $\epsilon_I$. We will use a Taylor series expansion for $\widehat{m}_{z,ij}^{(r)}$ of the following form:
\[
\widehat{m}_{z,ij}^{(r)}
=
m_{z,ij}^{(r)}
+ d_1(\bm{X}_{ij}, N_{ij}, \Theta^*)^\top (\Theta - \Theta^*)
+ (\Theta - \Theta^*)^\top
H_1(\bm{X}_{ij}, N_{ij}, \widetilde{\Theta})
(\Theta - \Theta^*),
\]
where $\Theta^*$ is the true value of $\Theta$, $\widetilde{\Theta}$ lies on the line segment between $\Theta$ and $\Theta^*$, and the outcome regression function has bounded first and second derivatives. Here and in Assumption~4, $\Theta$ denotes only the fixed-dimensional vector of regression coefficients that enter the conditional mean ${m}^{(r)}_{z,ij}$. The cluster and cluster-period random effects are integrated out in this conditional mean.

\noindent\textbf{Assumption 4 (Moments of the posterior distribution)}: (a) $\mathbb{E}_{\Theta|\bm D}[\|\Theta - \Theta^*\|_2^4] = o_p(I^{-1})$; (b) $\mathbb{E}_{\bm D}\big[\mathbb{E}_{\Theta|\bm D}\|\Theta - \Theta^*\|_2^4\big] = o(I^{-1})$ and $\mathbb{E}_{\bm D}\|\bm a_{\mathrm{diff}}(\bm D)\|_2^4 = O(I^{-2})$, where $\bm a_{\mathrm{diff}}$ is the centered derivative difference defined in Appendix~2. Part (a) controls the fourth posterior moment of the regression-parameter error at a rate sufficient for the Taylor remainder bounds; part (b) is an integrated version used only for the repeated-sampling variance in Appendix~2.

\noindent\textbf{Assumption 5 (Regularity conditions on cluster weights).}  
There exist constants \( 0 < b_{\omega} < B_{\omega} < \infty \), such that for all clusters \( i = 1, \ldots, I\) and periods \( j = 1, \ldots, J\),
\[
b_{\omega} \leq \omega_{ij}^{(r)} \leq B_{\omega}.
\]
This ensures that each cluster-period weight is bounded above and bounded away from zero. For subgroup analyses, Assumption 5 requires every cluster-period to contain at least one subgroup-$r$ member. The theoretical results below are established under this condition and do not cover the zero-weight convention for empty subgroup cells.

\noindent\textbf{Assumption 6 (Randomization and contemporaneous conditional mean).} Let $\bm{G}_i=\{(\bm X_{ij},N_{ij}):j=1,\ldots,J\}$ denote the covariate and cluster-period size history of cluster $i$, which determines the weights $\omega_{ij}^{(r)}$. (a) The treatment sequence $\bm Z_i$ is independent of $\bm G_i$ and of the potential outcomes, so that $\mathbb{I}(Z_{ij}=z)$ is independent of $\bm G_i$ with $P(Z_{ij}=z)=\pi_{z,j}$. (b) For $j=2,\ldots,J-1$ and $z\in\{0,1\}$, $\mathbb{E}[\overline{Y}_{ij}^{(r)}\mid Z_{ij}=z,\bm G_i]=\mathbb{E}[\overline{Y}_{ij}^{(r)}\mid Z_{ij}=z,\bm X_{ij},N_{ij}]=m_{z,ij}^{(r)}$.

Under Assumption 6, every tower-property step below that conditions on $(\bm X_{ij},N_{ij})$ remains valid with the conditioning set enlarged to $\bm G_i$, with respect to which $\omega_{ij}^{(r)}$ is measurable. Part (b) is the contemporaneous conditional mean restriction and is needed only for weights that depend on cluster-period sizes from other periods, such as those of the h-cATE.

\newpage
\clearpage

\section*{Appendix 1. Proof of Theorem 1: Root-$I$ Consistency of the Model-Robust Point Estimator}

Recall the decomposition $\widehat{\mu}_{\omega}^{\mathrm{mr},(r)}(z) - 
\mu_{\omega}^{(r)}(z) = A + B$ introduced above, where
\begin{align*}
A &= \left( \sum_{j=2}^{J-1}\omega_j^{(r)} \right)^{-1} \sum_{j=2}^{J-1}\omega_j^{(r)}  \left\{\frac{\sum_{i=1}^{I}\omega_{ij}^{(r)}\left(\widehat{m}^{(r)}_{z,ij} - {m}^{(r)}_{z,ij}\right)}
{\omega_j^{(r)}} -\frac{\sum_{i=1}^{I}\omega_{ij}^{(r)}\mathbb{I}(Z_{ij}=z)\left(\widehat{m}^{(r)}_{z,ij} - {m}^{(r)}_{z,ij}\right)}
{\sum_{i=1}^{I}\mathbb{I}(Z_{ij}=z)\omega_{ij}^{(r)}}\right\}, \\
B &= \left( \sum_{j=2}^{J-1}\omega_j^{(r)} \right)^{-1} \sum_{j=2}^{J-1}\omega_j^{(r)}  \left\{
\frac{\sum_{i=1}^{I}\omega_{ij}^{(r)}\mathbb{I}(Z_{ij}=z)\left(\overline{Y}_{ij}^{(r)}-{m}^{(r)}_{z,ij}\right)}
{\sum_{i=1}^{I}\mathbb{I}(Z_{ij}=z)\omega_{ij}^{(r)}} + \frac{\sum_{i=1}^{I}\omega_{ij}^{(r)}{m}^{(r)}_{z,ij}}
{\omega_j^{(r)}} - {\mu}^{(r)}_{\omega}(z)
\right\}.
\end{align*}
We examine the contraction rates for each component and show that, even under outcome model misspecification, the estimator remains root-$I$ consistent, that is, $\widehat{\mu}_{\omega}^{\mathrm{mr}, (r)}(z) - {\mu}^{(r)}_{\omega}(z)$ contracts at rate $I^{-1/2}$.

\paragraph{Contraction rate of component $B$}

We begin with component $B$, which does not depend on the posterior distribution of the outcome model. Decompose $B=B^{\star}+B^{\star\star}$, where 
\begin{align*}
B^{\star} &= \left( \sum_{j=2}^{J-1}\omega_j^{(r)} \right)^{-1} \sum_{j=2}^{J-1}\omega_j^{(r)}  \left\{
\frac{\sum_{i=1}^{I}\omega_{ij}^{(r)}\mathbb{I}(Z_{ij}=z)\left(\overline{Y}_{ij}^{(r)}-{m}^{(r)}_{z,ij}\right)}
{\sum_{i=1}^{I}\mathbb{I}(Z_{ij}=z)\omega_{ij}^{(r)}}
\right\},\\
B^{\star\star} &= \left( \sum_{j=2}^{J-1}\omega_j^{(r)} \right)^{-1} \sum_{j=2}^{J-1}\omega_j^{(r)}  \left\{\frac{\sum_{i=1}^{I}\omega_{ij}^{(r)}{m}^{(r)}_{z,ij}}
{\omega_j^{(r)}} - {\mu}^{(r)}_{\omega}(z)
\right\}.
\end{align*}

\begin{align*}
&\sup_{P_0} \mathbb{E}_{P_0} \mathbb{P}_I \left(|B^{\star}| > C\epsilon_I \mid \boldsymbol{D}\right) \\
&=\sup_{P_0} \mathbb{E}_{P_0} \mathbb{P}_I \left(\left|\left( \sum_{j=2}^{J-1}\omega_j^{(r)} \right)^{-1} \sum_{j=2}^{J-1}\omega_j^{(r)}  \left\{
\frac{\sum_{i=1}^{I}\omega_{ij}^{(r)}\mathbb{I}(Z_{ij}=z)\left(\overline{Y}_{ij}^{(r)}-{m}^{(r)}_{z,ij}\right)}
{\sum_{i=1}^{I}\mathbb{I}(Z_{ij}=z)\omega_{ij}^{(r)}}
\right\}\right| > C\epsilon_I \middle| \boldsymbol{D}\right)\\
&=\sup_{P_0} \mathbb{E}_{P_0} \mathbb{I} \left(\left|\left( \sum_{j=2}^{J-1}\omega_j^{(r)} \right)^{-1} \sum_{j=2}^{J-1}\omega_j^{(r)}  \left\{
\frac{\sum_{i=1}^{I}\omega_{ij}^{(r)}\mathbb{I}(Z_{ij}=z)\left(\overline{Y}_{ij}^{(r)}-{m}^{(r)}_{z,ij}\right)}
{\sum_{i=1}^{I}\mathbb{I}(Z_{ij}=z)\omega_{ij}^{(r)}}
\right\}\right| > C\epsilon_I \right)\\
&=\sup_{P_0} \mathbb{P}_{P_0} \left(\left|\left( \sum_{j=2}^{J-1}\omega_j^{(r)} \right)^{-1} \sum_{j=2}^{J-1}\omega_j^{(r)}  \left\{
\frac{\sum_{i=1}^{I}\omega_{ij}^{(r)}\mathbb{I}(Z_{ij}=z)\left(\overline{Y}_{ij}^{(r)}-{m}^{(r)}_{z,ij}\right)}
{\sum_{i=1}^{I}\mathbb{I}(Z_{ij}=z)\omega_{ij}^{(r)}}
\right\}\right| > C\epsilon_I \right)\\
& \leq \sup_{P_0} \frac{\mathbb{E}_{P_0}\left[\left|\left( \sum_{j=2}^{J-1}\omega_j^{(r)} \right)^{-1} \sum_{j=2}^{J-1}\omega_j^{(r)}  \left\{
\frac{\sum_{i=1}^{I}\omega_{ij}^{(r)}\mathbb{I}(Z_{ij}=z)\left(\overline{Y}_{ij}^{(r)}-{m}^{(r)}_{z,ij}\right)}
{\sum_{i=1}^{I}\mathbb{I}(Z_{ij}=z)\omega_{ij}^{(r)}}
\right\}\right|^2\right]}{C^2 \epsilon_I^2} \\
& \leq \sup_{P_0} \frac{1}{C^2 \epsilon_I^2 \cdot (J-2)^2 \cdot I^2 \cdot b_{\omega}^2} \mathbb{E}_{P_0} \left[\left| \sum_{j=2}^{J-1}\omega_j^{(r)}  \left\{
\frac{\sum_{i=1}^{I}\omega_{ij}^{(r)}\mathbb{I}(Z_{ij}=z)\left(\overline{Y}_{ij}^{(r)}-{m}^{(r)}_{z,ij}\right)}
{\sum_{i=1}^{I}\mathbb{I}(Z_{ij}=z)\omega_{ij}^{(r)}}
\right\} \right|^2\right]\\
&\le \sup_{P_0} \frac{1}{C^2 \epsilon_I^2 \cdot (J-2)^2 \cdot I^2 \cdot b_{\omega}^2} \cdot (J-2) \cdot C_1^2 \cdot \sum_{j=2}^{J-1}
\mathbb{E}_{P_0}\Biggl[\Biggl|
\sum_{i=1}^{I}\omega_{ij}^{(r)}\I(Z_{ij} =z) \left(\overline{Y}_{ij}^{(r)}-{m}^{(r)}_{z,ij}\right)
\Biggr|^2\Biggr] \\
&\quad \text{since } \left|\sum_{j=2}^{J-1} a_j\right|^2 \le (J-2)\sum_{j=2}^{J-1}|a_j|^2 \text{, and } 0<\frac{\omega_j^{(r)}}{\sum_{i=1}^{I}\I(Z_{ij} =z)\omega_{ij}^{(r)}} \le C_1\\
&\quad \text{for some constant } C_1 < \infty \text{ under the stepped-wedge randomization.}\\
&= \sup_{P_0} \frac{C_1^2}{C^2 \epsilon_I^2 \cdot (J-2) \cdot I^2 \cdot b_{\omega}^2} \cdot \sum_{j=2}^{J-1}
\sum_{i=1}^{I}
\mathbb{E}_{P_0}\!\left[\left\{
\omega_{ij}^{(r)}\I(Z_{ij}=z)\bigl(\overline Y_{ij}^{(r)}-m_{z,ij}^{(r)}\bigr)
\right\}^2\right] \\
&\quad \text{since cross terms vanish because the i.i.d.\ summands are mean-zero by Assumption~1(a) and the tower property} \\
&\quad \mathbb{E}_{P_0}[\omega_{ij}^{(r)}\,\I(Z_{ij}=z)\,(\overline Y_{ij}^{(r)} - m^{(r)}_{z,ij})] = \mathbb{E}_{P_0}\bigl[\omega_{ij}^{(r)}\, \I(Z_{ij}=z)\, \mathbb{E}_{P_0}[\overline Y_{ij}^{(r)} - m^{(r)}_{z,ij} \mid Z_{ij}=z, \bm X_{ij}, N_{ij}]\bigr] = 0\\
&\le \sup_{P_0} \frac{C_1^2}{C^2 \epsilon_I^2 \cdot (J-2) \cdot I^2 \cdot b_{\omega}^2} \cdot \sum_{j=2}^{J-1}
\sum_{i=1}^I
B_\omega^2\,
\mathbb E_{P_0}\!\left[
\bigl(\overline Y_{ij}^{(r)}-m_{z,ij}^{(r)}\bigr)^2
\right] \\
&\le \frac{C_1^2}{C^2 \epsilon_I^2 \cdot (J-2) \cdot I^2 \cdot b_{\omega}^2} \cdot \sum_{j=2}^{J-1}
\sum_{i=1}^I
B_\omega^2
K_y\\
& = \frac{C_1^2 B_\omega^2 K_y}{C^2 \epsilon_I^2 b_{\omega}^2} \cdot \frac{1}{I}.
\end{align*}
The second equality holds because $B^\star$ is a function of the observed data $\bm{D}$, so the conditional probability $\mathbb{P}_I(\cdot \mid \bm{D})$ reduces to an indicator, whose expectation is the marginal probability. The step from the tail probability to the second-moment bound applies Markov’s inequality $\mathbb{P}(|X| > t) \le \mathbb{E}[X^2]/t^2$. The remaining steps use $|\sum_j a_j|^2 \le (J-2)\sum_j |a_j|^2$, the bounded ratio $\omega_j^{(r)}/\sum_i \I(Z_{ij}=z)\omega_{ij}^{(r)} \le C_1$ under stepped-wedge randomization, the i.i.d.\ cluster structure under the super-population framework, the bounded weights (Assumption~5), and the residual second moment bound $\mathbb{E}_{P_0}[(\overline{Y}_{ij}^{(r)} - m^{(r)}_{z,ij})^2] \le K_y$ (Assumption~1(c)). Hence $B^\star$ contracts at rate $I^{-1/2}$, since the right-hand side $\to 0$ whenever $\sqrt{I}\epsilon_I \to \infty$.

\begin{align*}
&\sup_{P_0} \mathbb{E}_{P_0} \mathbb{P}_I \left(|B^{\star\star}| > C\epsilon_I \mid \boldsymbol{D}\right) \\
&=\sup_{P_0} \mathbb{E}_{P_0} \mathbb{P}_I \left(\left|\left( \sum_{j=2}^{J-1}\omega_j^{(r)} \right)^{-1} \sum_{j=2}^{J-1}\omega_j^{(r)}  \left\{ \frac{\sum_{i=1}^{I}\omega_{ij}^{(r)}{m}^{(r)}_{z,ij}}
{\omega_j^{(r)}} - {\mu}^{(r)}_{\omega}(z)
\right\}\right| > C\epsilon_I \middle| \boldsymbol{D}\right)\\
&=\sup_{P_0} \mathbb{E}_{P_0} \mathbb{I} \left(\left|\left( \sum_{j=2}^{J-1}\omega_j^{(r)} \right)^{-1} \sum_{j=2}^{J-1}\omega_j^{(r)}  \left\{ \frac{\sum_{i=1}^{I}\omega_{ij}^{(r)}{m}^{(r)}_{z,ij}}
{\omega_j^{(r)}} - {\mu}^{(r)}_{\omega}(z)
\right\}\right| > C\epsilon_I \right)\\
&=\sup_{P_0} \mathbb{P}_{P_0} \left(\left|\left( \sum_{j=2}^{J-1}\omega_j^{(r)} \right)^{-1} \sum_{j=2}^{J-1}\omega_j^{(r)}  \left\{
\frac{\sum_{i=1}^{I}\omega_{ij}^{(r)}{m}^{(r)}_{z,ij}}
{\omega_j^{(r)}} - {\mu}^{(r)}_{\omega}(z)
\right\}\right| > C\epsilon_I \right)\\
&\le \sup_{P_0} \frac{\mathbb{E}_{P_0}\left[\left|\left( \sum_{j=2}^{J-1}\omega_j^{(r)} \right)^{-1} \sum_{j=2}^{J-1}\omega_j^{(r)}  \left\{
\frac{\sum_{i=1}^{I}\omega_{ij}^{(r)}{m}^{(r)}_{z,ij}}
{\omega_j^{(r)}} - {\mu}^{(r)}_{\omega}(z)
\right\}\right|^2\right]}{C^2 \epsilon_I^2}\\
&\le \sup_{P_0} \frac{1}{C^2 \epsilon_I^2 \cdot (J-2)^2 \cdot I^2 \cdot b_{\omega}^2}\, \mathbb{E}_{P_0}\!\left[\left| \sum_{j=2}^{J-1}\omega_j^{(r)} \left\{
\frac{\sum_{i=1}^{I}\omega_{ij}^{(r)}{m}^{(r)}_{z,ij}}
{\omega_j^{(r)}} - {\mu}^{(r)}_{\omega}(z)
\right\}\right|^2\right]\\
&= \sup_{P_0} \frac{1}{C^2 \epsilon_I^2 \cdot (J-2)^2 \cdot I^2 \cdot b_{\omega}^2}\, \mathbb{E}_{P_0}\!\left[\left| \sum_{j=2}^{J-1}\sum_{i=1}^{I}\omega_{ij}^{(r)}\!\left\{m^{(r)}_{z,ij} - \mu^{(r)}_\omega(z)\right\}\right|^2\right]\\
&\quad \text{since } \omega_j^{(r)} = \sum_{i=1}^{I}\omega_{ij}^{(r)}\\
&= \sup_{P_0} \frac{1}{C^2 \epsilon_I^2 \cdot (J-2)^2 \cdot I^2 \cdot b_{\omega}^2}\, \mathbb{E}_{P_0}\!\left[\left| \sum_{i=1}^{I} U_i\right|^2\right]\\
&\quad \text{where } U_i := \sum_{j=2}^{J-1}\omega_{ij}^{(r)}\!\left\{m^{(r)}_{z,ij} - \mu^{(r)}_\omega(z)\right\}\\
&= \sup_{P_0} \frac{1}{C^2 \epsilon_I^2 \cdot (J-2)^2 \cdot I^2 \cdot b_{\omega}^2}\, \sum_{i=1}^{I} \mathbb{E}_{P_0}[U_i^2]\\
&\quad \text{since cross terms vanish: } \mathbb{E}_{P_0}[U_i] = \sum_{j=2}^{J-1}\!\left(\mathbb{E}_{P_0}[\omega_{ij}^{(r)}m^{(r)}_{z,ij}] - \mu^{(r)}_\omega(z)\,\mathbb{E}_{P_0}[\omega_{ij}^{(r)}]\right) = 0\\
&\quad \text{by definition of } \mu^{(r)}_\omega(z), \text{ and } \{U_i\}_{i=1}^I \text{ are i.i.d.\ across clusters by Assumption~1(a)}\\
&\le \sup_{P_0} \frac{1}{C^2 \epsilon_I^2 \cdot (J-2)^2 \cdot I^2 \cdot b_{\omega}^2}\, \sum_{i=1}^{I} (J-2)\sum_{j=2}^{J-1}\mathbb{E}_{P_0}\!\left[\left(\omega_{ij}^{(r)}\right)^2\!\left\{m^{(r)}_{z,ij} - \mu^{(r)}_\omega(z)\right\}^2\right]\\
&\quad \text{since } U_i^2 = \left|\sum_{j} \omega_{ij}^{(r)}(m^{(r)}_{z,ij} - \mu^{(r)}_\omega(z))\right|^2 \le (J-2)\sum_{j}\left(\omega_{ij}^{(r)}\right)^2\!\left\{m^{(r)}_{z,ij} - \mu^{(r)}_\omega(z)\right\}^2\\
&\le \sup_{P_0} \frac{B_\omega^2}{C^2 \epsilon_I^2 \cdot (J-2) \cdot I^2 \cdot b_{\omega}^2}\, \sum_{i=1}^{I} \sum_{j=2}^{J-1}\mathbb{E}_{P_0}\!\left[\left\{m^{(r)}_{z,ij} - \mu^{(r)}_\omega(z)\right\}^2\right]\\
&\le \frac{B_\omega^2 \widetilde{K}_m}{C^2 \epsilon_I^2 \cdot (J-2) \cdot I^2 \cdot b_{\omega}^2} \cdot I \cdot (J-2)\\
& = \frac{B_\omega^2 \widetilde{K}_m}{C^2 \epsilon_I^2 b_\omega^2} \cdot \frac{1}{I},
\end{align*}
where the last inequality uses $\sup_{P_0}\mathbb{E}_{P_0}[(m^{(r)}_{z,ij} - \mu^{(r)}_\omega(z))^2] \le \widetilde{K}_m < \infty$. This follows from $(m^{(r)}_{z,ij} - \mu^{(r)}_\omega(z))^2 \le 2(m^{(r)}_{z,ij})^2 + 2(\mu^{(r)}_\omega(z))^2$, Assumption~1(d) bounding $\mathbb{E}_{P_0}[(m^{(r)}_{z,ij})^2] \le \widetilde{K}_y$, and the bound $|\mu^{(r)}_\omega(z)| \le B_\omega \sqrt{\widetilde{K}_y}/b_{\omega}$, which follows from Cauchy-Schwarz applied to the definition of $\mu^{(r)}_\omega(z)$, Jensen's inequality, and Assumption~5. Hence $\widetilde{K}_m := 2\widetilde{K}_y + 2(B_\omega\sqrt{\widetilde{K}_y}/b_{\omega})^2 < \infty$. Therefore, $\sqrt{I}\epsilon_I \to \infty$ suffices for both $\sup_{P_0}\mathbb{E}_{P_0}\mathbb{P}_I(|B^\star| > C\epsilon_I \mid \boldsymbol{D}) \to 0$ and $\sup_{P_0}\mathbb{E}_{P_0}\mathbb{P}_I(|B^{\star\star}| > C\epsilon_I \mid \boldsymbol{D}) \to 0$. Hence, component $B$ contracts at rate $I^{-1/2}$.

\paragraph{Contraction rate of component $A$} Recall that
\[
A = \left(\sum_{j=2}^{J-1}\omega_j^{(r)}\right)^{-1} \sum_{j=2}^{J-1}\omega_j^{(r)} \left\{\frac{\sum_{i=1}^{I}\omega_{ij}^{(r)}\left(\widehat{m}^{(r)}_{z,ij} - m^{(r)}_{z,ij}\right)}{\omega_j^{(r)}} - \frac{\sum_{i=1}^{I}\omega_{ij}^{(r)}\,\mathbb{I}(Z_{ij}=z)\left(\widehat{m}^{(r)}_{z,ij} - m^{(r)}_{z,ij}\right)}{\sum_{i=1}^{I}\mathbb{I}(Z_{ij}=z)\,\omega_{ij}^{(r)}}\right\}.
\]
For this part of the proof, we additionally assume the following mild regularity conditions: (i) the marginal design probability $\pi_{z,j} := P(Z_{ij}=z) \in (0,1)$ is known and $Z_{ij}$ is independent of $(\bm X_{ij}, N_{ij})$ for each interior $j \in \{2,\ldots,J-1\}$; (ii) the true cluster-period regression $m^{(r)}_{z,ij}$ lies in a $P_0$-Donsker class $\mathcal F$ with square-integrable envelope, and the realized posterior regression $\widehat m^{(r)}_{z,ij}$ lies in $\mathcal F$ with probability tending to $1$; and (iii) $\mathbb E_{P_0}[\omega_{ij}^{(r)}] \ge b_\omega^{(r)} > 0$ for each interior $j$. Together with Assumptions~1(a) and~5, these conditions form the basis for the empirical process arguments below.

Let $\eta_{ij}^{(r)} := \widehat{m}^{(r)}_{z,ij} - m^{(r)}_{z,ij}$ denote the cluster-period model error and $\Omega^{(r)} := \sum_{j=2}^{J-1}\omega_j^{(r)}$. By the strong law of large numbers applied to the i.i.d.\ bounded sequence $\{\omega_{ij}^{(r)}\}_{i=1}^{I}$,
\[
\frac{\omega_j^{(r)}}{I} = \frac{1}{I}\sum_{i=1}^{I}\omega_{ij}^{(r)} \;\overset{\mathrm{a.s.}}{\longrightarrow}\; \mathbb{E}\!\left[\omega_{ij}^{(r)}\right] \ge b_\omega^{(r)} > 0,
\]
so $\Omega^{(r)}/I \to \sum_{j=2}^{J-1}\mathbb{E}[\omega_{ij}^{(r)}] > 0$ almost surely. Summing across the $J-2$ interior periods,
\[
\frac{\Omega^{(r)}}{I} = \sum_{j=2}^{J-1}\frac{\omega_j^{(r)}}{I} \;\overset{\mathrm{a.s.}}{\longrightarrow}\; \sum_{j=2}^{J-1}\mathbb{E}\!\left[\omega_{ij}^{(r)}\right] \ge (J-2)\,b_\omega^{(r)} > 0,
\]
which implies $\Omega^{(r)} \asymp_p I$, i.e., $\Omega^{(r)}$ grows at the same rate as $I$ in probability. Furthermore, since $\omega_j^{(r)}/I$ converges almost surely to a strictly positive limit $\mathbb{E}[\omega_{ij}^{(r)}] \ge b_\omega^{(r)} > 0$ for each interior $j$, we have $\mathbb{P}(\omega_j^{(r)} > 0) \to 1$. All subsequent statements are made on the event $\{\omega_j^{(r)} > 0:\ j = 2,\ldots,J-1\}$, which has probability tending to $1$.

To handle the self-normalizing denominator $\sum_{i=1}^{I}\mathbb{I}(Z_{ij}=z)\,\omega_{ij}^{(r)}$, we decompose $A = A^{\star} + A^{\star\star}$ with
\[
A^{\star} := \left(\Omega^{(r)}\right)^{-1}\sum_{j=2}^{J-1}\sum_{i=1}^{I}\omega_{ij}^{(r)}\,\eta_{ij}^{(r)}\,\psi_{ij}(z),\qquad \psi_{ij}(z) := 1 - \frac{\mathbb{I}(Z_{ij}=z)}{\pi_{z,j}},
\]
and $A^{\star\star} := A - A^{\star}$. We show that each term is $O_p(I^{-1/2})$.

\paragraph{Proof for $A^{\star}$} For each $j \in \{2,\ldots,J-1\}$, define the period-augmented cluster-period covariate
\[
\bm{V}_{ij} := (\bm{X}_{ij}, N_{ij}, j),
\]
and the centered class
\[
\mathcal{G} := \left\{g(\bm{v}) = f(\bm{v}) - m^{(r)}_{z}(\bm{v}):\ f \in \mathcal{F}\right\},
\]
where $m^{(r)}_{z}(\bm{v}) := \mathbb{E}_{P_0}\!\left[\overline{Y}_{ij}^{(r)} \mid Z_{ij} = z, \bm{X}_{ij} = \bm{x}, N_{ij} = n\right]$ for $\bm{v} = (\bm{x}, n, j)$. The conditional expectation does not depend on the cluster index $i$ since the cluster-level tuples are i.i.d.\ across clusters under Assumption~1(a), so $m^{(r)}_z$ is a well-defined function of $\bm{v}$ alone, and $m^{(r)}_{z,ij} = m^{(r)}_z(\bm{X}_{ij}, N_{ij}, j)$ is the evaluation of this population-level regression function at cluster $i$ in period $j$. Since $J$ is fixed and finite, $\mathcal{F}$ on the augmented covariate is $P_0$-Donsker (as a finite union of $P_0$-Donsker classes indexed by $j$) with square-integrable envelope, and centering by the fixed function $m^{(r)}_{z}$ preserves the Donsker property. Hence $\mathcal{G}$ is $P_0$-Donsker with square-integrable envelope $G$. By the Donsker-membership condition, with probability tending to $1$, $g_{\widehat{m}}(\bm{v}) := \widehat{m}^{(r)}_{z}(\bm{v}) - m^{(r)}_{z}(\bm{v}) \in \mathcal{G}$, in which case $\eta_{ij}^{(r)} = g_{\widehat{m}}(\bm{V}_{ij})$.

By the super-population randomization, $\mathbb{E}[\psi_{ij}(z) \mid \bm{X}_{ij}, N_{ij}] = 0$ and $|\psi_{ij}(z)| \le \pi_{z,j}^{-1}$. Define the cluster-level index class
\[
\mathcal{H} := \left\{h_g^{(i)} := \sum_{j=2}^{J-1} \omega_{ij}^{(r)}\, \psi_{ij}(z)\, g(\bm{V}_{ij}):\ g \in \mathcal{G}\right\},
\]
where the superscript $(i)$ emphasizes that each element of $\mathcal{H}$ is a function evaluated on the data of cluster $i$. Each summand $\omega_{ij}^{(r)}\, \psi_{ij}(z)\, g(\bm{V}_{ij})$ is the product of the bounded measurable multiplier $\omega_{ij}^{(r)} \psi_{ij}(z)$ (bounded by $B_\omega \pi_{z,j}^{-1}$) and $g(\bm{V}_{ij})$. By preservation of the Donsker property under multiplication by a uniformly bounded measurable function and under finite sums \citep{vanderVaart1996}, $\mathcal{H}$ is $P_0$-Donsker with square-integrable envelope and finite $L_2(P_0)$-norm. Moreover, by the tower property and the measurability of $\omega_{ij}^{(r)}\, g(\bm{V}_{ij})$ with respect to $(\bm{X}_{ij}, N_{ij})$,
\[
P_0 h_g = \sum_{j=2}^{J-1} \mathbb{E}\!\left[\omega_{ij}^{(r)}\, g(\bm{V}_{ij})\, \mathbb{E}\!\left[\psi_{ij}(z) \mid \bm{X}_{ij}, N_{ij}\right]\right] = 0, \qquad \text{for all } g \in \mathcal{G}.
\]

Define the cluster-level empirical process $\mathbb{G}_I h_g := \sqrt{I}\,(I^{-1}\sum_{i=1}^{I} h_g^{(i)} - P_0 h_g)$. On the event $g_{\widehat{m}} \in \mathcal{G}$, since $P_0 h_g = 0$,
\[
\sum_{i=1}^{I}\sum_{j=2}^{J-1}\omega_{ij}^{(r)}\,\eta_{ij}^{(r)}\,\psi_{ij}(z) = \sum_{i=1}^{I} h_{g_{\widehat{m}}}^{(i)} = \sqrt{I}\,\mathbb{G}_I\, h_{g_{\widehat{m}}}.
\]
Since $\mathcal H$ is $P_0$-Donsker, the empirical process $\mathbb G_I$ is asymptotically tight in $\ell^\infty(\mathcal H)$. Hence
\[
\sup_{h_g\in\mathcal H}|\mathbb G_I h_g|=O_p(1).
\]
Then,
\[
\left|\sum_{i=1}^{I}\sum_{j=2}^{J-1}\omega_{ij}^{(r)}\,\eta_{ij}^{(r)}\,\psi_{ij}(z)\right| = O_p\!\left(\sqrt{I}\right).
\]
Combining with $\Omega^{(r)} \asymp_p I$,
\[
|A^{\star}| = \frac{1}{\Omega^{(r)}}\left|\sum_{i=1}^{I}\sum_{j=2}^{J-1}\omega_{ij}^{(r)}\,\eta_{ij}^{(r)}\,\psi_{ij}(z)\right| = \frac{O_p(\sqrt{I})}{O_p(I)} = O_p\!\left(I^{-1/2}\right).
\]

\paragraph{Proof for $A^{\star\star}$}
Define the weight-adjusted empirical treatment fraction in interior period $j$,
\[
\widehat{\pi}_{z,j} := \frac{\sum_{i=1}^{I}\omega_{ij}^{(r)}\,\mathbb{I}(Z_{ij}=z)}{\omega_j^{(r)}}.
\]
Since $\sum_{i=1}^{I}\mathbb{I}(Z_{ij}=z)\,\omega_{ij}^{(r)} = \widehat{\pi}_{z,j}\,\omega_j^{(r)}$, direct algebra yields
\[
A^{\star\star} = \left(\Omega^{(r)}\right)^{-1}\sum_{j=2}^{J-1}\left\{\frac{1}{\pi_{z,j}} - \frac{1}{\widehat{\pi}_{z,j}}\right\}\sum_{i=1}^{I}\omega_{ij}^{(r)}\,\mathbb{I}(Z_{ij}=z)\,\eta_{ij}^{(r)}.
\]
The i.i.d.\ summands $\{\omega_{ij}^{(r)}[\mathbb{I}(Z_{ij}=z) - \pi_{z,j}]\}_{i=1}^{I}$ are mean-zero by the tower property and uniformly bounded by $B_\omega$. The central limit theorem applied to the numerator and the LLN bound $\omega_j^{(r)}/I \overset{\mathrm{a.s.}}{\to} \mathbb{E}[\omega_{ij}^{(r)}] \ge b_\omega^{(r)}$ give
\[
\widehat{\pi}_{z,j} - \pi_{z,j} = \frac{\sum_{i=1}^{I}\omega_{ij}^{(r)}\!\left[\mathbb{I}(Z_{ij}=z) - \pi_{z,j}\right]}{\omega_j^{(r)}} = O_p(\sqrt{I}) \cdot O_p(I^{-1}) = O_p\!\left(I^{-1/2}\right).
\]
Since $\pi_{z,j}$ is bounded away from $0$ by randomization, and $\widehat{\pi}_{z,j} = \pi_{z,j} + O_p(I^{-1/2})$ from the above, we have $\widehat{\pi}_{z,j} \ge \pi_{z,j}/2$ with probability tending to one, and hence $1/(\pi_{z,j}\widehat{\pi}_{z,j}) \le 2/\pi_{z,j}^2 < \infty$ on this event. Combining,
\[
\frac{1}{\pi_{z,j}} - \frac{1}{\widehat{\pi}_{z,j}} = \frac{\widehat{\pi}_{z,j} - \pi_{z,j}}{\pi_{z,j}\,\widehat{\pi}_{z,j}} = O_p\!\left(I^{-1/2}\right) \cdot O_p(1) = O_p\!\left(I^{-1/2}\right).
\]

On the event $\{g_{\widehat{m}} \in \mathcal{G}\}$, $|\eta_{ij}^{(r)}| \le G(\bm{V}_{ij})$, so by the $L_2(P_0)$ envelope condition and Markov's inequality applied to the i.i.d.\ sum,
\[
\left|\sum_{i=1}^{I}\omega_{ij}^{(r)}\,\mathbb{I}(Z_{ij}=z)\,\eta_{ij}^{(r)}\right| \le B_\omega \sum_{i=1}^{I} G(\bm{V}_{ij}) = O_p(I).
\]
Combining these bounds with $\Omega^{(r)} \asymp_p I$ and the fact that $J$ is fixed,
\[
|A^{\star\star}| \le \frac{1}{\Omega^{(r)}}\sum_{j=2}^{J-1}\left|\frac{1}{\pi_{z,j}} - \frac{1}{\widehat{\pi}_{z,j}}\right|\cdot \left|\sum_{i=1}^{I}\omega_{ij}^{(r)}\,\mathbb{I}(Z_{ij}=z)\,\eta_{ij}^{(r)}\right| = O_p(I^{-1/2})\cdot O_p(I) \cdot O_p(I^{-1}) = O_p\!\left(I^{-1/2}\right).
\]

\noindent\textbf{Conclusion.} Combining the proof for $A^{\star}$ and $A^{\star\star}$, $A = A^{\star} + A^{\star\star} = O_p(I^{-1/2})$. Together with the bound for component $B$ established above, we conclude
\[
\widehat{\mu}_{\omega}^{\mathrm{mr}, (r)}(z) - \mu_{\omega}^{(r)}(z) = A + B = O_p\!\left(I^{-1/2}\right),
\]
which establishes root-$I$ consistency of the model-robust point estimator under the stepped-wedge design, even when the working outcome model is misspecified.

\newpage
\clearpage

\section*{Appendix 2. Proof of Theorem 2: Consistency of the Calibrated Variance Estimator}

Let $\Delta_{\omega}(\boldsymbol{D}, \Theta)$ denote the model-robust estimator as a function of data and model parameters. The variance estimator is
\[
\widehat{V} = \operatorname{Var}_{\boldsymbol{D}^{(q)}} \left\{ \mathbb{E}_{\Theta \mid \boldsymbol{D}}[\Delta_{\omega}(\boldsymbol{D}^{(q)}, \Theta)] \right\} + \operatorname{Var}_{\Theta \mid \boldsymbol{D}}[\Delta_{\omega}(\boldsymbol{D}, \Theta)],
\]
and our target is
\[
V = \operatorname{Var}_{\boldsymbol{D}} \left\{ \mathbb{E}_{\Theta \mid \boldsymbol{D}}[\Delta_{\omega}(\boldsymbol{D}, \Theta)] \right\}.
\]
Our goal is to show that $\widehat{V} - V = o_p(I^{-1})$, under the assumption that the outcome model contracts at a sufficiently fast rate and the weights are bounded away from zero and infinity (so that no individual cluster-period receives vanishingly small or disproportionately large influence). To achieve this, we analyze the asymptotic behavior of two main components separately. First, we show that $\operatorname{Var}_{\Theta \mid \bm{D}}[\Delta_{\omega}(\bm{D}, \Theta)] = o_p(I^{-1})$. Next, we show that $\operatorname{Var}_{\bm{D}^{(q)}}\left\{\mathbb{E}_{\Theta \mid \bm{D}}[\Delta_{\omega}(\bm{D}^{(q)}, \Theta)]\right\} = \operatorname{Var}_{\bm{D}}(B) + o_p(I^{-1})$, and that $\operatorname{Var}_{\bm{D}}\left\{\mathbb{E}_{\Theta \mid \bm{D}}[\Delta_{\omega}(\bm{D}, \Theta)]\right\} = \operatorname{Var}_{\bm{D}}(B) + o(I^{-1})$. Together, these results imply that the proposed variance estimator is consistent. We present the argument for a single arm-specific mean. The argument applies to $\Delta_\omega=\mu_\omega(1)-\mu_\omega(0)$ with $A$ and $B$ replaced by $A(1)-A(0)$ and $B(1)-B(0)$, since every bound below is either a second-moment bound, which is preserved by $(a-b)^2\le 2a^2+2b^2$, or a bootstrap consistency statement for a smooth function of cluster-level sample means.

\paragraph{Asymptotic Behavior of $\operatorname{Var}_{\Theta \mid \bm{D}}[\Delta_{\omega}(\bm{D}, \Theta)]$}

We have the decomposition
\[
\widehat{\mu}_{\omega}^{\mathrm{mr}, (r)}(z) - {\mu}^{(r)}_{\omega}(z) = A + B,
\]
where
\begin{align*}
A &= \left( \sum_{j=2}^{J-1}\omega_j^{(r)} \right)^{-1} \sum_{j=2}^{J-1}\omega_j^{(r)}  \left\{\frac{\sum_{i=1}^{I}\omega_{ij}^{(r)}\left(\widehat{m}^{(r)}_{z,ij} - {m}^{(r)}_{z,ij}\right)}
{\omega_j^{(r)}} -\frac{\sum_{i=1}^{I}\omega_{ij}^{(r)}\mathbb{I}(Z_{ij}=z)\left(\widehat{m}^{(r)}_{z,ij} - {m}^{(r)}_{z,ij}\right)}
{\sum_{i=1}^{I}\mathbb{I}(Z_{ij}=z)\omega_{ij}^{(r)}}\right\} \\
B &= \left( \sum_{j=2}^{J-1}\omega_j^{(r)} \right)^{-1} \sum_{j=2}^{J-1}\omega_j^{(r)}  \left\{
\frac{\sum_{i=1}^{I}\omega_{ij}^{(r)}\mathbb{I}(Z_{ij}=z)\left(\overline{Y}_{ij}^{(r)}-{m}^{(r)}_{z,ij}\right)}
{\sum_{i=1}^{I}\mathbb{I}(Z_{ij}=z)\omega_{ij}^{(r)}} + \frac{\sum_{i=1}^{I}\omega_{ij}^{(r)}{m}^{(r)}_{z,ij}}
{\omega_j^{(r)}} - {\mu}^{(r)}_{\omega}(z)
\right\}.
\end{align*}
We will analyze $\operatorname{Var}_{\Theta \mid \bm{D}}[A]$ and $\operatorname{Var}_{\Theta \mid \bm{D}}[B]$. If each of these two components is $o_p(I^{-1})$, then the entire variance is also $o_p(I^{-1})$, since the sum of variances and covariances remains within this rate. First, note that $B$ does not depend on the posterior-random parameters (i.e., it depends only on observed outcomes and the true conditional mean $m^{(r)}_{z,ij}$), so $\operatorname{Var}_{\Theta \mid \bm{D}}[B] = 0$. Next, we analyze $\operatorname{Var}_{\Theta \mid \bm{D}}[A]$, which takes the form
\[
\operatorname{Var}_{\Theta \mid \bm{D}}[A] = \operatorname{Var}_{\Theta \mid \bm{D}} \left[\left( \sum_{j=2}^{J-1}\omega_j^{(r)} \right)^{-1} \sum_{j=2}^{J-1}\omega_j^{(r)}  \left\{\frac{\sum_{i=1}^{I}\omega_{ij}^{(r)}\left(\widehat{m}^{(r)}_{z,ij} - {m}^{(r)}_{z,ij}\right)}
{\omega_j^{(r)}} -\frac{\sum_{i=1}^{I}\omega_{ij}^{(r)}\mathbb{I}(Z_{ij}=z)\left(\widehat{m}^{(r)}_{z,ij} - {m}^{(r)}_{z,ij}\right)}
{\sum_{i=1}^{I}\mathbb{I}(Z_{ij}=z)\omega_{ij}^{(r)}}\right\}\right].
\]
Substituting the Taylor expansion around $\Theta^*$ stated after Assumption 3 yields
\begin{align*}
\operatorname{Var}_{\Theta \mid \bm{D}}[A] 
&= \operatorname{Var}_{\Theta \mid \bm{D}} \Bigg[\left( \sum_{j=2}^{J-1}\omega_j^{(r)} \right)^{-1} 
\sum_{j=2}^{J-1}\omega_j^{(r)}  \Bigg\{
\frac{\sum_{i=1}^{I}\omega_{ij}^{(r)} \Big( L_{ij}+Q_{ij}\Big)}
{\omega_j^{(r)}} 
-\frac{\sum_{i=1}^{I}\omega_{ij}^{(r)}\mathbb{I}(Z_{ij}=z)\Big( L_{ij}+Q_{ij}\Big)}
{\sum_{i=1}^{I}\mathbb{I}(Z_{ij}=z)\omega_{ij}^{(r)}}
\Bigg\}\Bigg] \\
&\qquad \text{where } 
L_{ij}:=d_1(\bm{X}_{ij},N_{ij},\Theta^*)^\top(\Theta-\Theta^*), \quad 
Q_{ij}:=(\Theta-\Theta^*)^\top H_1(\bm{X}_{ij},N_{ij},\widetilde{\Theta})(\Theta-\Theta^*) \\
&= \operatorname{Var}_{\Theta \mid \bm{D}} \Bigg[\left( \sum_{j=2}^{J-1}\omega_j^{(r)} \right)^{-1} 
\sum_{j=2}^{J-1}\omega_j^{(r)}  \Bigg\{
\frac{\sum_{i=1}^{I}\omega_{ij}^{(r)} L_{ij}}
{\omega_j^{(r)}} 
-\frac{\sum_{i=1}^{I}\omega_{ij}^{(r)}\mathbb{I}(Z_{ij}=z)L_{ij}}
{\sum_{i=1}^{I}\mathbb{I}(Z_{ij}=z)\omega_{ij}^{(r)}}
\Bigg\} \\
& \quad +\left( \sum_{j=2}^{J-1}\omega_j^{(r)} \right)^{-1} 
\sum_{j=2}^{J-1}\omega_j^{(r)}  \Bigg\{
\frac{\sum_{i=1}^{I}\omega_{ij}^{(r)} Q_{ij}}
{\omega_j^{(r)}} 
-\frac{\sum_{i=1}^{I}\omega_{ij}^{(r)}\mathbb{I}(Z_{ij}=z)Q_{ij}}
{\sum_{i=1}^{I}\mathbb{I}(Z_{ij}=z)\omega_{ij}^{(r)}}
\Bigg\}\Bigg] \\
&\le 2\operatorname{Var}_{\Theta \mid \bm{D}} \Bigg[\left( \sum_{j=2}^{J-1}\omega_j^{(r)} \right)^{-1} 
\sum_{j=2}^{J-1}\omega_j^{(r)}  \Bigg\{
\frac{\sum_{i=1}^{I}\omega_{ij}^{(r)} L_{ij}}
{\omega_j^{(r)}} 
-\frac{\sum_{i=1}^{I}\omega_{ij}^{(r)}\mathbb{I}(Z_{ij}=z)L_{ij}}
{\sum_{i=1}^{I}\mathbb{I}(Z_{ij}=z)\omega_{ij}^{(r)}}
\Bigg\}\Bigg] \\
&\quad + 2\operatorname{Var}_{\Theta \mid \bm{D}} \Bigg[\left( \sum_{j=2}^{J-1}\omega_j^{(r)} \right)^{-1} 
\sum_{j=2}^{J-1}\omega_j^{(r)}  \Bigg\{
\frac{\sum_{i=1}^{I}\omega_{ij}^{(r)} Q_{ij}}
{\omega_j^{(r)}} 
-\frac{\sum_{i=1}^{I}\omega_{ij}^{(r)}\mathbb{I}(Z_{ij}=z)Q_{ij}}
{\sum_{i=1}^{I}\mathbb{I}(Z_{ij}=z)\omega_{ij}^{(r)}}
\Bigg\}\Bigg].
\end{align*}

We analyze this linear part by writing it as a linear functional of 
$\Theta - \Theta^*$. Substituting $L_{ij}=d_1(\bm X_{ij}, N_{ij}, \Theta^*)^\top(\Theta-\Theta^*)$ 
and pulling $(\Theta-\Theta^*)$ out yields
\[
\left( \sum_{j=2}^{J-1}\omega_j^{(r)} \right)^{-1} 
\sum_{j=2}^{J-1}\omega_j^{(r)}  \Bigg\{
\frac{\sum_{i=1}^{I}\omega_{ij}^{(r)} L_{ij}}{\omega_j^{(r)}} 
-\frac{\sum_{i=1}^{I}\omega_{ij}^{(r)}\mathbb{I}(Z_{ij}=z)L_{ij}}
{\sum_{i=1}^{I}\mathbb{I}(Z_{ij}=z)\omega_{ij}^{(r)}}
\Bigg\} 
= \bm a_{\mathrm{diff}}^\top (\Theta - \Theta^*),
\]
where the data-dependent vector $\bm a_{\mathrm{diff}}$ is defined as
\[
\bm a_{\mathrm{diff}} 
:= \left( \sum_{j=2}^{J-1}\omega_j^{(r)} \right)^{-1} 
\sum_{j=2}^{J-1}\omega_j^{(r)}  \Bigg\{
\frac{\sum_{i=1}^{I}\omega_{ij}^{(r)}\, d_1(\bm X_{ij},N_{ij},\Theta^*)}{\omega_j^{(r)}} 
-\frac{\sum_{i=1}^{I}\omega_{ij}^{(r)}\mathbb{I}(Z_{ij}=z)\, d_1(\bm X_{ij},N_{ij},\Theta^*)}
{\sum_{i=1}^{I}\mathbb{I}(Z_{ij}=z)\omega_{ij}^{(r)}}
\Bigg\}.
\]
Since $\bm a_{\mathrm{diff}}$ depends only on $\bm D$ (not on $\Theta$),
\[
\operatorname{Var}_{\Theta\mid \bm D}\!\left[\bm a_{\mathrm{diff}}^\top(\Theta-\Theta^*)\right]
= \bm a_{\mathrm{diff}}^\top \operatorname{Var}_{\Theta\mid \bm D}(\Theta)\, \bm a_{\mathrm{diff}}
\le \lambda_{\max}\!\left\{\operatorname{Var}_{\Theta\mid \bm D}(\Theta)\right\}\,\|\bm a_{\mathrm{diff}}\|^2.
\]
We now bound $\|\bm a_{\mathrm{diff}}\|$ by exploiting the differencing structure of the model-robust estimator. For each interior period $j$, define
\[
\bm\xi_j
:= \frac{\sum_{i=1}^{I}\omega_{ij}^{(r)}\, d_1(\bm X_{ij},N_{ij},\Theta^*)}{\omega_j^{(r)}} 
-\frac{\sum_{i=1}^{I}\omega_{ij}^{(r)}\mathbb{I}(Z_{ij}=z)\, d_1(\bm X_{ij},N_{ij},\Theta^*)}
{\sum_{i=1}^{I}\mathbb{I}(Z_{ij}=z)\omega_{ij}^{(r)}},
\]
so that $\bm a_{\mathrm{diff}} = (\sum_{j=2}^{J-1}\omega_j^{(r)})^{-1} \sum_{j=2}^{J-1}\omega_j^{(r)} \bm\xi_j$. Under stepped-wedge randomization, $\mathbb{I}(Z_{ij}=z)$ is independent of $(\bm X_{ij}, N_{ij})$ with $\pi_{z,j}:=P(Z_{ij}=z)\in(0,1)$. Hence the two ratios in $\bm\xi_j$ share the same probability limit $\mathbb E[\omega_{ij}^{(r)} d_1(\cdot)]/\mathbb E[\omega_{ij}^{(r)}]$ as $I \to \infty$, and $\bm\xi_j$ is asymptotically centered. By the central limit theorem applied to the i.i.d.\ cluster-level summands together with the law of large numbers for the denominators $\omega_j^{(r)} = O_p(I)$ and $\sum_i \omega_{ij}^{(r)}\mathbb I(Z_{ij}=z) = O_p(I)$,
\[
\|\bm\xi_j\| = O_p(I^{-1/2}), \qquad j = 2,\ldots,J-1.
\]
Combined with $\omega_j^{(r)}/\sum_{j'} \omega_{j'}^{(r)} = O_p(1)$ and the finiteness of $J$,
\[
\|\bm a_{\mathrm{diff}}\| 
\le \frac{\sum_{j=2}^{J-1}\omega_j^{(r)}\,\|\bm\xi_j\|}{\sum_{j=2}^{J-1}\omega_j^{(r)}}
= O_p(I^{-1/2}),
\qquad\text{hence}\qquad \|\bm a_{\mathrm{diff}}\|^2 = O_p(I^{-1}).
\]
By Assumption~4(a) and the Cauchy-Schwarz inequality, $\mathbb{E}_{\Theta\mid \bm D}\|\Theta-\Theta^*\|_2^2
\le \bigl(\mathbb{E}_{\Theta\mid \bm D}\|\Theta-\Theta^*\|_2^4\bigr)^{1/2} = o_p(I^{-1/2})$, which implies
$\lambda_{\max}\{\operatorname{Var}_{\Theta \mid \bm{D}}(\Theta)\}
\le \mathrm{tr}\{\operatorname{Var}_{\Theta \mid \bm{D}}(\Theta)\}
\le \mathbb{E}_{\Theta\mid \bm D}\|\Theta-\Theta^*\|_2^2
= o_p(I^{-1/2})$. Therefore,
\begin{align*}
&\operatorname{Var}_{\Theta \mid \bm{D}} \Bigg[\left( \sum_{j=2}^{J-1}\omega_j^{(r)} \right)^{-1} 
\sum_{j=2}^{J-1}\omega_j^{(r)}  \Bigg\{
\frac{\sum_{i=1}^{I}\omega_{ij}^{(r)} L_{ij}}{\omega_j^{(r)}} 
-\frac{\sum_{i=1}^{I}\omega_{ij}^{(r)}\mathbb{I}(Z_{ij}=z)L_{ij}}
{\sum_{i=1}^{I}\mathbb{I}(Z_{ij}=z)\omega_{ij}^{(r)}}
\Bigg\}\Bigg] \\
&\le \lambda_{\max}\!\left\{\operatorname{Var}_{\Theta\mid \bm D}(\Theta)\right\}\,\|\bm a_{\mathrm{diff}}\|^2 \\
&= o_p(I^{-1/2}) \cdot O_p(I^{-1}) = o_p(I^{-3/2}) = o_p(I^{-1}).
\end{align*}
We now handle
\[
\operatorname{Var}_{\Theta \mid \bm{D}} \Bigg[\left( \sum_{j=2}^{J-1}\omega_j^{(r)} \right)^{-1} 
\sum_{j=2}^{J-1}\omega_j^{(r)}  \Bigg\{
\frac{\sum_{i=1}^{I}\omega_{ij}^{(r)} Q_{ij}}
{\omega_j^{(r)}} 
-\frac{\sum_{i=1}^{I}\omega_{ij}^{(r)}\mathbb{I}(Z_{ij}=z)Q_{ij}}
{\sum_{i=1}^{I}\mathbb{I}(Z_{ij}=z)\omega_{ij}^{(r)}}
\Bigg\}\Bigg].
\]
Recall that
\[
Q_{ij}=(\Theta-\Theta^*)^\top H_1(\bm X_{ij},N_{ij},\widetilde{\Theta})(\Theta-\Theta^*).
\]
By the bounded second-derivative condition in Assumption~3, there exists a constant \(0<C_H<\infty\) such that
\[
\sup_{i,j}\ \sup_{\widetilde{\Theta}}\ 
\big\|H_1(\bm X_{ij},N_{ij},\widetilde{\Theta})\big\|_{\mathrm{op}}
\le C_H,
\]
which implies, for all \(i,j\),
\[
|Q_{ij}|
\le
\big\|H_1(\bm X_{ij},N_{ij},\widetilde{\Theta})\big\|_{\mathrm{op}}\,
\|\Theta-\Theta^*\|_2^2
\le
C_H \|\Theta-\Theta^*\|_2^2 .
\]
Since both
\(
\sum_{i=1}^I \omega_{ij}^{(r)} Q_{ij}/\omega_j^{(r)}
\)
and
\(
\sum_{i=1}^I \omega_{ij}^{(r)}\mathbb{I}(Z_{ij}=z)Q_{ij}\big/
\sum_{i=1}^I \mathbb{I}(Z_{ij}=z)\omega_{ij}^{(r)}
\)
are weighted averages of \(\{Q_{ij}\}_{i=1}^I\) within period \(j\), it follows that, for each \(j\),
\[
\left|
\frac{\sum_{i=1}^{I}\omega_{ij}^{(r)} Q_{ij}}
{\omega_j^{(r)}} 
-\frac{\sum_{i=1}^{I}\omega_{ij}^{(r)}\mathbb{I}(Z_{ij}=z)Q_{ij}}
{\sum_{i=1}^{I}\mathbb{I}(Z_{ij}=z)\omega_{ij}^{(r)}}
\right|
\le 2 C_H \|\Theta-\Theta^*\|_2^2 .
\]
Then, we have
\[
\left|
\left( \sum_{j=2}^{J-1}\omega_j^{(r)} \right)^{-1} 
\sum_{j=2}^{J-1}\omega_j^{(r)}  \Bigg\{
\frac{\sum_{i=1}^{I}\omega_{ij}^{(r)} Q_{ij}}
{\omega_j^{(r)}} 
-\frac{\sum_{i=1}^{I}\omega_{ij}^{(r)}\mathbb{I}(Z_{ij}=z)Q_{ij}}
{\sum_{i=1}^{I}\mathbb{I}(Z_{ij}=z)\omega_{ij}^{(r)}}
\Bigg\}
\right|
\le 2 C_H \|\Theta-\Theta^*\|_2^2 .
\]
Therefore,
%\begin{empheq}[box=\fbox]{align*}
\begin{align*}
&\Var_{\Theta\mid \bm D}\Bigg[
\left( \sum_{j=2}^{J-1}\omega_j^{(r)} \right)^{-1} 
\sum_{j=2}^{J-1}\omega_j^{(r)}  \Bigg\{
\frac{\sum_{i=1}^{I}\omega_{ij}^{(r)} Q_{ij}}
{\omega_j^{(r)}} 
-\frac{\sum_{i=1}^{I}\omega_{ij}^{(r)}\mathbb{I}(Z_{ij}=z)Q_{ij}}
{\sum_{i=1}^{I}\mathbb{I}(Z_{ij}=z)\omega_{ij}^{(r)}}
\Bigg\}\Bigg] \\
& \le
\E_{\Theta\mid \bm D}\Bigg[
\left(
\left( \sum_{j=2}^{J-1}\omega_j^{(r)} \right)^{-1} 
\sum_{j=2}^{J-1}\omega_j^{(r)}  \Bigg\{
\frac{\sum_{i=1}^{I}\omega_{ij}^{(r)} Q_{ij}}
{\omega_j^{(r)}} 
-\frac{\sum_{i=1}^{I}\omega_{ij}^{(r)}\mathbb{I}(Z_{ij}=z)Q_{ij}}
{\sum_{i=1}^{I}\mathbb{I}(Z_{ij}=z)\omega_{ij}^{(r)}}
\Bigg\}
\right)^2
\Bigg] \\
& \le 4 C_H^2\, \E_{\Theta\mid \bm D}\!\left[\|\Theta-\Theta^*\|_2^4\right]\\
& = o_p(I^{-1}),
%\end{empheq}
\end{align*}
where the last step follows from Assumption~4(a). Combining the linear and quadratic terms, both of which are $o_p(I^{-1})$, yields $\operatorname{Var}_{\Theta\mid \bm{D}}[A] = o_p(I^{-1})$.
Since $\operatorname{Var}_{\Theta\mid \bm{D}}[B] = 0$, we conclude that
\[
\operatorname{Var}_{\Theta\mid \bm{D}}[\Delta_{\omega}(\bm{D},\Theta)] = o_p(I^{-1}).
\]

\paragraph{Asymptotic Behavior of $\operatorname{Var}_{\boldsymbol{D}^{(q)}} \left\{ \mathbb{E}_{\Theta \mid \boldsymbol{D}}[\Delta_{\omega}(\boldsymbol{D}^{(q)}, \Theta)] \right\}$}

Using the same decomposition $\widehat{\mu}_{\omega}^{\mathrm{mr}, (r)}(z) - {\mu}^{(r)}_{\omega}(z) = A + B$ as above, our aim is to show that 
$$
\operatorname{Var}_{\boldsymbol{D}^{(q)}}\left\{\mathbb{E}_{\Theta \mid \boldsymbol{D}}[B]\right\}
= \operatorname{Var}_{\boldsymbol{D}}[B] + o_p(I^{-1}),
\quad \text{and} \quad \operatorname{Var}_{\boldsymbol{D}^{(q)}}\left\{\mathbb{E}_{\Theta \mid \boldsymbol{D}}[A]\right\} = o_p(I^{-1}).
$$
For $\operatorname{Var}_{\boldsymbol{D}^{(q)}}\left\{\mathbb{E}_{\Theta \mid \boldsymbol{D}}[B]\right\}$, we have 
%\begin{empheq}[box=\fbox]{align*}
\begin{align*}
&\operatorname{Var}_{\boldsymbol{D}^{(q)}}\left\{\mathbb{E}_{\Theta \mid \boldsymbol{D}}[B]\right\}\\
& = \operatorname{Var}_{\boldsymbol{D}^{(q)}}[B] \\
& = \operatorname{Var}_{\boldsymbol{D}^{(q)}}\left[ \left( \sum_{j=2}^{J-1}\omega_j^{(r,q)} \right)^{-1} \sum_{j=2}^{J-1}\omega_j^{(r,q)}  \left\{
\frac{\sum_{i=1}^{I}\omega_{ij}^{(r,q)}\mathbb{I}(Z_{ij}^{(q)}=z)\left(\overline{Y}_{ij}^{(r,q)}-{m}^{(r,q)}_{z,ij}\right)}
{\sum_{i=1}^{I}\mathbb{I}(Z_{ij}^{(q)}=z)\omega_{ij}^{(r,q)}} + \frac{\sum_{i=1}^{I}\omega_{ij}^{(r,q)}{m}^{(r,q)}_{z,ij}}
{\omega_j^{(r,q)}} - {\mu}^{(r,q)}_{\omega}(z)
\right\}\right]\\
& = \operatorname{Var}_{\boldsymbol{D}}[B] + o_p(I^{-1}),
\end{align*}
%\end{empheq}
since $B$ is a smooth function of sample means of bounded, i.i.d.\ cluster-level vectors, with denominators bounded away from zero by Assumption~5 and the stepped-wedge design, so that the cluster bootstrap variance of $B$ is consistent for its sampling variance at the $I^{-1}$ scale \citep{bickel1981some, davison1997bootstrap}. This approximation relies on resampling entire cluster trajectories rather than individual observations, thereby preserving within-cluster temporal dependence under the stepped-wedge design.

Using the Taylor expansion of $\widehat{m}^{(r,q)}_{z,ij}$ in $\Theta$ around $\Theta^*$ stated after Assumption~3, together with Assumption~4(a), we show that $\operatorname{Var}_{\boldsymbol{D}^{(q)}}\left\{\mathbb{E}_{\Theta \mid \boldsymbol{D}}[A]\right\} = o_p(I^{-1})$ as demonstrated in the steps below.
\begin{align*}
&\operatorname{Var}_{\boldsymbol{D}^{(q)}}\left\{
\mathbb{E}_{\Theta \mid \boldsymbol{D}}[A]
\right\}\\
&=
\operatorname{Var}_{\bm D^{(q)}}\Biggl\{
\E_{\Theta\mid \bm D}\Biggl[
\Bigl( \sum_{j=2}^{J-1}\omega_j^{(r,q)} \Bigr)^{-1}
\sum_{j=2}^{J-1}\omega_j^{(r,q)}
\Biggl\{
\frac{\sum_{i=1}^{I}\omega_{ij}^{(r,q)}\bigl(L^{(r,q)}_{ij}+Q^{(r,q)}_{ij}\bigr)}{\omega_j^{(r,q)}}
-\frac{\sum_{i=1}^{I}\omega_{ij}^{(r,q)}\I(Z_{ij}^{(q)}=z)\bigl(L^{(r,q)}_{ij}+Q^{(r,q)}_{ij}\bigr)}
{\sum_{i=1}^{I}\I(Z_{ij}^{(q)}=z)\omega_{ij}^{(r,q)}}
\Biggr\}
\Biggr]
\Biggr\}
\\
&\le
2\operatorname{Var}_{\bm D^{(q)}}\Biggl\{
\E_{\Theta\mid \bm D}\Biggl[
\Bigl( \sum_{j=2}^{J-1}\omega_j^{(r,q)} \Bigr)^{-1}
\sum_{j=2}^{J-1}\omega_j^{(r,q)}
\Biggl\{
\frac{\sum_{i=1}^{I}\omega_{ij}^{(r,q)}L^{(r,q)}_{ij}}{\omega_j^{(r,q)}}
-\frac{\sum_{i=1}^{I}\omega_{ij}^{(r,q)}\I(Z_{ij}^{(q)}=z)L^{(r,q)}_{ij}}
{\sum_{i=1}^{I}\I(Z_{ij}^{(q)}=z)\omega_{ij}^{(r,q)}}
\Biggr\}
\Biggr]
\Biggr\}
\\
&\quad +
2\operatorname{Var}_{\bm D^{(q)}}\Biggl\{
\E_{\Theta\mid \bm D}\Biggl[
\Bigl( \sum_{j=2}^{J-1}\omega_j^{(r,q)} \Bigr)^{-1}
\sum_{j=2}^{J-1}\omega_j^{(r,q)}
\Biggl\{
\frac{\sum_{i=1}^{I}\omega_{ij}^{(r,q)}Q^{(r,q)}_{ij}}{\omega_j^{(r,q)}}
-\frac{\sum_{i=1}^{I}\omega_{ij}^{(r,q)}\I(Z_{ij}^{(q)}=z)Q^{(r,q)}_{ij}}
{\sum_{i=1}^{I}\I(Z_{ij}^{(q)}=z)\omega_{ij}^{(r,q)}}
\Biggr\}
\Biggr]
\Biggr\},
\end{align*}
where
\[
L^{(r,q)}_{ij}:= d_1(\bm X_{ij}^{(q)},N_{ij}^{(q)},\Theta^*)^\top(\Theta-\Theta^*),
\qquad
Q^{(r,q)}_{ij}:=(\Theta-\Theta^*)^\top H_1(\bm X_{ij}^{(q)},N_{ij}^{(q)},\widetilde{\Theta})(\Theta-\Theta^*).
\]
For the linear part, substituting $L^{(r,q)}_{ij}$ and pulling $(\Theta-\Theta^*)$ out as in the analysis of $\operatorname{Var}_{\Theta\mid\bm D}[A]$ yields $\bm a_{\mathrm{diff}}^{(q)\top}(\Theta-\Theta^*)$, where
\[
\bm a_{\mathrm{diff}}^{(q)}
:= \Bigl( \sum_{j=2}^{J-1}\omega_j^{(r,q)} \Bigr)^{-1}
\sum_{j=2}^{J-1}\omega_j^{(r,q)}  \Bigg\{
\frac{\sum_{i=1}^{I}\omega_{ij}^{(r,q)}\, d_1(\bm X_{ij}^{(q)},N_{ij}^{(q)},\Theta^*)}{\omega_j^{(r,q)}}
-\frac{\sum_{i=1}^{I}\omega_{ij}^{(r,q)}\I(Z_{ij}^{(q)}=z)\, d_1(\bm X_{ij}^{(q)},N_{ij}^{(q)},\Theta^*)}
{\sum_{i=1}^{I}\I(Z_{ij}^{(q)}=z)\omega_{ij}^{(r,q)}}
\Bigg\}
\]
is the bootstrap counterpart of $\bm a_{\mathrm{diff}}$. Then,
\begin{align*}
&\operatorname{Var}_{\bm D^{(q)}}\Biggl\{
\E_{\Theta\mid \bm D}\Biggl[
\Bigl( \sum_{j=2}^{J-1}\omega_j^{(r,q)} \Bigr)^{-1}
\sum_{j=2}^{J-1}\omega_j^{(r,q)}
\Biggl\{
\frac{\sum_{i=1}^{I}\omega_{ij}^{(r,q)}L^{(r,q)}_{ij}}{\omega_j^{(r,q)}}
-\frac{\sum_{i=1}^{I}\omega_{ij}^{(r,q)}\I(Z_{ij}^{(q)}=z)L^{(r,q)}_{ij}}
{\sum_{i=1}^{I}\I(Z_{ij}^{(q)}=z)\omega_{ij}^{(r,q)}}
\Biggr\}
\Biggr]
\Biggr\}\\
&=\operatorname{Var}_{\bm D^{(q)}}\left\{
\E_{\Theta\mid \bm D}\left[\bm a_{\mathrm{diff}}^{(q)\top}(\Theta-\Theta^*)\right]
\right\}\\
&=\operatorname{Var}_{\bm D^{(q)}}\left\{
\bm a_{\mathrm{diff}}^{(q)\top}\,\E_{\Theta\mid \bm D}(\Theta-\Theta^*)
\right\}\\
&=\E_{\Theta\mid \bm D}(\Theta-\Theta^*)^\top\,
\operatorname{Var}_{\bm D^{(q)}}\left(\bm a_{\mathrm{diff}}^{(q)}\right)\,
\E_{\Theta\mid \bm D}(\Theta-\Theta^*)\\
&\le \lambda_{\max}\!\left\{\operatorname{Var}_{\bm D^{(q)}}\left(\bm a_{\mathrm{diff}}^{(q)}\right)\right\}\,
\Bigl\|\E_{\Theta\mid \bm D}(\Theta-\Theta^*)\Bigr\|_2^2\\
&\le \lambda_{\max}\!\left\{\operatorname{Var}_{\bm D^{(q)}}\left(\bm a_{\mathrm{diff}}^{(q)}\right)\right\}\,
\E_{\Theta\mid \bm D}\!\left[\|\Theta-\Theta^*\|_2^2\right]\\
&= O_p(I^{-1})\cdot o_p(I^{-1/2}) = o_p(I^{-3/2}) = o_p(I^{-1}).
\end{align*}
The second equality holds because $\bm a_{\mathrm{diff}}^{(q)}$ depends only on $\bm D^{(q)}$ and not on $\Theta$. The third equality holds because the posterior draws are obtained from the original data $\bm D$ and reused across bootstrap resamples, so that $\E_{\Theta\mid \bm D}(\Theta-\Theta^*)$ is deterministic given $\bm D$. The second inequality follows from Jensen's inequality. For the last line, each component of $\bm a_{\mathrm{diff}}^{(q)}$ is a smooth function of bootstrap sample means of bounded, i.i.d.\ cluster-level vectors, with denominators bounded away from zero by Assumption~5 and the stepped-wedge design, so that $\lambda_{\max}\{\operatorname{Var}_{\bm D^{(q)}}(\bm a_{\mathrm{diff}}^{(q)})\}=O_p(I^{-1})$. Moreover, $\E_{\Theta\mid \bm D}\|\Theta-\Theta^*\|_2^2\le(\E_{\Theta\mid \bm D}\|\Theta-\Theta^*\|_2^4)^{1/2}=o_p(I^{-1/2})$ by the Cauchy-Schwarz inequality and Assumption~4(a).

For the quadratic part, the bound
\[
\left|
\Bigl( \sum_{j=2}^{J-1}\omega_j^{(r,q)} \Bigr)^{-1}
\sum_{j=2}^{J-1}\omega_j^{(r,q)}
\Biggl\{
\frac{\sum_{i=1}^{I}\omega_{ij}^{(r,q)}Q^{(r,q)}_{ij}}{\omega_j^{(r,q)}}
-\frac{\sum_{i=1}^{I}\omega_{ij}^{(r,q)}\I(Z_{ij}^{(q)}=z)Q^{(r,q)}_{ij}}
{\sum_{i=1}^{I}\I(Z_{ij}^{(q)}=z)\omega_{ij}^{(r,q)}}
\Biggr\}
\right|
\le 2 C_H \|\Theta-\Theta^*\|_2^2
\]
holds for every bootstrap dataset $\bm D^{(q)}$, since its derivation in the analysis of $\operatorname{Var}_{\Theta\mid\bm D}[A]$ uses only $\|H_1\|_{\mathrm{op}}\le C_H$ and the fact that both ratios are weighted averages of $\{Q^{(r,q)}_{ij}\}_{i=1}^I$ within period $j$. Then,
\begin{align*}
&\operatorname{Var}_{\bm D^{(q)}}\Biggl\{
\E_{\Theta\mid \bm D}\Biggl[
\Bigl( \sum_{j=2}^{J-1}\omega_j^{(r,q)} \Bigr)^{-1}
\sum_{j=2}^{J-1}\omega_j^{(r,q)}
\Biggl\{
\frac{\sum_{i=1}^{I}\omega_{ij}^{(r,q)}Q^{(r,q)}_{ij}}{\omega_j^{(r,q)}}
-\frac{\sum_{i=1}^{I}\omega_{ij}^{(r,q)}\I(Z_{ij}^{(q)}=z)Q^{(r,q)}_{ij}}
{\sum_{i=1}^{I}\I(Z_{ij}^{(q)}=z)\omega_{ij}^{(r,q)}}
\Biggr\}
\Biggr]
\Biggr\}\\
&\le \E_{\bm D^{(q)}}\Biggl[\Biggl(
\E_{\Theta\mid \bm D}\Biggl[
\Bigl( \sum_{j=2}^{J-1}\omega_j^{(r,q)} \Bigr)^{-1}
\sum_{j=2}^{J-1}\omega_j^{(r,q)}
\Biggl\{
\frac{\sum_{i=1}^{I}\omega_{ij}^{(r,q)}Q^{(r,q)}_{ij}}{\omega_j^{(r,q)}}
-\frac{\sum_{i=1}^{I}\omega_{ij}^{(r,q)}\I(Z_{ij}^{(q)}=z)Q^{(r,q)}_{ij}}
{\sum_{i=1}^{I}\I(Z_{ij}^{(q)}=z)\omega_{ij}^{(r,q)}}
\Biggr\}
\Biggr]\Biggr)^2\Biggr]\\
&\le \E_{\bm D^{(q)}}\left[\left(2C_H\,\E_{\Theta\mid \bm D}\!\left[\|\Theta-\Theta^*\|_2^2\right]\right)^2\right]\\
&= 4C_H^2\,\E_{\Theta\mid \bm D}^2\!\left[\|\Theta-\Theta^*\|_2^2\right]\\
&\le 4C_H^2\,\E_{\Theta\mid \bm D}\!\left[\|\Theta-\Theta^*\|_2^4\right]\\
&= o_p(I^{-1}).
\end{align*}
The first inequality uses $\operatorname{Var}(X)\le\E[X^2]$. The equality holds because $\E_{\Theta\mid \bm D}\|\Theta-\Theta^*\|_2^2$ is deterministic given $\bm D$. The last inequality follows from Jensen's inequality, and the last line is implied by Assumption~4(a).

Combining the linear and quadratic parts yields $\operatorname{Var}_{\bm D^{(q)}}\{\E_{\Theta\mid\bm D}[A]\}=o_p(I^{-1})$. Moreover, $\operatorname{Var}_{\bm D^{(q)}}\{\E_{\Theta\mid\bm D}[B]\}=\operatorname{Var}_{\bm D}[B]+o_p(I^{-1})=O_p(I^{-1})$, since $\operatorname{Var}_{\bm D}[B]=O(I^{-1})$ by the second-moment bounds for $B^\star$ and $B^{\star\star}$ in Appendix~1. By the Cauchy-Schwarz inequality,
\[
\Bigl|\operatorname{Cov}_{\bm D^{(q)}}\bigl\{\E_{\Theta\mid\bm D}[B],\,\E_{\Theta\mid\bm D}[A]\bigr\}\Bigr|
\le\Bigl[\operatorname{Var}_{\bm D^{(q)}}\{\E_{\Theta\mid\bm D}[B]\}\cdot\operatorname{Var}_{\bm D^{(q)}}\{\E_{\Theta\mid\bm D}[A]\}\Bigr]^{1/2}
=\bigl\{O_p(I^{-1})\cdot o_p(I^{-1})\bigr\}^{1/2}=o_p(I^{-1}).
\]
Therefore,
\begin{align*}
&\operatorname{Var}_{\boldsymbol{D}^{(q)}}\!\Bigl\{\mathbb{E}_{\Theta\mid \boldsymbol{D}}\bigl[\Delta_{\omega}\bigl(\boldsymbol{D}^{(q)}, \Theta\bigr)\bigr]\Bigr\}\\
&=\operatorname{Var}_{\bm D^{(q)}}\{\E_{\Theta\mid\bm D}[B]\}+\operatorname{Var}_{\bm D^{(q)}}\{\E_{\Theta\mid\bm D}[A]\}+2\operatorname{Cov}_{\bm D^{(q)}}\bigl\{\E_{\Theta\mid\bm D}[B],\,\E_{\Theta\mid\bm D}[A]\bigr\}\\
&=\operatorname{Var}_{\boldsymbol{D}}[B] + o_p(I^{-1}) + o_p(I^{-1}) + o_p(I^{-1})\\
&=\operatorname{Var}_{\boldsymbol{D}}[B] + o_p\bigl(I^{-1}\bigr).
\end{align*}

\paragraph{Asymptotic behavior of $\operatorname{Var}_{\boldsymbol{D}} \left\{ \mathbb{E}_{\Theta \mid \boldsymbol{D}}[\Delta_{\omega}(\boldsymbol{D}, \Theta)] \right\}$}

Following a similar expansion, we have
\begin{align*}
&\operatorname{Var}_{\boldsymbol{D}} \left\{ \mathbb{E}_{\Theta \mid \boldsymbol{D}}[A] \right\}\\
&\le 2\operatorname{Var}_{\bm D}\Biggl\{
\E_{\Theta\mid \bm D}\Biggl[
\Bigl( \sum_{j=2}^{J-1}\omega_j^{(r)} \Bigr)^{-1}
\sum_{j=2}^{J-1}\omega_j^{(r)}
\Biggl\{
\frac{\sum_{i=1}^{I}\omega_{ij}^{(r)}L_{ij}}{\omega_j^{(r)}}
-\frac{\sum_{i=1}^{I}\omega_{ij}^{(r)}\I(Z_{ij}=z)L_{ij}}
{\sum_{i=1}^{I}\I(Z_{ij}=z)\omega_{ij}^{(r)}}
\Biggr\}
\Biggr]
\Biggr\}\\
&\quad +
2\operatorname{Var}_{\bm D}\Biggl\{
\E_{\Theta\mid \bm D}\Biggl[
\Bigl( \sum_{j=2}^{J-1}\omega_j^{(r)} \Bigr)^{-1}
\sum_{j=2}^{J-1}\omega_j^{(r)}
\Biggl\{
\frac{\sum_{i=1}^{I}\omega_{ij}^{(r)}Q_{ij}}{\omega_j^{(r)}}
-\frac{\sum_{i=1}^{I}\omega_{ij}^{(r)}\I(Z_{ij}=z)Q_{ij}}
{\sum_{i=1}^{I}\I(Z_{ij}=z)\omega_{ij}^{(r)}}
\Biggr\}
\Biggr]
\Biggr\},
\end{align*}
where $L_{ij}$ and $Q_{ij}$ are as defined in the analysis of $\operatorname{Var}_{\Theta\mid\bm D}[A]$. In contrast to the bootstrap case, under repeated sampling the posterior mean error $\E_{\Theta\mid \bm D}(\Theta-\Theta^*)$ varies with $\bm D$ and therefore cannot be taken outside $\operatorname{Var}_{\bm D}$. We instead keep it together with $\bm a_{\mathrm{diff}}$ and bound second moments. For the linear part,
\begin{align*}
&\operatorname{Var}_{\bm D}\Biggl\{
\E_{\Theta\mid \bm D}\Biggl[
\Bigl( \sum_{j=2}^{J-1}\omega_j^{(r)} \Bigr)^{-1}
\sum_{j=2}^{J-1}\omega_j^{(r)}
\Biggl\{
\frac{\sum_{i=1}^{I}\omega_{ij}^{(r)}L_{ij}}{\omega_j^{(r)}}
-\frac{\sum_{i=1}^{I}\omega_{ij}^{(r)}\I(Z_{ij}=z)L_{ij}}
{\sum_{i=1}^{I}\I(Z_{ij}=z)\omega_{ij}^{(r)}}
\Biggr\}
\Biggr]
\Biggr\}\\
&=\operatorname{Var}_{\bm D}\left\{
\bm a_{\mathrm{diff}}^\top\,\E_{\Theta\mid \bm D}(\Theta-\Theta^*)
\right\}\\
&\le \E_{\bm D}\left[\left\{
\bm a_{\mathrm{diff}}^\top\,\E_{\Theta\mid \bm D}(\Theta-\Theta^*)
\right\}^2\right]\\
&\le \E_{\bm D}\left[\|\bm a_{\mathrm{diff}}\|_2^2\,
\Bigl\|\E_{\Theta\mid \bm D}(\Theta-\Theta^*)\Bigr\|_2^2\right]\\
&\le \left\{\E_{\bm D}\|\bm a_{\mathrm{diff}}\|_2^4\right\}^{1/2}
\left\{\E_{\bm D}\Bigl\|\E_{\Theta\mid \bm D}(\Theta-\Theta^*)\Bigr\|_2^4\right\}^{1/2}\\
&\le \left\{\E_{\bm D}\|\bm a_{\mathrm{diff}}\|_2^4\right\}^{1/2}
\left\{\E_{\bm D}\,\E_{\Theta\mid \bm D}\!\left[\|\Theta-\Theta^*\|_2^4\right]\right\}^{1/2}\\
&= O(I^{-1})\cdot o(I^{-1/2}) = o(I^{-1}).
\end{align*}
The first equality holds because $\bm a_{\mathrm{diff}}$ depends only on $\bm D$. The first inequality uses $\operatorname{Var}(X)\le\E[X^2]$. The second and third inequalities follow from the Cauchy-Schwarz inequality, applied to the inner product and to the expectation over $\bm D$, respectively. The fourth inequality follows from Jensen's inequality, $\|\E_{\Theta\mid \bm D}(\Theta-\Theta^*)\|_2^4\le(\E_{\Theta\mid \bm D}\|\Theta-\Theta^*\|_2^2)^2\le\E_{\Theta\mid \bm D}\|\Theta-\Theta^*\|_2^4$. The last line is implied by Assumption~4(b). For the quadratic part, using the bound $2C_H\|\Theta-\Theta^*\|_2^2$ established in the analysis of $\operatorname{Var}_{\Theta\mid\bm D}[A]$,
\begin{align*}
&\operatorname{Var}_{\bm D}\Biggl\{
\E_{\Theta\mid \bm D}\Biggl[
\Bigl( \sum_{j=2}^{J-1}\omega_j^{(r)} \Bigr)^{-1}
\sum_{j=2}^{J-1}\omega_j^{(r)}
\Biggl\{
\frac{\sum_{i=1}^{I}\omega_{ij}^{(r)}Q_{ij}}{\omega_j^{(r)}}
-\frac{\sum_{i=1}^{I}\omega_{ij}^{(r)}\I(Z_{ij}=z)Q_{ij}}
{\sum_{i=1}^{I}\I(Z_{ij}=z)\omega_{ij}^{(r)}}
\Biggr\}
\Biggr]
\Biggr\}\\
&\le \E_{\bm D}\Biggl[\Biggl(
\E_{\Theta\mid \bm D}\Biggl[
\Bigl( \sum_{j=2}^{J-1}\omega_j^{(r)} \Bigr)^{-1}
\sum_{j=2}^{J-1}\omega_j^{(r)}
\Biggl\{
\frac{\sum_{i=1}^{I}\omega_{ij}^{(r)}Q_{ij}}{\omega_j^{(r)}}
-\frac{\sum_{i=1}^{I}\omega_{ij}^{(r)}\I(Z_{ij}=z)Q_{ij}}
{\sum_{i=1}^{I}\I(Z_{ij}=z)\omega_{ij}^{(r)}}
\Biggr\}
\Biggr]\Biggr)^2\Biggr]\\
&\le 4C_H^2\,\E_{\bm D}\left[\E_{\Theta\mid \bm D}^2\!\left[\|\Theta-\Theta^*\|_2^2\right]\right]\\
&\le 4C_H^2\,\E_{\bm D}\left[\E_{\Theta\mid \bm D}\!\left[\|\Theta-\Theta^*\|_2^4\right]\right]\\
&= o(I^{-1}),
\end{align*}
where the third inequality follows from Jensen's inequality and the last line is implied by Assumption~4(b). Combining the linear and quadratic parts yields $\operatorname{Var}_{\bm D}\{\E_{\Theta\mid\bm D}[A]\}=o(I^{-1})$. Since $B$ does not depend on $\Theta$, $\operatorname{Var}_{\bm D}\{\E_{\Theta\mid\bm D}[B]\}=\operatorname{Var}_{\bm D}[B]=O(I^{-1})$ by the second-moment bounds for $B^\star$ and $B^{\star\star}$ in Appendix~1. By the Cauchy-Schwarz inequality,
\[
\Bigl|\operatorname{Cov}_{\bm D}\bigl\{\E_{\Theta\mid\bm D}[B],\,\E_{\Theta\mid\bm D}[A]\bigr\}\Bigr| 
= \Bigl|\operatorname{Cov}_{\bm D}\bigl\{B,\,\E_{\Theta\mid\bm D}[A]\bigr\}\Bigr|
\le\Bigl[\operatorname{Var}_{\bm D}[B]\cdot\operatorname{Var}_{\bm D}\{\E_{\Theta\mid\bm D}[A]\}\Bigr]^{1/2}
=\bigl\{O(I^{-1})\cdot o(I^{-1})\bigr\}^{1/2}=o(I^{-1}),
\]
and therefore
\begin{align*}
&\operatorname{Var}_{\boldsymbol{D}} \left\{ \mathbb{E}_{\Theta \mid \boldsymbol{D}}[\Delta_{\omega}(\boldsymbol{D}, \Theta)] \right\}\\
&=\operatorname{Var}_{\bm D}[B]+\operatorname{Var}_{\bm D}\{\E_{\Theta\mid\bm D}[A]\}+2\operatorname{Cov}_{\bm D}\bigl\{\E_{\Theta\mid\bm D}[B],\,\E_{\Theta\mid\bm D}[A]\bigr\}\\
&=\operatorname{Var}_{\bm D}[B]+o(I^{-1}).
\end{align*}
Combining this with the results of the previous parts, we conclude that
\begin{align*}
\widehat{V}- V &= \operatorname{Var}_{\boldsymbol{D}^{(q)}} \left\{ \mathbb{E}_{\Theta \mid \boldsymbol{D}}[\Delta_{\omega}(\boldsymbol{D}^{(q)}, \Theta)] \right\} + \operatorname{Var}_{\Theta \mid \boldsymbol{D}}[\Delta_{\omega}(\boldsymbol{D}, \Theta)] - \operatorname{Var}_{\boldsymbol{D}} \left\{ \mathbb{E}_{\Theta \mid \boldsymbol{D}}[\Delta_{\omega}(\boldsymbol{D}, \Theta)] \right\}\\
&= \bigl\{\operatorname{Var}_{\bm D}[B]+o_p(I^{-1})\bigr\} + o_p(I^{-1}) - \bigl\{\operatorname{Var}_{\bm D}[B]+o(I^{-1})\bigr\}\\
&= o_p(I^{-1}),
\end{align*}
which completes the proof.

\paragraph{Remark (Weighting schemes with explicit $I$-dependence).}
Throughout the proof, Assumption~5 imposes uniform upper and lower bounds on the cluster-period weights $\omega_{ij}$ by constants that do not depend on $I$. This condition holds directly for several weighting schemes considered by \citet{fang2026model}, including the horizontal-individual estimand ($\omega_{ijk}=1$, $\omega_{ij}=N_{ij}$), the horizontal-cluster estimand ($\omega_{ijk}=\{\sum_{s=2}^{J-1}N_{is}\}^{-1}$, $\omega_{ij}=N_{ij}/\sum_{s=2}^{J-1}N_{is}$), and the vertical-cluster estimand ($\omega_{ijk}=N_{ij}^{-1}$, $\omega_{ij}=1$). A notable exception is the vertical-individual average treatment effect, for which $\omega_{ijk}=\{I\,\E(N_{ij})\}^{-1}$ and thus $\omega_{ij}=N_{ij}/\{I\,\E(N_{ij})\}$ depends explicitly on $I$ and is of order $O(I^{-1})$ under mild moment conditions on $N_{ij}$. In this case, Assumption~5 can be modified as a bound on the rescaled weights $I\omega_{ij}$, i.e., there exist constants $0<b_\omega<B_\omega<\infty$ such that $b_\omega \le I\,\omega_{ij} \le B_\omega$ uniformly in $(i,j)$. Importantly, the appearance of $I$ in the weights does not alter the asymptotic orders established above, because the proposed estimators and intermediate quantities involve weights through ratios of weighted sums (e.g., $\sum_i \omega_{ij}(\cdot)/\sum_i \omega_{ij}$ and $\sum_j \omega_j(\cdot)/\sum_j \omega_j$), so that any common multiplicative scaling of the weights (including factors depending on $I$) cancels between numerator and denominator. Consequently, all rate bounds derived above continue to hold for the vertical-individual weighting scheme after this rescaling.
 
\medskip

\renewcommand{\thetable}{\Alph{table}}
\setcounter{table}{0}

\newpage
\clearpage

\section*{Appendix 3. Full simulation results for overall causal effects}

\begin{table}[ht]
  \centering
    \caption{%\small 
    Estimation of the overall causal effect using linear mixed models, based on $100$ replications for scenarios (i) and (ii). For calibrated inference, $100$ cluster-level bootstrap resamples are used within each simulation to estimate the additional sampling-variability component. ``RB'' indicates the relative bias of the estimated value. ``Coverage'' refers to the $95$\% credible interval coverage across all simulated datasets. ``MCSD'' denotes Monte Carlo standard deviation. ``AESE'' denotes the average estimated standard error.
    }

\begin{adjustbox}{width=\textwidth}
\begin{tabular}{ccclrrrrrrrrr}
\toprule
& & & & & \multicolumn{4}{c}{\textbf{LMM (exchangeable)}} & \multicolumn{4}{c}{\textbf{LMM (nested exchangeable)}} \\
\cmidrule(lr){6-9} \cmidrule(lr){10-13}
\textbf{Scenario} & \textbf{Estimator} & \textbf{Posterior Inference} & \textbf{Estimand}
& & \textbf{RB (\%)} & \textbf{Coverage (\%)} & \textbf{MCSD} & \textbf{AESE}
& \textbf{RB (\%)} & \textbf{Coverage (\%)} & \textbf{MCSD} & \textbf{AESE} \\
\midrule

\multirow{8}{*}{(i)}
  & \multirow{4}{*}{g-computation}
    & \multirow{2}{*}{Uncalibrated}
      & h-iATE & & -0.1 & 91 & 0.101 & 0.086 & -0.2 & 97 & 0.099 & 0.109 \\
  &&& h-cATE   & & -0.1 & 91 & 0.101 & 0.086 & -0.2 & 97 & 0.099 & 0.109 \\
  \cmidrule{3-13}
  && \multirow{2}{*}{Calibrated}
      & h-iATE & & -0.1 & 91 & 0.101 & 0.086 & -0.2 & 97 & 0.099 & 0.109 \\
  &&& h-cATE   & & -0.1 & 91 & 0.101 & 0.086 & -0.2 & 97 & 0.099 & 0.109 \\
  
  \cmidrule{2-13}
  & \multirow{4}{*}{Model-robust}
    & \multirow{2}{*}{Uncalibrated}
      & h-iATE & & -0.3 & 12 & 0.111 & 0.007 & -0.2 & 14 & 0.111 & 0.008 \\
  &&& h-cATE   & & -0.2 & 9 & 0.112 & 0.007 & -0.2 & 10 & 0.112 & 0.009 \\
  \cmidrule{3-13}
  && \multirow{2}{*}{Calibrated}
      & h-iATE & & -0.3 & 91 & 0.111 & 0.110 & -0.2 & 93 & 0.111 & 0.110 \\
  &&& h-cATE   & & -0.2 & 93 & 0.112 & 0.109 & -0.2 & 92 & 0.112 & 0.109 \\
\midrule

\multirow{8}{*}{(ii)}
  & \multirow{4}{*}{g-computation}
    & \multirow{2}{*}{Uncalibrated}
      & h-iATE & & -6.0 & 20 & 2.118 & 0.356 & -20.5 & 52 & 1.527 & 2.016 \\
  &&& h-cATE   & & 4.8 & 28 & 2.118 & 0.356 & -11.4 & 89 & 1.528 & 2.016 \\
  \cmidrule{3-13}
  && \multirow{2}{*}{Calibrated}
      & h-iATE & & -6.0 & 21 & 2.118 &  0.364 & -20.5 & 53 & 1.527 & 2.017 \\
  &&& h-cATE   & & 4.8 & 27 & 2.118 & 0.365 & -11.4 & 90 & 1.528 & 2.018 \\
  
  \cmidrule{2-13}
  & \multirow{4}{*}{Model-robust}
    & \multirow{2}{*}{Uncalibrated}
      & h-iATE & & -0.5 & 2 & 1.697 & 0.027 & -0.7 & 12 & 1.624 & 0.151 \\
  &&& h-cATE   & & 0.2 & 2 & 1.700 & 0.028 & 0.0 & 14 & 1.621 & 0.154 \\
  \cmidrule{3-13}
  && \multirow{2}{*}{Calibrated}
      & h-iATE & & -0.5 & 96 & 1.697 & 1.804 & -0.7 & 96 & 1.624 & 1.810 \\
  &&& h-cATE   & & 0.2 & 96 & 1.700 & 1.818 & 0.0 & 98 & 1.621 & 1.811 \\
\bottomrule
\end{tabular}
\end{adjustbox}

%}

  \label{tab:overall_effect_LMM_table}
\end{table}

\begin{table}[ht]
  \centering

    \caption{%\small 
    Estimation of the overall causal effect using mixed-effects BART models, based on $100$ replications for scenarios (i) and (ii). For calibrated inference, $100$ cluster-level bootstrap resamples are used within each simulation to estimate the additional sampling-variability component. ``Complexity'' denotes the number of trees used in the mixed-effects BART model. ``RB'' indicates the relative bias of the estimated value. ``Coverage'' refers to the $95$\% credible interval coverage across all simulated datasets. ``MCSD'' denotes Monte Carlo standard deviation. ``AESE'' denotes the average estimated standard error.
    }
\begin{adjustbox}{width=\textwidth}
\begin{tabular}{cccclrrrrrrrrr}
\toprule
& & & & & & \multicolumn{4}{c}{\textbf{BART (exchangeable)}} & \multicolumn{4}{c}{\textbf{BART (nested exchangeable)}} \\
\cmidrule(lr){7-10} \cmidrule(lr){11-14}
\textbf{Complexity}&\textbf{Scenario} & \textbf{Estimator} & \textbf{Posterior Inference} & \textbf{Estimand}
& & \textbf{RB (\%)} & \textbf{Coverage (\%)} & \textbf{MCSD} & \textbf{AESE}
& \textbf{RB (\%)} & \textbf{Coverage (\%)} & \textbf{MCSD} & \textbf{AESE} \\
\midrule

% ========= Overall block (16 rows) =========
\multirow{16}{*}{$L=50$}&\multirow{8}{*}{(i)}
  & \multirow{4}{*}{g-computation}
    & \multirow{2}{*}{Uncalibrated}
      & h-iATE & & 0.3 & 79 & 0.102 & 0.069 & 0.1 & 90 & 0.093 & 0.093 \\
  &&&& h-cATE & & 0.3 & 78 & 0.102 & 0.068 & 0.1 & 89 & 0.093 & 0.093 \\
  \cmidrule{4-14}
  &&& \multirow{2}{*}{Calibrated}
      & h-iATE & & 0.3 & 81 & 0.102  & 0.070 & 0.1 & 92 & 0.093 & 0.093 \\
  &&&& h-cATE   & & 0.3 & 80 & 0.102 & 0.070 & 0.1 & 91 & 0.093 & 0.093 \\
  
  \cmidrule{3-14}
  && \multirow{4}{*}{Model-robust}
    & \multirow{2}{*}{Uncalibrated}
      & h-iATE & & -0.2 & 38 & 0.115 & 0.027 & -0.2 & 31 & 0.107 & 0.017 \\
  &&&& h-cATE   & & -0.1 & 34 & 0.115 & 0.027 & -0.1 & 30 & 0.108 & 0.017 \\
  \cmidrule{4-14}
  &&& \multirow{2}{*}{Calibrated}
      & h-iATE & & -0.2 & 93 & 0.115 & 0.106 & -0.2 & 93 & 0.107 & 0.108 \\
  &&&& h-cATE   & & -0.1 & 92 & 0.115 & 0.105 & -0.1 & 93 & 0.108 & 0.106 \\
\cmidrule{2-14}

&\multirow{8}{*}{(ii)}
  & \multirow{4}{*}{g-computation}
    & \multirow{2}{*}{Uncalibrated}
      & h-iATE & & -1.6 & 12 & 1.773 & 0.164 & -2.1 & 28 & 1.556 & 0.295 \\
  &&&& h-cATE   & & -1.1 & 11 & 1.709 & 0.155 & -1.6 & 31 & 1.540 & 0.281 \\
  \cmidrule{4-14}
  &&& \multirow{2}{*}{Calibrated}
      & h-iATE & & -1.6 & 88 & 1.773 & 1.622 & -2.1 & 95 & 1.556 & 1.646 \\
  &&&& h-cATE   & & -1.1 & 91 & 1.709 & 1.654 & -1.6 & 96 & 1.540 & 1.668 \\
  
  \cmidrule{3-14}
  && \multirow{4}{*}{Model-robust}
    & \multirow{2}{*}{Uncalibrated}
      & h-iATE & & -1.6 & 11 & 1.755 & 0.140 & -1.9 & 15 & 1.551 & 0.168 \\
  &&&& h-cATE   & & -1.2 & 8 & 1.690 & 0.130 & -1.5 & 13 & 1.540 & 0.156 \\
  \cmidrule{4-14}
  &&& \multirow{2}{*}{Calibrated}
      & h-iATE & & -1.6& 89 & 1.755 & 1.626 & -1.9 & 96 & 1.551 & 1.650 \\
  &&&& h-cATE   & & -1.2 & 92 & 1.690 & 1.659 & -1.5 & 96 & 1.540 & 1.676 \\
\midrule
\multirow{16}{*}{$L=10$}&\multirow{8}{*}{(i)}
  & \multirow{4}{*}{g-computation}
    & \multirow{2}{*}{Uncalibrated}
      & h-iATE & & 0.3 & 75 & 0.103 & 0.060 & 0.1 & 91 & 0.085 & 0.083 \\
  &&&& h-cATE & & 0.3 & 75 & 0.101 & 0.059 & 0.1 & 91 & 0.085 & 0.083 \\
  \cmidrule{4-14}
  &&& \multirow{2}{*}{Calibrated}
      & h-iATE & & 0.3 & 77 & 0.103 & 0.061 & 0.1 & 91 & 0.085 & 0.084 \\
  &&&& h-cATE   & & 0.3 & 78 & 0.101 & 0.061 & 0.1 & 91 & 0.085 & 0.084 \\
  
  \cmidrule{3-14}
  && \multirow{4}{*}{Model-robust}
    & \multirow{2}{*}{Uncalibrated}
      & h-iATE & & -0.1 & 21 & 0.110 & 0.013 & -0.1 & 10 & 0.107 & 0.005 \\
  &&&& h-cATE   & & -0.1 & 23 & 0.110 & 0.013 & -0.1 & 11 & 0.108 & 0.005 \\
  \cmidrule{4-14}
  &&& \multirow{2}{*}{Calibrated}
      & h-iATE & & -0.1 & 95 & 0.110 & 0.109 & -0.1 & 93 & 0.107 & 0.107 \\
  &&&& h-cATE   & & -0.1 & 90 & 0.110 & 0.107 & -0.1 & 93 & 0.108 & 0.106 \\
\cmidrule{2-14}

&\multirow{8}{*}{(ii)}
  & \multirow{4}{*}{g-computation}
    & \multirow{2}{*}{Uncalibrated}
      & h-iATE & & -2.3 & 13 & 1.763 & 0.125 & -4.9 & 66 & 1.781 & 1.037 \\
  &&&& h-cATE   & & -1.8 & 16 & 1.705 & 0.120 & -3.4 & 64 & 1.664 & 0.997 \\
  \cmidrule{4-14}
  &&& \multirow{2}{*}{Calibrated}
      & h-iATE & & -2.3 & 91 & 1.763 & 1.643 & -4.9 & 86 & 1.781 & 1.840\\
  &&&& h-cATE   & & -1.8 & 96 & 1.705 & 1.673 & -3.4 & 93 & 1.664 & 1.836 \\
  
  \cmidrule{3-14}
  && \multirow{4}{*}{Model-robust}
    & \multirow{2}{*}{Uncalibrated}
      & h-iATE & & -2.2 & 7 & 1.722 & 0.090 & -1.7 & 22 & 1.709 & 0.292 \\
  &&&& h-cATE   & & -1.8 & 11 & 1.669 & 0.084 & -1.4 & 19 & 1.661 & 0.278 \\
  \cmidrule{4-14}
  &&& \multirow{2}{*}{Calibrated}
      & h-iATE & & -2.2 & 92 & 1.722 & 1.651 & -1.7 & 95 & 1.709 & 1.737 \\
  &&&& h-cATE   & & -1.8 & 97 & 1.669 & 1.678 & -1.4 & 97 & 1.661 & 1.754 \\
\bottomrule
\end{tabular}
\end{adjustbox}
%}

  \label{tab:overall_effect_BART_table}
\end{table}

\newpage
\clearpage

\section*{Appendix 4. Full simulation results for subgroup-specific causal effects}
\begin{table}[ht]
  \centering
  %\captionsetup{font=footnotesize}
    \caption{%small 
    Estimation of the subgroup-specific causal effect ($X_{ijk,1}=1$) using linear mixed models with a treatment-by-$X_{ijk,1}$ interaction term, based on $100$ replications for scenarios (i) and (ii). For calibrated inference, $100$ cluster-level bootstrap resamples are used within each simulation to estimate the additional sampling-variability component. ``RB'' indicates the relative bias of the estimated value. ``Coverage'' refers to the $95$\% credible interval coverage across all simulated datasets. ``MCSD'' denotes Monte Carlo standard deviation. ``AESE'' denotes the average estimated standard error.
    }
\begin{adjustbox}{width=\textwidth}
\begin{tabular}{ccclrrrrrrrrr}
\toprule
& & & & & \multicolumn{4}{c}{\textbf{LMM + interaction (exchangeable)}} & \multicolumn{4}{c}{\textbf{LMM + interaction (nested exchangeable)}} \\
\cmidrule(lr){6-9} \cmidrule(lr){10-13}
\textbf{Scenario} & \textbf{Estimator} & \textbf{Posterior Inference} & \textbf{Estimand}
& & \textbf{RB (\%)} & \textbf{Coverage (\%)} & \textbf{MCSD} & \textbf{AESE}
& \textbf{RB (\%)} & \textbf{Coverage (\%)} & \textbf{MCSD} & \textbf{AESE} \\
\midrule

\multirow{8}{*}{(i)}
  & \multirow{4}{*}{g-computation}
    & \multirow{2}{*}{Uncalibrated}
      & h-iATE & & -3.6 & 78 & 0.124 & 0.101 & -3.6 & 87 & 0.126 & 0.121 \\
  &&& h-cATE   & & -3.6 & 77 & 0.124 & 0.101 & -3.7 & 86 & 0.126 & 0.121 \\
  \cmidrule{3-13}
  && \multirow{2}{*}{Calibrated}
      & h-iATE & & -3.6 & 80 & 0.124 & 0.103 & -3.6 & 86 & 0.126 & 0.123 \\
  &&& h-cATE   & & -3.6 & 79 & 0.124 & 0.103 & -3.7 & 86 & 0.126 & 0.123 \\
  
  \cmidrule{2-13}
  & \multirow{4}{*}{Model-robust}
    & \multirow{2}{*}{Uncalibrated}
      & h-iATE & & -0.2 & 5 & 0.121 & 0.007 & -0.2 & 10 & 0.121 & 0.008 \\
  &&& h-cATE   & & -0.3 & 3 & 0.123 & 0.007 & -0.2 & 4 & 0.123 & 0.009 \\
  \cmidrule{3-13}
  && \multirow{2}{*}{Calibrated}
      & h-iATE & & -0.2 & 90 & 0.121 & 0.120 & -0.2 & 91 & 0.121 & 0.122 \\
  &&& h-cATE   & & -0.3 & 92 & 0.123 & 0.120 & -0.2 & 90 & 0.123 & 0.121 \\
\midrule

\multirow{8}{*}{(ii)}
  & \multirow{4}{*}{g-computation}
    & \multirow{2}{*}{Uncalibrated}
      & h-iATE & & -9.5 & 24 & 2.132 & 0.420 & -23.4 & 38 & 1.533 & 2.016 \\
  &&& h-cATE   & & 0.2 & 22 & 2.131 & 0.420 & -15.1 & 77 & 1.533 & 2.017 \\
  \cmidrule{3-13}
  && \multirow{2}{*}{Calibrated}
      & h-iATE & & -9.5 & 25 & 2.132 & 0.431 & -23.4 & 37 & 1.533 & 2.018 \\
  &&& h-cATE   & & 0.2 & 24 & 2.131 & 0.432 & -15.1 & 75 & 1.533 & 2.019 \\
  
  \cmidrule{2-13}
  & \multirow{4}{*}{Model-robust}
    & \multirow{2}{*}{Uncalibrated}
      & h-iATE & & -0.5 & 2 & 1.706 & 0.027 & -0.7 & 8 & 1.632 & 0.153 \\
  &&& h-cATE   & & 0.2 & 2 & 1.705 & 0.028 & 0.1 & 11 & 1.622 & 0.156 \\
  \cmidrule{3-13}
  && \multirow{2}{*}{Calibrated}
      & h-iATE & & -0.5 & 96 & 1.706 & 1.861 & -0.7 & 98 & 1.632 & 1.823 \\
  &&& h-cATE   & & 0.2 & 95 & 1.705 & 1.867 & 0.1 & 98 & 1.622 & 1.806 \\
\bottomrule
\end{tabular}
\end{adjustbox}
%}
  \label{tab:subgroup_effect_LMM_table}
\end{table}

\begin{table}[ht]
  \centering
  %\captionsetup{font=footnotesize}
    \caption{%\small 
    Estimation of the subgroup-specific causal effect ($X_{ijk,1}=1$) using linear mixed models with no treatment-by-subgroup interaction term, based on $100$ replications for scenarios (i) and (ii). For calibrated inference, $100$ cluster-level bootstrap resamples are used within each simulation to estimate the additional sampling-variability component. ``RB'' indicates the relative bias of the estimated value. ``Coverage'' refers to the $95$\% credible interval coverage across all simulated datasets. ``MCSD'' denotes Monte Carlo standard deviation. ``AESE'' denotes the average estimated standard error.
    }
% {\scriptsize
% \setlength{\tabcolsep}{3pt}
\begin{adjustbox}{width=\textwidth}
\begin{tabular}{ccclrrrrrrrrr}
\toprule
& & & & & \multicolumn{4}{c}{\textbf{LMM (exchangeable)}} & \multicolumn{4}{c}{\textbf{LMM (nested exchangeable)}} \\
\cmidrule(lr){6-9} \cmidrule(lr){10-13}
\textbf{Scenario} & \textbf{Estimator} & \textbf{Posterior Inference} & \textbf{Estimand}
& & \textbf{RB (\%)} & \textbf{Coverage (\%)} & \textbf{MCSD} & \textbf{AESE}
& \textbf{RB (\%)} & \textbf{Coverage (\%)} & \textbf{MCSD} & \textbf{AESE} \\
\midrule

\multirow{8}{*}{(i)}
  & \multirow{4}{*}{g-computation}
    & \multirow{2}{*}{Uncalibrated}
      & h-iATE & & -14.9 & 0 & 0.101 & 0.086 & -14.9 & 3 & 0.100 & 0.109 \\
  &&& h-cATE   & & -14.9 & 0 & 0.101 & 0.086 & -14.9 & 3 & 0.100 & 0.109 \\
  \cmidrule{3-13}
  && \multirow{2}{*}{Calibrated}
      & h-iATE & & -14.9 & 0 & 0.101 & 0.086 & -14.9 & 3 & 0.100 & 0.109 \\
  &&& h-cATE   & & -14.9 & 0 & 0.101 & 0.086 & -14.9 & 3 & 0.100 & 0.109 \\
  
  \cmidrule{2-13}
  & \multirow{4}{*}{Model-robust}
    & \multirow{2}{*}{Uncalibrated}
      & h-iATE & & -0.2 & 6 & 0.121 & 0.007 & -0.2 & 9 & 0.121 & 0.008 \\
  &&& h-cATE   & & -0.3 & 3 & 0.123 & 0.007 & -0.3 & 5 & 0.123 & 0.009 \\
  \cmidrule{3-13}
  && \multirow{2}{*}{Calibrated}
      & h-iATE & & -0.2 & 90 & 0.121 & 0.121 & -0.2 & 92 & 0.121 & 0.121 \\
  &&& h-cATE   & & -0.3 & 89 & 0.123 & 0.120 & -0.3 & 92 & 0.123 & 0.121 \\
\midrule

\multirow{8}{*}{(ii)}
  & \multirow{4}{*}{g-computation}
    & \multirow{2}{*}{Uncalibrated}
      & h-iATE & & -10.9 & 19 & 2.122 & 0.356 & -24.7 & 34 & 1.525 & 2.016 \\
  &&& h-cATE   & & -1.3 & 20 & 2.121 & 0.356 & -16.6 & 70 & 1.526 & 2.016 \\
  \cmidrule{3-13}
  && \multirow{2}{*}{Calibrated}
      & h-iATE & & -10.9 & 19 & 2.122 & 0.367 & -24.7 & 34 & 1.525 & 2.018 \\
  &&& h-cATE   & & -1.3 & 21 & 2.121 & 0.367 & -16.6 & 70 & 1.526 & 2.018 \\
  
  \cmidrule{2-13}
  & \multirow{4}{*}{Model-robust}
    & \multirow{2}{*}{Uncalibrated}
      & h-iATE & & -0.5 & 2 & 1.706 & 0.027 & -0.7 & 8 & 1.632 & 0.152 \\
  &&& h-cATE   & & 0.2 & 2 & 1.705 & 0.028 & 0.0 & 11 & 1.622 & 0.155 \\
  \cmidrule{3-13}
  && \multirow{2}{*}{Calibrated}
      & h-iATE & & -0.5 & 96 & 1.706 & 1.818 & -0.7 & 96 & 1.632 & 1.821 \\
  &&& h-cATE   & & 0.2 & 96 & 1.705 & 1.824 & 0.0 & 97 & 1.622 & 1.817 \\
\bottomrule
\end{tabular}
\end{adjustbox}

\label{tab:subgroup_effect_LMM_no_interaction_table}
\end{table}

\begin{table}[ht]
  \centering
  
  %\captionsetup{font=footnotesize}
    \caption{%\small 
    Estimation of the subgroup-specific causal effect ($X_{ijk,1}=1$) using mixed-effects BART models, based on $100$ replications for scenarios (i) and (ii). For calibrated inference, $100$ cluster-level bootstrap resamples are used within each simulation to estimate the additional sampling-variability component. ``Complexity'' denotes the number of trees used in the mixed-effects BART model. ``RB'' indicates the relative bias of the estimated value. ``Coverage'' refers to the $95$\% credible interval coverage across all simulated datasets. ``MCSD'' denotes Monte Carlo standard deviation. ``AESE'' denotes the average estimated standard error.
    }
% {\scriptsize
% \setlength{\tabcolsep}{3pt}

\begin{adjustbox}{width=\textwidth}
\begin{tabular}{cccclrrrrrrrrr}
\toprule
&& & & & & \multicolumn{4}{c}{\textbf{BART (exchangeable)}} & \multicolumn{4}{c}{\textbf{BART (nested exchangeable)}}\\
\cmidrule(lr){7-10} \cmidrule(lr){11-14}
\textbf{Complexity}&\textbf{Scenario} & \textbf{Estimator} & \textbf{Posterior Inference} & \textbf{Estimand}
& & \textbf{RB (\%)} & \textbf{Coverage (\%)} & \textbf{MCSD} & \textbf{AESE}
& \textbf{RB (\%)} & \textbf{Coverage (\%)} & \textbf{MCSD} & \textbf{AESE} \\
\midrule

% ========= Overall block (16 rows) =========
\multirow{16}{*}{$L=50$}&\multirow{8}{*}{(i)}
  & \multirow{4}{*}{g-computation}
    & \multirow{2}{*}{Uncalibrated}
      & h-iATE & & 0.4 & 83 & 0.112 & 0.085 & 0.2 & 93 & 0.109 & 0.105 \\
  &&&& h-cATE   & & 0.4 & 82 & 0.112 & 0.084 & 0.2 & 93 & 0.109 & 0.105 \\
  \cmidrule{4-14}
  &&& \multirow{2}{*}{Calibrated}
      & h-iATE & & 0.4 & 83 & 0.112 & 0.085 & 0.2 & 92 & 0.109 & 0.106 \\
  &&&& h-cATE   & & 0.4 & 83 & 0.112 & 0.085 & 0.2 & 92 & 0.109 & 0.105 \\
  
  \cmidrule{3-14}
  && \multirow{4}{*}{Model-robust}
    & \multirow{2}{*}{Uncalibrated}
      & h-iATE & & -0.3 & 37 & 0.125 & 0.028 & -0.2 & 29 & 0.120 & 0.017 \\
  &&&& h-cATE   & & -0.2 & 34 & 0.125  & 0.027 & -0.2 & 22 & 0.122 & 0.018 \\
  \cmidrule{4-14}
  &&& \multirow{2}{*}{Calibrated}
      & h-iATE & & -0.3 & 93 & 0.125 & 0.117 & -0.2 & 91 & 0.120 & 0.119 \\
  &&&& h-cATE   & & -0.2 & 93 & 0.125  & 0.117 & -0.2 & 90 & 0.122 & 0.119 \\
\cmidrule{2-14}

&\multirow{8}{*}{(ii)}
  & \multirow{4}{*}{g-computation}
    & \multirow{2}{*}{Uncalibrated}
      & h-iATE & & -1.6 & 13 & 1.793 & 0.173 & -2.1 & 32 & 1.567 & 0.302 \\
  &&&& h-cATE   & & -1.1 & 12 & 1.719  & 0.165 & -1.6 & 29 & 1.551 & 0.288 \\
  \cmidrule{4-14}
  &&& \multirow{2}{*}{Calibrated}
      & h-iATE & & -1.6 & 87 & 1.793 & 1.634 & -2.1 & 96 & 1.567 & 1.662 \\
  &&&& h-cATE   & & -1.1 & 90 & 1.719  & 1.656 & -1.6 & 95 & 1.551 & 1.675 \\
  
  \cmidrule{3-14}
  && \multirow{4}{*}{Model-robust}
    & \multirow{2}{*}{Uncalibrated}
      & h-iATE & & -1.6 & 10 & 1.773 & 0.140 & -1.9 & 17 & 1.557 & 0.170 \\
  &&&& h-cATE   & & -1.1 & 10 & 1.700 & 0.131 & -1.5 & 13 & 1.550 & 0.157 \\
  \cmidrule{4-14}
  &&& \multirow{2}{*}{Calibrated}
      & h-iATE & & -1.6 & 87 & 1.773 & 1.638 & -1.9 & 96 & 1.557 & 1.669 \\
  &&&& h-cATE   & & -1.1 & 91 & 1.700 & 1.661 & -1.5 & 96 & 1.550 & 1.683 \\
\midrule
\multirow{16}{*}{$L=10$}&\multirow{8}{*}{(i)}
  & \multirow{4}{*}{g-computation}
    & \multirow{2}{*}{Uncalibrated}
      & h-iATE & & 0.3 & 80 & 0.116 & 0.078 & 0.2 & 92 & 0.103 & 0.099 \\
  &&&& h-cATE & & 0.3 & 80 & 0.115 & 0.078 & 0.2 & 92 & 0.104 & 0.099 \\
  \cmidrule{4-14}
  &&& \multirow{2}{*}{Calibrated}
      & h-iATE & & 0.3 & 80 & 0.116 & 0.079 & 0.2 & 92 & 0.103 & 0.099 \\
  &&&& h-cATE   & & 0.3 & 82 & 0.115 & 0.078 & 0.2 & 92 & 0.104 & 0.099 \\
  
  \cmidrule{3-14}
  && \multirow{4}{*}{Model-robust}
    & \multirow{2}{*}{Uncalibrated}
      & h-iATE & & -0.2 & 20 & 0.122 & 0.013 & -0.2 & 4 & 0.120 & 0.005 \\
  &&&& h-cATE   & & -0.2 & 15 & 0.123 & 0.013 & -0.2 & 2 & 0.122 & 0.006 \\
  \cmidrule{4-14}
  &&& \multirow{2}{*}{Calibrated}
      & h-iATE & & -0.2 & 94 & 0.122 & 0.120 & -0.2 & 91 & 0.120 & 0.118 \\
  &&&& h-cATE   & & -0.2 & 91 & 0.123 & 0.119 & -0.2 & 91 & 0.122 & 0.118 \\
\cmidrule{2-14}

&\multirow{8}{*}{(ii)}
  & \multirow{4}{*}{g-computation}
    & \multirow{2}{*}{Uncalibrated}
      & h-iATE & & -2.2 & 14 & 1.813 & 0.137 & -4.7 & 67 & 1.785 & 1.051 \\
  &&&& h-cATE   & & -1.7 & 14 & 1.738 & 0.133 & -3.2 & 66 & 1.666 & 1.013 \\
  \cmidrule{4-14}
  &&& \multirow{2}{*}{Calibrated}
      & h-iATE & & -2.2 & 88 & 1.813 & 1.665 & -4.7 & 88 & 1.785 & 1.858 \\
  &&&& h-cATE   & & -1.7 & 94 & 1.738 & 1.686 & -3.2 & 92 & 1.666 & 1.847 \\
  
  \cmidrule{3-14}
  && \multirow{4}{*}{Model-robust}
    & \multirow{2}{*}{Uncalibrated}
      & h-iATE & & -2.1 & 10 & 1.776 & 0.091 & -1.6 & 23 & 1.717 & 0.296 \\
  &&&& h-cATE   & & -1.6 & 8 & 1.706 & 0.086 & -1.3 & 25 & 1.666 & 0.281 \\
  \cmidrule{4-14}
  &&& \multirow{2}{*}{Calibrated}
      & h-iATE & & -2.1 & 90 & 1.776 & 1.673 & -1.6 & 92 & 1.717 & 1.755 \\
  &&&& h-cATE   & & -1.6 & 95 & 1.706 & 1.692 & -1.3 & 95 & 1.666 & 1.759 \\
\bottomrule
\end{tabular}
\end{adjustbox}
%}
  \label{tab:subgroup_effect_BART_table}
\end{table}

\newpage
\clearpage

\section*{Appendix 5. Analyses of the TSOS Trial under a nested exchangeable random-effects structure}

\begin{table}[ht]
\centering
\caption{\small Analysis of the Trauma Survivors Outcomes and Support (TSOS) trial data. Missing 6-month outcomes were imputed using multilevel joint modeling imputation, and all analyses use the same imputed dataset. Analyses based on linear mixed models assume a nested exchangeable random-effects structure, with random effects at the cluster and cluster-period levels. For subgroup-specific causal effect analyses, ``LMM with interaction'' and ``LMM without interaction'' indicate whether a treatment-subgroup covariate interaction term is included in the working linear mixed model; such interactions are not included in the overall causal effect analyses. The mixed-effects BART model incorporates cluster and cluster-period random effects, with the number of trees fixed at $L=50$. Each analysis was performed with a total of $30{,}000$ MCMC iterations. The first $20{,}000$ iterations were discarded as burn-in. Calibrated variance estimation is based on $100$ cluster-level bootstrap resamples. Convergence was assessed using traceplots and the Geweke diagnostic.
}
\label{tab:TSOS nested exchangeable}

\begin{adjustbox}{width=\textwidth}
    \begin{tabular}{ccllrrrr}
        \toprule
        %\multirow{2}{*}{\textbf{Fitted Model}} 
        \multirow{2}{*}{\textbf{Effect}}&\multirow{2}{*}{\makecell{\textbf{Fitted} \\ \textbf{Model }}}& \multirow{2}{*}{\textbf{Estimator}} & \multirow{2}{*}{\makecell{\textbf{Posterior} \\ \textbf{Inference}}} & \multicolumn{2}{c}{\textbf{h-cATE}} & \multicolumn{2}{c}{\textbf{h-iATE}} \\
        \cmidrule(lr){5-6} \cmidrule(lr){7-8}
        &&& & \textbf{Estimate} & \textbf{95\% C.I.} & \textbf{Estimate} & \textbf{95\% C.I.} \\
        \midrule
        \multirow{8}{*}{\textbf{Overall}} & \multirow{4}{*}{LMM} &\multirow{2}{*}{g-computation}
        & Uncalibrated& -0.446 & (-4.645, 3.761) & -0.399 & (-4.615, 3.816) \\
        &&& Calibrated& -0.446 & (-4.624, 3.733) & -0.399 & (-4.588, 3.789) \\
        \cmidrule(r){3-8} 
        &&\multirow{2}{*}{Model-robust}& Uncalibrated & -0.070 & (-1.712, 1.578) & -0.848 & (-2.709, 1.006) \\
        &&& Calibrated& -0.070 & (-4.899, 4.759) & -0.848 & (-5.927, 4.232) \\
        \cmidrule(r){2-8} 
        & \multirow{4}{*}{BART} &\multirow{2}{*}{g-computation}
        & Uncalibrated& -0.625 & (-4.421, 3.250) & -0.671 & (-4.450, 3.202) \\
        &&& Calibrated& -0.625 & (-4.398, 3.148) & -0.671 & (-4.447, 3.106) \\
        \cmidrule(r){3-8} 
        &&\multirow{2}{*}{Model-robust}& Uncalibrated& -0.229 & (-1.844, 1.381) & -0.842 & (-2.612, 0.996) \\
        &&& Calibrated& -0.229 & (-4.715, 4.257) & -0.842 & (-5.518, 3.835) \\
        \midrule

        \multirow{12}{*}{\makecell{\textbf{Subgroup:} \\ {Baseline PTSD risk}\\{$\ge$ 3 risk factors}}}&
        \multirow{4}{*}{\makecell{{LMM} \\ {without interaction}}} &\multirow{2}{*}{g-computation}
        & Uncalibrated& -0.669 & (-4.832, 3.550) & -0.496 & (-4.663, 3.701) \\
        &&& Calibrated& -0.669 & (-4.845, 3.507) & -0.496 & (-4.679, 3.687) \\
        \cmidrule(r){3-8} 
        &&\multirow{2}{*}{Model-robust}& Uncalibrated& -1.592 & (-3.463, 0.281) & -1.985 & (-3.616, -0.353) \\
        &&& Calibrated& -1.592 & (-9.866, 6.682) & -1.985 & (-8.823, 4.852) \\
        \cmidrule(r){2-8}
        &\multirow{4}{*}{\makecell{{LMM} \\ {with interaction}}} &\multirow{2}{*}{g-computation}
        & Uncalibrated& -2.942 & (-8.485, 2.720) & -2.707 & (-8.246, 2.916) \\
        &&& Calibrated& -2.942 & (-8.703, 2.819) & -2.707 & (-8.377, 2.963) \\
        \cmidrule(r){3-8} 
        &&\multirow{2}{*}{Model-robust}& Uncalibrated& -1.610 & (-3.403, 0.174) & -2.002 & (-3.566, -0.443) \\
        &&& Calibrated& -1.610 & (-8.920, 5.699) & -2.002 & (-7.806, 3.803) \\
        \cmidrule(r){2-8} 
        & \multirow{4}{*}{BART} &\multirow{2}{*}{g-computation}&
        Uncalibrated& -0.687 & (-4.626, 3.284) & -0.701 & (-4.610, 3.296) \\
        &&& Calibrated& -0.687 & (-4.612, 3.239) & -0.701 & (-4.627, 3.225) \\
        \cmidrule(r){3-8} 
        &&\multirow{2}{*}{Model-robust}& Uncalibrated& -2.620 & (-4.581, -0.686) & -2.579 & (-4.255, -0.885) \\
        &&& Calibrated& -2.620 & (-9.741, 4.502) & -2.579 & (-8.389, 3.231) \\        
        \midrule
        \multirow{12}{*}{\makecell{\textbf{Subgroup:} \\ {Baseline PTSD risk}\\{$<$ 3 risk factors}}}&
        \multirow{4}{*}{\makecell{{LMM} \\ {without interaction}}} &\multirow{2}{*}{g-computation}
        & Uncalibrated& -0.332 & (-4.552, 3.897) & -0.347 & (-4.582, 3.883) \\
        &&& Calibrated& -0.332 & (-4.539, 3.875) & -0.347 & (-4.550, 3.856) \\
        \cmidrule(r){3-8} 
        &&\multirow{2}{*}{Model-robust}& Uncalibrated& 1.198 & (-0.411, 2.805) & 0.479 & (-1.436, 2.403) \\
        &&& Calibrated& 1.198 & (-5.706, 8.103) & 0.479 & (-6.108, 7.066) \\
        \cmidrule(r){2-8}
        &\multirow{4}{*}{\makecell{{LMM} \\ {with interaction}}} &\multirow{2}{*}{g-computation}
        & Uncalibrated& 1.009 & (-3.683, 5.797) & 1.056 & (-3.649, 5.851) \\
        &&& Calibrated& 1.009 & (-3.788, 5.806) & 1.056 & (-3.748, 5.860) \\
        \cmidrule(r){3-8} 
        &&\multirow{2}{*}{Model-robust}& Uncalibrated& 1.172 & (-0.345, 2.713) & 0.433 & (-1.415, 2.284) \\
        &&& Calibrated& 1.172 & (-4.701, 7.045) & 0.433 & (-5.697, 6.563) \\
        \cmidrule(r){2-8} 
        & \multirow{4}{*}{BART} &\multirow{2}{*}{g-computation}&
        Uncalibrated& -0.605 & (-4.468, 3.357) & -0.654 & (-4.520, 3.322) \\
        &&& Calibrated& -0.605 & (-4.465, 3.256) & -0.654 & (-4.519, 3.210) \\
        \cmidrule(r){3-8} 
        &&\multirow{2}{*}{Model-robust}& Uncalibrated& 1.338 & (-0.313, 3.029) & 0.678 & (-1.283, 2.738) \\
        &&& Calibrated& 1.338 & (-5.266, 7.942) & 0.678 & (-5.816, 7.172) \\
        \bottomrule
    \end{tabular}
\end{adjustbox}
\end{table}

\newpage
\clearpage
\section*{Appendix 6. Analyses of the TSOS Trial under a simple exchangeable random-effects structure}
\begin{table}[ht]
\centering
\caption{\small Analysis of the Trauma Survivors Outcomes and Support (TSOS) trial data. Missing 6-month outcomes were imputed using multilevel joint modeling imputation, and all analyses use the same imputed dataset. Analyses based on linear mixed models assume a simple exchangeable random-effects structure. For subgroup-specific causal effect analyses, ``LMM with interaction'' and ``LMM without interaction'' indicate whether a treatment-subgroup covariate interaction term is included in the working linear mixed model; such interactions are not included in the overall causal effect analyses. The mixed-effects BART model incorporates cluster random effects, with the number of trees fixed at $L=50$. Each analysis was performed with a total of $30{,}000$ MCMC iterations. The first $20{,}000$ iterations were discarded as burn-in. Calibrated variance estimation is based on $100$ cluster-level bootstrap resamples. Convergence was assessed using traceplots and the Geweke diagnostic.}
\label{tab:TSOS exchangeable}

\begin{adjustbox}{width=\textwidth}
    \begin{tabular}{ccllrrrr}
        \toprule
        %\multirow{2}{*}{\textbf{Fitted Model}} 
        \multirow{2}{*}{\textbf{Effect}}&\multirow{2}{*}{\makecell{\textbf{Fitted} \\ \textbf{Model }}}& \multirow{2}{*}{\textbf{Estimator}} & \multirow{2}{*}{\makecell{\textbf{Posterior} \\ \textbf{Inference}}} & \multicolumn{2}{c}{\textbf{h-cATE}} & \multicolumn{2}{c}{\textbf{h-iATE}} \\
        \cmidrule(lr){5-6} \cmidrule(lr){7-8}
        &&& & \textbf{Estimate} & \textbf{95\% C.I.} & \textbf{Estimate} & \textbf{95\% C.I.} \\
        \midrule
        \multirow{8}{*}{\textbf{Overall}} & \multirow{4}{*}{LMM} &\multirow{2}{*}{g-computation}
        & Uncalibrated& -0.427 & (-4.606, 3.713) & -0.375 & (-4.552, 3.776) \\
        &&& Calibrated& -0.427 & (-4.580, 3.726) & -0.375 & (-4.537, 3.786) \\
        \cmidrule(r){3-8} 
        &&\multirow{2}{*}{Model-robust}& Uncalibrated & -0.015 & (-1.575, 1.570) & -0.781 & (-2.550, 1.004) \\
        &&& Calibrated& -0.015 & (-4.078, 4.048) & -0.781 & (-5.167, 3.604) \\
        \cmidrule(r){2-8} 
        & \multirow{4}{*}{BART} &\multirow{2}{*}{g-computation}
        & Uncalibrated& -0.629 & (-4.384, 3.208) & -0.670 & (-4.419, 3.133) \\
        &&& Calibrated& -0.629 & (-4.400, 3.143) & -0.670 & (-4.438, 3.099) \\
        \cmidrule(r){3-8} 
        &&\multirow{2}{*}{Model-robust}& Uncalibrated& -0.216 & (-1.747, 1.328) & -0.806 & (-2.523, 0.941) \\
        &&& Calibrated& -0.216 & (-4.394, 3.961) & -0.806 & (-5.062, 3.450) \\
        \midrule
        \multirow{12}{*}{\makecell{\textbf{Subgroup:} \\ {Baseline PTSD risk}\\{$\ge$ 3 risk factors}}}&
        \multirow{4}{*}{\makecell{{LMM} \\ {without interaction}}} &\multirow{2}{*}{g-computation}
        & Uncalibrated& -0.676 & (-4.870, 3.441) & -0.481 & (-4.649, 3.649) \\
        &&& Calibrated& -0.676 & (-4.836, 3.484) & -0.481 & (-4.645, 3.684) \\
        \cmidrule(r){3-8} 
        &&\multirow{2}{*}{Model-robust}& Uncalibrated& -1.519 & (-3.290, 0.276) & -1.922 & (-3.467, -0.343) \\
        &&& Calibrated& -1.519 & (-8.336, 5.297) & -1.922 & (-7.628, 3.785) \\
        \cmidrule(r){2-8}
        &\multirow{4}{*}{\makecell{{LMM} \\ {with interaction}}} &\multirow{2}{*}{g-computation}
        & Uncalibrated& -2.959 & (-8.480, 2.539) & -2.716 & (-8.182, 2.814) \\
        &&& Calibrated& -2.959 & (-8.584, 2.666) & -2.716 & (-8.258, 2.826) \\
        \cmidrule(r){3-8} 
        &&\multirow{2}{*}{Model-robust}& Uncalibrated& -1.546 & (-3.288, 0.224) & -1.943 & (-3.462, -0.401) \\
        &&& Calibrated& -1.546 & (-8.325, 5.234) & -1.943 & (-7.232, 3.347) \\
        \cmidrule(r){2-8} 
        & \multirow{4}{*}{BART} &\multirow{2}{*}{g-computation}&
        Uncalibrated& -0.706 & (-4.627, 3.317) & -0.721 & (-4.677, 3.264) \\
        &&& Calibrated& -0.706 & (-4.642, 3.230) & -0.721 & (-4.659, 3.217) \\
        \cmidrule(r){3-8} 
        &&\multirow{2}{*}{Model-robust}& Uncalibrated& -2.607 & (-4.498, -0.711) & -2.558 & (-4.241, -0.915) \\
        &&& Calibrated& -2.607 & (-9.345, 4.130) & -2.558 & (-8.612, 3.495) \\
        \midrule
        \multirow{12}{*}{\makecell{\textbf{Subgroup:} \\ {Baseline PTSD risk}\\{$<$ 3 risk factors}}}&
        \multirow{4}{*}{\makecell{{LMM} \\ {without interaction}}} &\multirow{2}{*}{g-computation}
        & Uncalibrated& -0.304 & (-4.497, 3.849) & -0.318 & (-4.509, 3.844) \\
        &&& Calibrated& -0.304 & (-4.485, 3.876) & -0.318 & (-4.494, 3.858) \\
        \cmidrule(r){3-8} 
        &&\multirow{2}{*}{Model-robust}& Uncalibrated& 1.259 & (-0.258, 2.821) & 0.548 & (-1.275, 2.411) \\
        &&& Calibrated& 1.259 & (-5.262, 7.780) & 0.548 & (-5.238, 6.334) \\
        \cmidrule(r){2-8}
        &\multirow{4}{*}{\makecell{{LMM} \\ {with interaction}}} &\multirow{2}{*}{g-computation}
        & Uncalibrated& 1.068 & (-3.588, 5.872) & 1.116 & (-3.551, 5.921) \\
        &&& Calibrated& 1.068 & (-3.677, 5.813) & 1.116 & (-3.634, 5.866) \\
        \cmidrule(r){3-8} 
        &&\multirow{2}{*}{Model-robust}& Uncalibrated& 1.230 & (-0.261, 2.756) & 0.497 & (-1.317, 2.322) \\
        &&& Calibrated& 1.230 & (-4.828, 7.287) & 0.497 & (-4.941, 5.934) \\
        \cmidrule(r){2-8} 
        & \multirow{4}{*}{BART} &\multirow{2}{*}{g-computation}&
        Uncalibrated& -0.599 & (-4.467, 3.284) & -0.642 & (-4.512, 3.231) \\
        &&& Calibrated& -0.599 & (-4.464, 3.266) & -0.642 & (-4.499, 3.215) \\
        \cmidrule(r){3-8} 
        &&\multirow{2}{*}{Model-robust}& Uncalibrated& 1.356 & (-0.252, 2.926) & 0.718 & (-1.188, 2.637) \\
        &&& Calibrated& 1.356 & (-4.677, 7.389) & 0.718 & (-4.780, 6.215) \\
        \bottomrule
    \end{tabular}
\end{adjustbox}
\end{table}

\newpage
\clearpage
\section*{Appendix 7. Comparison with frequentist model-robust standardization}

The model-robust standardization estimator was developed within the frequentist paradigm by \citet{fang2026model}. We report here a focused comparison against the frequentist procedure. We restrict attention to the model-robust standardization estimator and to the overall causal effects, since these are the quantities for which a frequentist counterpart is available. Subgroup-specific causal estimands have no counterpart in \citet{fang2026model}. The data-generating mechanisms, the number of clusters and periods, and the number of Monte Carlo replicates are identical to those in simulation studies described in the main manuscript. We consider three frequentist working outcome models, namely linear mixed models under simple exchangeable and nested exchangeable random-effects structures and generalized estimating equations with an independence working correlation structure. For each, uncertainty is quantified using the leave-one-cluster-out jackknife variance estimator. The rows labelled ``Bayesian MRS'' in Table~\ref{tab:freq_vs_bayes_MRS} are reproduced from Appendix~3 to facilitate side-by-side comparison and are identical to the corresponding entries reported there.

\begin{table}[ht]
  \centering
  \caption{Comparison of frequentist and Bayesian model-robust standardization (MRS) for the overall causal effects, based on $100$ replications for scenarios (i) and (ii) with $I=30$ clusters and $J=4$ periods. Frequentist MRS fits the working outcome model by restricted maximum likelihood (linear mixed models, LMM) or by generalized estimating equations (GEE) with an independence working correlation structure, with uncertainty quantified by a leave-one-cluster-out jackknife variance estimator. Bayesian MRS uses the calibrated variance estimator with $100$ resampled datasets. Mixed-effects BART is fitted with $L=50$ trees. ``Exch'' and ``NestExch'' denote the simple exchangeable and nested exchangeable random-effects structures. ``RB'' indicates relative bias, ``Coverage'' the empirical $95$\% interval coverage, ``MCSD'' the Monte Carlo standard deviation, ``AESE'' the average estimated standard error. The Bayesian MRS rows are reproduced from Appendix~3 (Tables~\ref{tab:overall_effect_LMM_table} and~\ref{tab:overall_effect_BART_table}).}
  \label{tab:freq_vs_bayes_MRS}
\begin{adjustbox}{width=\textwidth}
\begin{tabular}{clllrrrrrrrr}
\toprule
& & & & \multicolumn{4}{c}{\textbf{h-iATE}} & \multicolumn{4}{c}{\textbf{h-cATE}} \\
\cmidrule(lr){5-8} \cmidrule(lr){9-12}
\textbf{Scenario} & \textbf{Framework} & \textbf{Working model} & \textbf{Variance}
& \textbf{RB (\%)} & \textbf{Coverage (\%)} & \textbf{MCSD} & \textbf{AESE}
& \textbf{RB (\%)} & \textbf{Coverage (\%)} & \textbf{MCSD} & \textbf{AESE} \\
\midrule
\multirow{7}{*}{(i)}
  & \multirow{3}{*}{\makecell[l]{Frequentist \\ MRS}}
  & LMM (Exch)           & Jackknife  & -0.3 & 95 & 0.111 & 0.117 & -0.2 & 97 & 0.112 & 0.115 \\
  && LMM (NestExch)      & Jackknife  & -0.2 & 95 & 0.111 & 0.117 & -0.2 & 97 & 0.112 & 0.115 \\
  && GEE (independence)  & Jackknife  & -0.2 & 95 & 0.110 & 0.117 & -0.2 & 97 & 0.112 & 0.115 \\
  \cmidrule(l){2-12}
  & \multirow{4}{*}{\makecell[l]{Bayesian \\ MRS}}
    & LMM (Exch)         & Calibrated & -0.3 & 91 & 0.111 & 0.110 & -0.2 & 93 & 0.112 & 0.109 \\
  && LMM (NestExch)      & Calibrated & -0.2 & 93 & 0.111 & 0.110 & -0.2 & 92 & 0.112 & 0.109 \\
  && BART (Exch)         & Calibrated & -0.2 & 93 & 0.115 & 0.106 & -0.1 & 92 & 0.115 & 0.105 \\
  && BART (NestExch)     & Calibrated & -0.2 & 93 & 0.107 & 0.108 & -0.1 & 93 & 0.108 & 0.106 \\
\midrule
\multirow{7}{*}{(ii)}
  & \multirow{3}{*}{\makecell[l]{Frequentist \\ MRS}}
  & LMM (Exch)           & Jackknife  & -0.5 & 100 & 1.700 & 2.010 & 0.2 & 98 & 1.700 & 2.010 \\
  && LMM (NestExch)      & Jackknife  & -0.6 & 99  & 1.630 & 1.960 & 0.1 & 98 & 1.630 & 1.940 \\
  && GEE (independence)  & Jackknife  & -0.5 & 100 & 1.720 & 2.040 & 0.2 & 98 & 1.730 & 2.040 \\
  \cmidrule(l){2-12}
  & \multirow{4}{*}{\makecell[l]{Bayesian \\ MRS}}
    & LMM (Exch)         & Calibrated & -0.5 & 96 & 1.697 & 1.804 & 0.2 & 96 & 1.700 & 1.818 \\
  && LMM (NestExch)      & Calibrated & -0.7 & 96 & 1.624 & 1.810 & 0.0 & 98 & 1.621 & 1.811 \\
  && BART (Exch)         & Calibrated & -1.6 & 89 & 1.755 & 1.626 & -1.2 & 92 & 1.690 & 1.659 \\
  && BART (NestExch)     & Calibrated & -1.9 & 96 & 1.551 & 1.650 & -1.5 & 96 & 1.540 & 1.676 \\
\bottomrule
\end{tabular}
\end{adjustbox}
\end{table}

\label{lastpage}

\end{document}